\documentclass[aps,epsf,amsmath,superscriptaddress,citeautoscript,onecolumn,
showpacs,floatfix]{revtex4-2}
\usepackage{bm}
\usepackage{amssymb}
\usepackage{graphicx}
\usepackage{amsmath, amsthm, amssymb, graphicx}
\usepackage[usenames]{color}
\graphicspath{{figures/}} %Setting the graphicspath
\usepackage{ulem}

\usepackage{amsmath,amsthm,amsfonts,amscd,amssymb}
\usepackage{eucal} 	 	% Euler fonts
\usepackage{verbatim}      	% Allows quoting source with commands.
\usepackage{makeidx}       	% Package to make an index.
\usepackage{braket}

\usepackage{natbib}
\usepackage{bibentry}
\usepackage{hyperref}

\usepackage{mathrsfs}
\usepackage{graphicx}% Include figure files
\usepackage{dcolumn}% Align table columns on decimal point
\usepackage{bm}% bold math
\begin{document}
	
	\preprint{}
	
	\title{The Infra-Red Road to Quantum Gravity }

\author{S.Tajik}
\affiliation{Physics and Astronomy, University of British Columbia, 6224 Agricultural
Rd. Vancouver, B.C., Canada, V6T 1Z1}
\author{M.J. Desrochers}
\affiliation{Physics and Astronomy, University of British Columbia, 6224 Agricultural
Rd. Vancouver, B.C., Canada, V6T 1Z1}
\author{P.C.E. Stamp}
\affiliation{Physics and Astronomy, University of British Columbia, 6224 Agricultural
Rd. Vancouver, B.C., Canada, V6T 1Z1}
\affiliation{Pacific Institute of Theoretical Physics, Univ. of British Columbia, 6224
Agricultural Rd. Vancouver, B.C., Canada, V6T 1Z1}

\begin{abstract}

We review work in areas ranging from condensed matter physics to quantum gravity, with the following interconnected questions in mind: (i) what is the nature of the vacuum in condensed matter systems, in quantum field theory, and in classical and quantum gravity; (ii) how do analogies between these systems work, how well do they work, and how useful are they; (iii) what modifications can we make to quantum mechanics to deal with quantum gravity, and (iv) how and why low-energy theories of quantum gravity are, in our view, the right way to make progress in this field. We use many different examples to illustrate our arguments.

\end{abstract}

\maketitle

% Preview source code from paragraph 57 to 115

%%%%%%%%%%%%%%%%%%%%%%%%%%%%%%%%%%%%%%%%%%%%%%%%%%%%%%%%%%%%%%%%%%%%%%%%%%%%%%%%%%%%%%
%%%%%%%%%%%%%%%%%%%%%%%%%%%%%%%%%%%%%%%%%%%%%%%%%%%%%%%%%%%%%%%%%%%%%%%%%%%%%%%%%%%%%%

\section*{1) Introduction}

%%%%%%%%%%%%%%%%%%%%%%%%%%%%%%%%%%%%%%%%%%%%%%%%%%%%%%%%%%%%%%%%%%%%%%%%%%%%%%%%%%%%%%
%%%%%%%%%%%%%%%%%%%%%%%%%%%%%%%%%%%%%%%%%%%%%%%%%%%%%%%%%%%%%%%%%%%%%%%%%%%%%%%%%%%%%%

This article is written to honour A.Starobinsky \cite{staro16} whose remarkable work in cosmology has fundamentally influenced the subject. Indeed, it now appears that the Starobinsky version \cite{staro-infl} of inflation theory – which was the very first version, and which thus started the whole field – may in fact be the only version that has survived the verdict of observational test. The inflation scenario is one of the more remarkable examples of the application – to the entire universe – of a more fundamental theory that we don’t yet have, viz. a widely accepted theory of quantum gravity. Such a theory is supposed to somehow unify General Relativity (GR) with Quantum Field Theory (QFT), in a third theory which reduces to one or another in the appropriate limits.

Starobinksy himself brought to bear on these questions a very interesting
background. His training followed the remarkable Soviet scientific
ethos of that time - a broad understanding of the whole of physics,
exemplified in the famous Landau-Lifshitz course of 10 volumes in
theoretical physics. In fact Starobinsky's first research
experience \cite{staro16} was in work with the well-known
solid-state theorist Ilya M.Lifshitz (brother of Evgenii M. Lifshitz),
on the theory of disordered conductors (a subject developed in the
Soviet Union by Lifshitz in the early 1960's, in work following that
of P.W. Anderson in 1958).

Starobinsky then fell under the influence of Ya. B. Zeldovich, a theorist who was himself largely self-taught, having
left school at 16 years old. Zeldovich began as a theoretical chemist,
before doing pioneering work in theoretical physics, relativistic astrophysics,
and cosmology (and on the way, developing with A.D. Sakharov the theory
behind Soviet nuclear weapons).

Starobinsky's early work in quantum gravity and cosmology anticipated
much of the work that is commonly highlighted in the West. This
included work on ``acceleration radiation''
by rotating objects, along with Zeldovich \cite{staro+YaZ}, and of course his invention
of inflation theory (before the work of Guth and Linde). In
recorded discussions one of us had \cite{staro16}
with Starobinsky and several Russian colleagues in 2016, Starobinsky
expressed a long-standing interest in
the physics of decoherence processes in both condensed matter and
cosmology, including their role in the early universe. He planned
with PCES an extended visit to UBC in 2020. Unfortunately and tragically, COVID
intervened to prevent this.

This article involves the interests of all three co-authors in topics
ranging from superfluids on earth, all the way to quantum gravity
and cosmology - all of these topics interested Starobinsky.

The common theme running through all the topics addressed here is
the application of quantum mechanics (QM) and quantum field
theory (QFT) to macroscopic systems, from the lab to the universe.
We begin by asking about
the nature of the vacuum in physics. Our discussion of this question
will range from 2d and 3d superfluids, and disordered insulators,
up to systems at the cosmological scale. Along the way we decided to address questions about (a) analogies in physics between phenomena in condensed matter systems and in quantum and classical gravity; and (b) the foundations of QM. This leads finally  to our original intention, viz., (c) to discuss the search for a consistent theory of quantum gravity (QGr). We argue that most theory, which has been looking at Planck-scale physics, has simply been looking in the wrong place, and that the key is to look for a low-energy replacement for quantum mechanics, in which QM breaks down at macroscopic scales, because of gravity; and we examine a consistent theory of this kind (the CWL theory).

\vspace{4mm}

%%%%%%%%%%%%%%%%%%%%%%%%%%%%%%%%%%%%%%%%%%%%%%%%%%%%%%%%%%%%%%%%%%%%%%%%%%%
%%%%%%%%%%%%%%%%%%%%%%%%%%%%%%%%%%%%%%%%%%%%%%%%%%%%%%%%%%%%%%%%%%%%%%%%%%%

\section*{2) Nature of the Vacuum in Physics}

%%%%%%%%%%%%%%%%%%%%%%%%%%%%%%%%%%%%%%%%%%%%%%%%%%%%%%%%%%%%%%%%%%%%%%%%%%%
%%%%%%%%%%%%%%%%%%%%%%%%%%%%%%%%%%%%%%%%%%%%%%%%%%%%%%%%%%%%%%%%%%%%%%%%%%%

We begin with a simple physical question. Suppose we have a large
cubical box of volume $L^{3}$, containing just a single H atom, which sits in one of
the huge voids between galactic superclusters. What is the vacuum state of
this system?

\begin{figure}
		\includegraphics[width=7.2in]{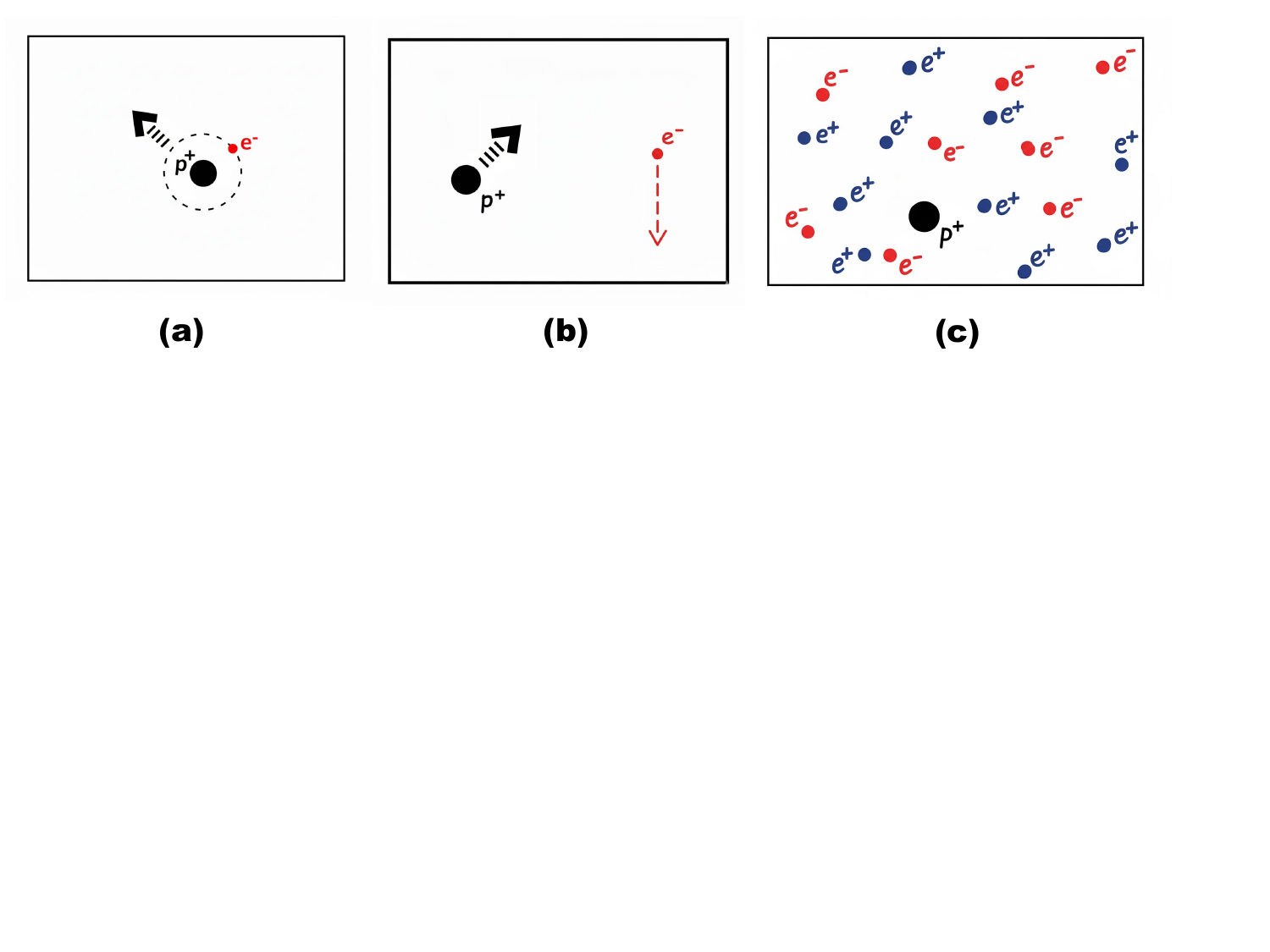}
	\vspace{-9.0cm}
		\caption{\label{fig:TDS1-box} Thought experiment for a $H$ atom in a box. In (a) we show the physical vacuum at $T=0$; one has a simple atom in its ground state. In (b) we shown the same vacuum at $T=2.7$K; for a large enough box the atom dissociates. In (c) we show the vacuum when $T-10^{10}$K; it is now full of thermally excited $e^+/e^-$ pairs. Finally, in (d) we show the system at $T=0$ but with a very strong electric field $E_o \sim 10^{18}~V/m$ applied. The vacuum is now unstable, with $e^+/e^-$ pairs dissociating out of the vacuum, the $H$ atom completely dissociated, and all particle accelerating to infinity.   }
			\vspace{4mm}
	\end{figure}

The answer seems obvious. If the box is at $T=0$, one has a bound $H$ atom in its ground
state, with orbital state being in the lowest energy state of translational
motion (with energy $3\hbar^{2}\pi^{2}/2ML^2$, where
$M$ is the H atom mass). Suppose however, the box is $T=2.7$K (as it will be, since we assume it is in thermal equilibrium with teh microwave background)?
Then, apart from low-energy photons in tghe box, in equilibrium with the $H$ atom, we will find that the atom in a mixture of states, in which it has
partially dissociated into separate $\text{e}^{+}$ and $e^{-}$
particles, according to the Saha equation. What
now if the temperature is $10^{10}$K? Then the QED vacuum will
be full of $e^{+}/e^{-}$ pairs, which will strongly modify its dielectric
properties, along with a dense sea of very hot photons. The situation is depicted in Fig. \ref{fig:TDS1-box}.

	%%%%%%%%%%%%%%%%%%%%%%%%%%%%%
	
	%%%%%%%%%%%%%%%%%%%%%%%%%%%%%

One may object that while we are clearly describing the \uline{physical} vacuum here, we are not describing the theoretical construct which we also call the ``vacuum'' or ``ground'' state, unless we are at $T=0$.
Fine; but suppose we do now go back to $T=0$, and now apply an electric field $E$? Then, according to the Schwinger 1-loop formula \cite{schwing51}, one
sees $e^{+}/e^{-}$ pairs being spontaneously produced by vacuum tunneling, so that there is no well-defined vacuum at all. So, what do we mean by
the vacuum?

In what follows we discuss the vacuum in several condensed matter
systems, finishing with a discussion of the vacuum in classical General
Relativity (GR), and in relativistic Quantum Field Theory (QFT). We will discuss some of the key conceptual questions for each vacuum, and discuss how one can expose these vacua, and see some of its hidden properties, using various probes. These probes include external current flow (for superfluids), strong electric fields (for QED), and a combination of singularities (like black holes) and low-energy experiments (for GR). In the next section (section 3), we spend some time discussing the role of analogies in physics, using the examples discussed in this section (section 2).

%%%%%%%%%%%%%%%%%%%%%%%%%%%%%%%%%%%%%%%%%%%%%%%%%%%%%%%%%%%%%%%%%%%%%%%%%%%%%%%%%%%
\subsection*{2.1) Vacuum in Condensed Matter Systems}
%%%%%%%%%%%%%%%%%%%%%%%%%%%%%%%%%%%%%%%%%%%%%%%%%%%%%%%%%%%%%%%%%%%%%%%%%%%%%%%%%%%

The above discussion shows that even a simple 1-particle problem becomes
a many-body problem when one looks at it properly. So what does condensed
matter theory say about all of this?

%%%%%%%%%%%%%%%%%%%%%%%%%%%%%%%%%%%%%%%%%%%%
\subsubsection*{2.1(a): General Remarks}
%%%%%%%%%%%%%%%%%%%%%%%%%%%%%%%%%%%%%%%%%%%%

Let's make a few general remarks - these are by and large well-knwn to any condensed matter physicist. In condensed matter systems, we can say that:

(i) One never deals with the true N-body states: they are far too
complicated, and there are far too many of them. Thus the density of
states $D\left(E\right)$ for a very small grain of diameter $d\sim0.4~$mm,
and mass $M\sim M_{\text{p}}=2\times10^{-8}$kg (so $N\sim10^{18}$),
goes like $D\left(E\right)\sim e^{S\left(E\right)}$, where the entropy
$S\left(E\right)\sim N\left(E/E_{0}\right)^{\frac{3}{4}}$ and $E_{0}\sim3\pi^{2}\hbar^{2}/2M_{\text{p}}d^{2}$.
Even for a small molecule, the number of states is beyond astronomical, even at very low energies.

(ii): Instead, one deals with an ``effective'' theory and an ``effective
Hamiltonian'' $H_{\text{eff}},$ written in terms of ``quasiparticles''
or some other kind of effective excitation. The effective Hilbert
space is drastically restricted, and the effective vacuum $\left|0\right>_{\text{eff}}$
is the ground state of $H_{\text{eff}}$.

(iii): Unfortunately, the real behaviour of the system may be quite
different from that of $H_{\text{eff}}$. A quick glance shows that
we are surrounded, not by perfect fluids or perfect crystals, but
instead by highly disordered systems. For such systems any putative
ground state is physically meaningless - the system would need a time
$\sim\mathcal{O}\left(\tau_{0}\exp\left(N^{\alpha}\right)\right)$ to reach it, if it exists at all (here $\alpha\simeq2-3$ , and $\tau_{0}$ is a microscopic timescale). On any meaningful timescale, the
physical behaviour is entirely controlled by excitations in a narrow
band of many-body energies, usually far above the ``ground state" in energy.

(iv): Only two condensed matter systems exist where one can meaningfully
discuss, and do experiments on, the vacuum state. These are superfluid
$^{4}$He and superfluid $^{3}$He, with the superfluidity discovered in 1937 and 1972 respectively.
Because only $^{3}$He dissolves in $^{4}$He , one can flush $^{4}$He
to fantastic purity; it has been understood since the 1950's that
it can be treated at very low $T$ as, for all practical purposes,
a perfect vacuum (the mean free path of quasiparticles, at the lowest
attainable temperatures in these two superfluids, is many light years).

The example of superfluid $^3$He is interesting because it has often been claimed
that it can provide an analogy for processes in cosmology. Analogies between radically differing physical systems have been employed throughout the
history of physics. Famous examples include Maxwell's analogy between the EM
field and different mechanical systems; the analogies which were employed,
in the early history of GR, between the spacetime metric field and
various fluid, and superfluid, and solid media; and an enormous number of
analogies made between various condensed matter media and the fields in
QFT - perhaps the most famous being the Anderson-Higgs boson \cite{PWA63}.

%%%%%%%%%%%%%%%%%%%%%%%%%%%%%%%%%%%%%%%%%%%%%
\subsubsection*{2.1(b): Key Points about Superfluids}
%%%%%%%%%%%%%%%%%%%%%%%%%%%%%%%%%%%%%%%%%%%%%

Here we recall some key points about (uncharged) superfluids that have a bearing on the topics in this article.

\vspace{2mm}

\begin{figure}
		\includegraphics[width=7.2in]{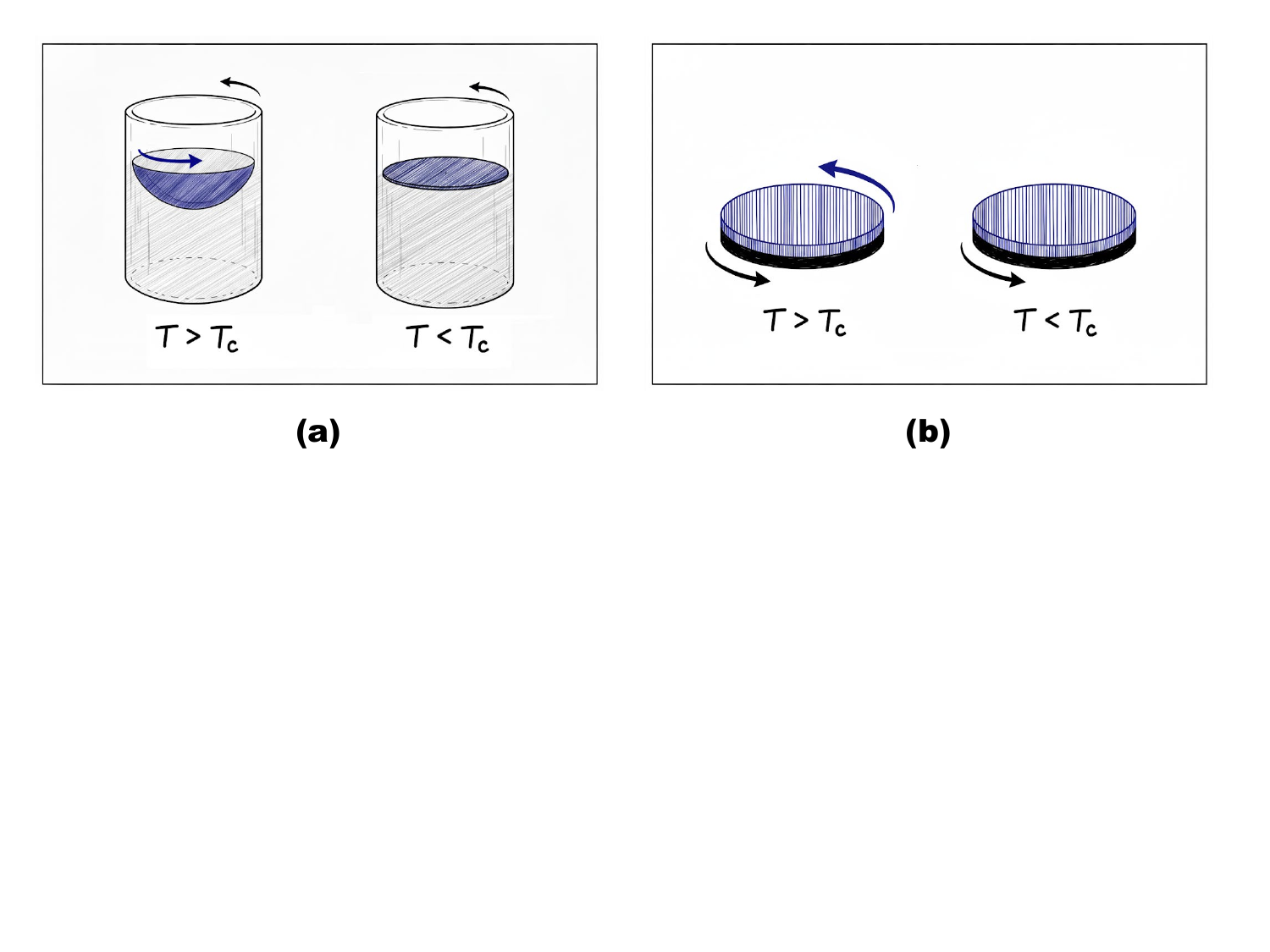}
\vspace{-7.6cm}
		\caption{\label{fig:TDS2-rotB} Operational definition of a superfluid. In (a) we show the equilibrium state of a superfluid 3-d rotating cylindrical bucket. If $T > T_c$, the superfluid transition temperature, then the normal fluid rotated with the cylinder (and shows a parabolic depression of the fluid surface). Below $T_c$, if the angular velocity $\Omega < \Omega_c$, where $\Omega_c$ is a critical angular velocity, then the superfluid is stationary in a local inertial frame. When $\Omega > \Omega_c$, the equilibrium state has single quantized vortices present. In (b) we show the same situation for a 2-d superfluid on a rotating disc, where the same argument applies, \uline{{\bf provided}} no vortex/anti-vortex pairs are excited - these pairs destabilize the superfluid state, even at $T=0$. Note that in (b), we do not show the parabolic depression of the film surface that occurs for a rotating disc above $T_c$.     }
	\vspace{4mm}
	\end{figure}

{\bf (i) Definition of Superfluidity}:  Operationally, this is provided by the
``Hess-Fairbank" effect \cite{hessF67}, which is the analogue for a neutral superfluid of the Meissner effect. One imagines cooling a rotating container of normal fluid (a cylinder in 3D, a circular disc in 2D), which while still in the normal state is precisely equivalent to Newton's famous ``rotating bucket" thought experiment. The 3D normal fluid will rotate with the bucket, and show a parabolic depression on the liquid surface. The 2 superfluid will rotate with the disc. However, as soon as we go through the superfluid transition, the fluid ``expels rotation" (see Fig. \ref{fig:TDS2-rotB}), and thereby comes to rest in a local inertial frame, even while the containers continue to rotate (indeed their angular velocity increases!).

	%%%%%%%%%%%%%%%%%%%%%%%%%%%%%
	
	%%%%%%%%%%%%%%%%%%%%%%%%%%%%%

We note that ``superflow", ie., the flow of the superfluid without viscosity, is not
part of the fundamental definition of a superfluid, but rather an ancillary concomitant
of it (and when vortices are in the system, one has dissipative superflow - see below). For many years superfluidity was thought to be a consequence of ``Off-Diagonal
Long- Range Order" (ODLRO), along with Bose condensation (BEC), in the superfluid density
matrix \cite{onsP56,yang61}. However ODLRO and BEC are actually unnecessary for superfluidity - indeed  2D superfluids have neither ODLRO nor BEC at any temperature. The definition of
superfluidity is more subtle for a 2D superfluid - we return to this below.

So far we have not discussed vortices. The system can lower its energy, above a critical rotation velocity, by having quantized vortices. These enter the superfluid from the walls, either by thermal activation (for higher $T$), or by tunneling (at very low $T$). This is a process we will call ``extrinsic vortex nucleation". The system is still superfluid, but it no longer shows superflow - the flow is now dissipative, because the superflow drives vortex motion, taking energy from the superflow, and unless one drives the superflow externally, it decays with time.  

Much more difficult is the process of ``intrinsic vortex nucleation", in which the vortex nucleates in the bulk superfluid, far from any containing boundary \cite{PNAS25}. In fact it has apparently never been seen - we will have a lot to say about it in this article, as well as its analogues in particle physics and quantum gravity.

\vspace{2mm}

\begin{figure}
		\includegraphics[width=7.4in]{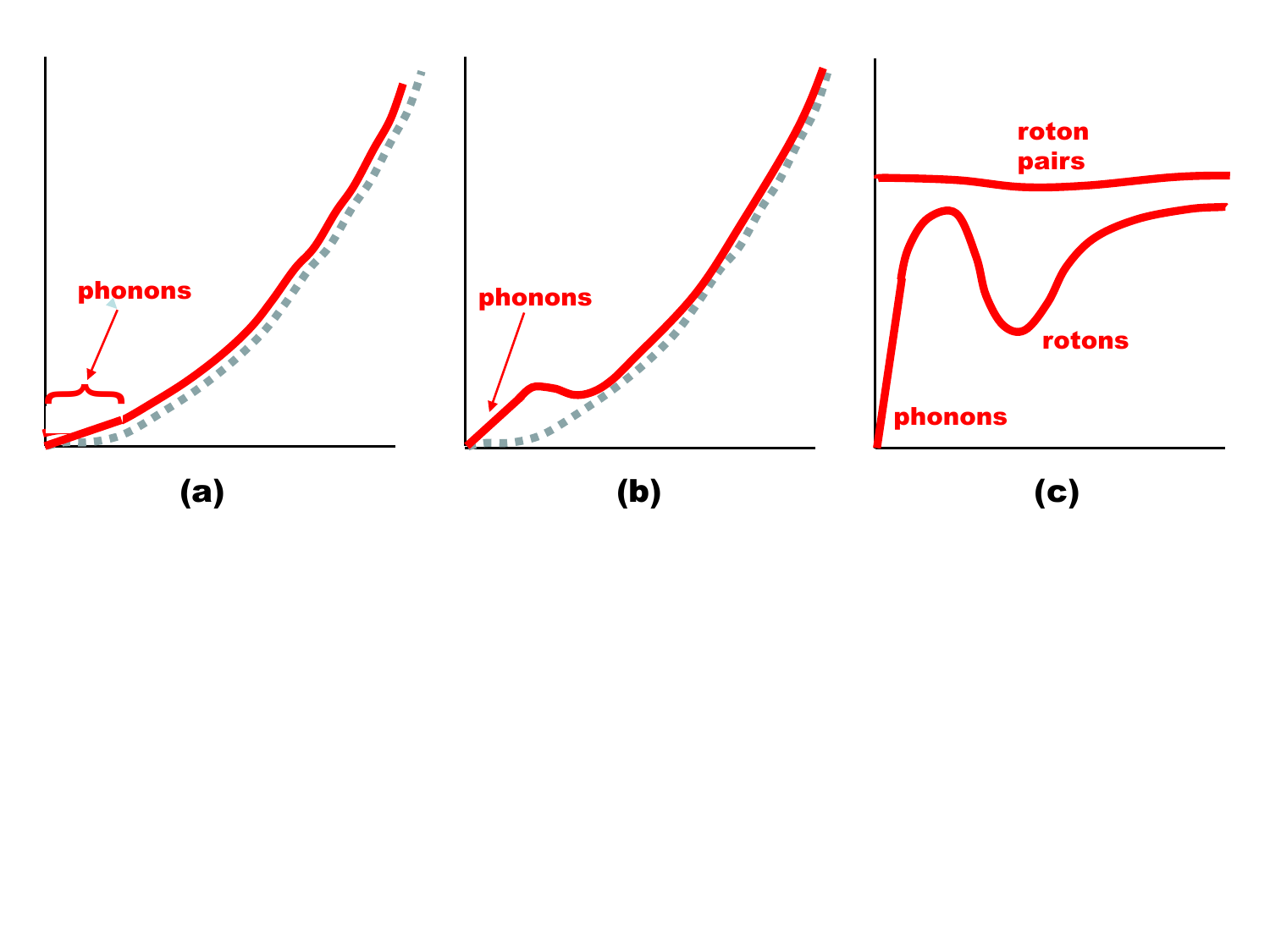}
	\vspace{-7.0cm}
		\caption{\label{fig:TDS3-QPspec} The form of the quasiparticle spectrum in Bose superfluids. In (a) we show the result for weak coupling, along with the free particle spectrum. In (b) we show what happens when one increases the coupling strength for a dilute 3D Bose gas; and In (c) we show the result for strongly-coupled 3D $^4$He superfluid. In (c) we show the main quasiparticle spectrum, and just one of the multiparticle branches (the bound roton pair states). In none of these three cases do we show the broad band of incoherent multi-quasiparticle excitations.   }
	\vspace{4mm}
	\end{figure}

{\bf (ii) Excitation spectra}: In 3D superfluids like $^4$He or several BECs, the excitation spectrum is well understood (see Fig. \ref{fig:TDS3-QPspec}). For weak coupling (typically this means a low-density BEC) one uses the results of Bogoliubov \cite{bog47}. We assume a low-density gas of bosonic particles, each of mass $M$, with short range repulsive interactions; then only $s$-wave scattering is important, with scattering length $a_s$. This defines a coupling constant $g$, given in 3D by $g = 4\pi \hbar a_s/M$.

Then the unbounded 3D system has single-particle quasiparticle excitations with dispersion
\begin{equation}
\epsilon_p^2 \;=\; \hbar^2 \omega_k^2 \;=\;  c_s^2 \hbar^2 k^2 + \frac{\hbar^4 k^4}{4M^2} \;\;\equiv \;\; c_s^2p^2 + (\epsilon_p^o)^2,  \qquad\qquad c_s = \sqrt{\frac{g\rho_0}{M}} ,    \qquad\qquad\qquad  \text{(weak coupling)}
\label{GPE}
\end{equation}
where $\epsilon_p^o = p^2/2m$ is the free particle dispersion, and $\rho_0$ is the density of particles in the Bose condensate; for weak coupling $\rho_0 \sim \rho$, the total mass density. Then for momenta $p \equiv \hbar k$ such that $p \ll 2Mc_s$, we have simple phonons with energy $\epsilon_p = c_sp$ moving at the rather low sound velocity $c_s$, whereas for $p \gg 2Mc_s$, this crosses over to the free particle form $\epsilon_p = \epsilon_p^o + Mc_s^2$. This latter motion is simply that of a free particle in a uniform medium with a refractive index $n > 1$.  We can write a very simple effective Hamiltonian for this system in two different ways. First, we define the condensate wave-function $\psi({\bf x}, t)$, and write \cite{grossP}
\begin{equation}
\hat{H}^{BEC}_{eff} \;=\; -\frac{\hbar^2}{2M}\nabla^2 + V(\mathbf{r}) + g|\psi({\bf r}, t)|^2        \qquad\qquad\qquad  \text{(weak coupling)}
 \label{H-GP}
\end{equation}
which gives the Gross-Pitaevski equation of motion $i\hbar \partial_t \psi(t,\mathbf{r}) = \hat{H}_{eff} \, \psi(t,\mathbf{r})$. Here  $V({\bf x})$ is some external potential acting on the superfluid (for cold BEC gases this would be a trapping potential). Alternatively, since $\epsilon_p$ simply gives the eigenvalues of $H_{eff}$, we write \cite{bog47}
\begin{equation}
H^{BEC}_{eff} \;=\; E_0 + \sum_p \epsilon_p \hat{a}_p^{\dagger} \hat{a}_p       \qquad\qquad\qquad  \text{(weak coupling)}  
 \label{BoseG}
 \end{equation}
where $\hat{a}_p^{\dagger}, \hat{a}_p$ are second quantized operators for the quasiparticles. The Hamiltonian $\hat{H}_{eff}$ is of course extremely simple - it is just a quadratic form describing a set of non-interacting quasiparticles.

	%%%%%%%%%%%%%%%%%%%%%%%%%%%%%
	
	%%%%%%%%%%%%%%%%%%%%%%%%%%%%%

For very strong repulsive coupling things are very different (Fig. \ref{fig:TDS3-QPspec}(c)). The well-known exemplar of this is 3D superfluid $^4$He, in which the interactions induce $\epsilon_p$ to turn over completely \cite{rotonP}, and descend to form roton quasiparticles at a momentum $p_o \sim 1/a_o$, where $a_o$ is the mean interparticle spacing. Phonons and rotons exist in both 3D and 2D superfluids; in 3D they can be created in many ways, by injecting everything from neutrons to picosecond laser pulses \cite{rotonP,PNAS23}. In 2D, this is more difficult \cite{hallock}. One also has quantized surface capillary waves (3rd sound), and rotons are strongly attracted to the surface and to each other \cite{wilks,tilley,rotonP,PNAS23,hallock}.

The interactions between the quasiparticles are now not negligible; we must write, following Landau \cite{LL-FLT}
\begin{equation}
{\cal H}_{eff} \;\;=\;\; E_0 \;+\; \sum_p \epsilon_p \hat{a}_p^{\dagger} \hat{a}_p  \;+\; {1 \over 2} \sum_{p,p'} \sum_q V(p,p',q)\, \hat{a}_{p+q}^{\dagger} \hat{a}_p^{\dagger} \hat{a}_{p'-q} \hat{a}_{p'}         \qquad\qquad\qquad  \text{(strong coupling)}
 \label{Bose-FLT}
\end{equation}
where the effect of the interaction $V(p,p',q)$ is to strongly renormalize all the quasiparticle properties, and to cause scattering between them. In superfluid $^4$He the effect of these interactions is very important - it makes, {\it inter alia}, the quasiparticle properties very strongly temperature dependent. The interactions between the quasiparticles also lead in 3D to bound roton pair states \cite{rotonP}, and even bound triplet roton states \cite{roton3}.

It is extremely hard to calculate the parameters in the strongly-coupled
effective Hamiltonian in any reliable way. For the weakly-coupled
Bose gas it is of course straightforward, using perturbation theory: then the BEC fraction $N_0/N \sim O(1)$ at $T=0$ (in this weak interaction limit, $1 - (N_o/N) = (8/3 \pi^{1/2})(na_s^3)^{1/2}$, where $n = N/V$ is the density of particles). For the strongly-coupled system $N_o/N \ll 1$ (for $^4$He, $N_o/N \sim 0.08$ at $T=0$). Of course in both systems the superfluid fraction $\rho_s/\rho \rightarrow 1$ as $T \rightarrow 0$; but $\rho_s/\rho$ has no connection to $N_0/N$.

In the intermediate coupling range (often called strong coupling by people working on BECs), the quasiparticle spectrum begins to turn over (see Fig \ref{fig:TDS3-QPspec}(b)). As far as we are aware there are no direct measurements of the quasiparticle spectrum in BECs; in 3D $^4$He it has been looked at in great detail. 

Well-known kinematic arguments of Landau \cite{landau41} predicted that the form of the excitation spectrum would yield a critical fluid velocity $v_c$ below which one would see superflow, and above which quasiparticles are excited from the vacuum. This has been verified in detail in 3D superfluid $^4$He and $^3$He, by moving ions rapidly through the superfluid at high pressure \cite{bowley} (at low pressure in $^4$He, the moving ions nucleate vortex rings \cite{reif,PCES79}).

All of the above remarks about weakly- and strongly coupled bosons and their excitation spectra apply to 3D boson systems. Many of them do {\it not} apply to 2D superfluids, which have to be treated differently.

%%%%%%%%%%%%%%%%%%%%%%%%%%%%%%%%%%%%%%%%%%%%%%%%%%%%%%%%%%%%%%%%%%%%%%%%%%%%%%%%
\subsection*{2.2. The Vacuum in Superfluid $^{4}$He Films}
%%%%%%%%%%%%%%%%%%%%%%%%%%%%%%%%%%%%%%%%%%%%%%%%%%%%%%%%%%%%%%%%%%%%%%%%%%%%%%%%

We now look at our first key example. Two-dimensional $^{4}$He superfluid films behave very differently from bulk 3d superfluids, and the property of
superfluidity in 2d has to be understood quite differently. As shown
by Berezinskii \cite{berez72} and Kosterlitz and Thouless \cite{KT},
there exists a critical temperature $T_{\text{K}}$ (the KT transition temperature), below which vortex/anti-vortex
pairs bind, leaving only quasiparticle excitations in a 2d superfluid
with short-range ordering, but without long-range order. However above $T_{\text{K}}$, the vortex/anti-vortex pairs unbind in profusion, destroying the superfluid state, and destroying even the short-range order. 

This argument, amply confirmed by
later experiments on strongly-coupled $^4$He films \cite{hallock}, is a thermodynamic one: the free energy
$F=U-TS$ is minimized in each phase, because  both $U$ and $S$
depend logarithmically on the distance between the nearest vortex/anti-vortex
pairs. In 2D BECs the physics is rather subtle \cite{dalibard}, because the interparticle interactions are weaker, making quantum fluctuations even more important; however these can be suppressed somewhat by a trapping potential.

For these and other reasons, the superfluid vacuum state is quite
different in the two cases. The 3d vacuum state has ODLRO, ie., it Bose condenses, with a macroscopic
occupation of zero-momentum
particle states (${\bf k}=0$ modes). Neither quantum nor thermal fluctuations
of the finite-${\bf k}$ quasiparticle modes can destroy this order
until we reach the superfluid transition temperature $T_{\text{C}}$.
Thus the 3d vacuum state, although fantastically complex because of
the strong interparticle interactions, is ordered in a fundamental
way.

The 2d vacuum state, on the other hand, is \uline{not} ordered. Right
down to $T=0$, long-wavelength quantum fluctuations of the gapless
phonon modes disorder the system: there is no ODLRO even at $T=0$. There
are instead strong quantum fluctuations of the vortex/anti-vortex modes;
and all of this feeds into the final vacuum state of the 2d superfluid. Basically, the vacuum state of the 2-d superfluid is quantum disordered, and although it can show superfluidity, there is no Bose condensate (the appelation ``BEC" given to cold 2D superfluid atomic gases is thus completely misleading - they are superfluid, but there is no condensate!). Instead one defines a ``quasi-condensate" (just another name for short-range order); if the system size is less than the length scale beyond which order disappears, then the idea of a quasi-condensate works. 

Many of the theoretical approaches that work for 3D Bose systems fail in 2D. Thus the Gross-Pitaevski equation works poorly, and ``small fluctuation" expansions of density modes about a (non-existent) condensate, or even about a quasi-condensate, are clearly doubtful unless the fluctuations are ``pinned" by the boundaries. So one needs to find alternative approaches.

%%%%%%%%%%%%%%%%%%%%%%%%%%%%%%%%%%%%%%%%%%%%%%%%%%%%%%%%%%%
\subsubsection*{2.2 (a): Long-wavelength analysis}
%%%%%%%%%%%%%%%%%%%%%%%%%%%%%%%%%%%%%%%%%%%%%%%%%%%%%%%%%%%

One way we can look at this theoretically \underline{{\bf in the long wavelength limit}} is by writing a long-wavelength ``quantum
hydrodynamic'' form for the action of the system as \cite{PNAS25,TS12},
\begin{equation}
S=\int dt\int d^{2}r\,\left[\frac{\hbar}{m_{\text{4}}}\rho\left(\boldsymbol{r},t\right)\dot{\Phi}\left(\boldsymbol{r},t\right)-\mathcal{H}\left(\boldsymbol{r},t\right)\right]
 \label{S-hydro}
\end{equation}
where $m_{4}$ is the mass of the $^{4}$He atom, $\rho\left(\boldsymbol{r},t\right)$
is the areal mass density (proportional to the film thickness $L_{z}$),
$\Phi\left(\boldsymbol{r},t\right)$ is the local superfluid phase
(so that the local superfluid velocity is $\boldsymbol{v}_{\text{s}}$$\left(\boldsymbol{r},t\right)=\left(\hbar/m_{4}\right)\nabla\Phi\left(\boldsymbol{r},t\right)$,
and $\mathcal{H}\left(\boldsymbol{r},t\right)$ is the Hamiltonian
density
\begin{equation}
\mathcal{H}\left(\boldsymbol{r},t\right)=\left[\frac{\rho\left(\boldsymbol{r},t\right)}{2m_{4}^{2}}\left(\hbar\nabla\Phi\left(\boldsymbol{r},t\right)\right)^{2}-\epsilon\left(\boldsymbol{r},t\right)\right]
 \label{Ham-2Dsfl}
\end{equation}
where $\epsilon\left(\boldsymbol{r},t\right)$ parametrizes energy
fluctuations around the homogeneous state (ie., around the vacuum
state if $T=0$). For a film of uniform thickness $\rho\left(\boldsymbol{r},t\right)\rightarrow\rho_{\text{s}}(T)$,
the $T$-dependent superfluid density. Since this is a 2D system, $\rho_s(T) \rightarrow \rho$ as $T \rightarrow 0$, but the BEC fraction $N_0/N = 0$ always.

Quite generally we can write
$\rho\left(\boldsymbol{r},t\right)=\rho_{\text{s}}+\eta\left(\boldsymbol{r},t\right)$,
where $\eta\left(\boldsymbol{r},t\right)$ parametrizes density fluctuations,
and $\Phi\left(\boldsymbol{r},t\right)=\Phi_{\text{S}}\left(\boldsymbol{r}\right)+\phi\left(\boldsymbol{r},t\right)$, where $\phi\left(\boldsymbol{r},t\right)$ parametrizes phase fluctuations
around the time-independent background phase  $\Phi_S\left(\boldsymbol{r}\right)$. One
assumes a background superflow $\boldsymbol{v}_{\text{S}}^{0}\left(\boldsymbol{r}
\right)=\left(\hbar/m_{4}\right)\nabla\Phi_S\left(\boldsymbol{r},t\right)$, created by
the geometry of the system and by any external driving flow (which is often imposed by
boundary conditions). As already emphasized, in a 2-d superfluid, the phase and density
fluctuations are \underline{{\bf not small}}! As one lowers the
dimensionality, quantum fluctuations become more important, and 2D is the critical
dimension in this context. The most serious are the phase fluctuations, and these occur
between
multiple vacuum states, between which the system is constantly fluctuating. These states have different numbers of topological excitations in the system, these being the quantum solitonic vortices, which have quantized circulation \cite{Ons50,RPF-vort}.

The flow field of the vortex falls off like $1/\left|\boldsymbol{r}-\boldsymbol{R}_{\text{V}}\left(t\right)\right|$, where ${\bf R}_V(t)$ is the vortex coordinate.
The total flow energy of the system, and the vacuum state, must then depend
on the behaviour of the system at any boundary, no matter how far
away. This dependence on boundary conditions is common to any system
(and the QFT that describes it) in which some effective long-range
gauge field exists, ie., for almost any macroscopic system in Nature!
For this reason, no vacuum state can be defined independently of the
boundaries, and indeed, the whole idea of a thermodynamic limit is
quite meaningless \cite{SHPMP06}. Two well-known examples are:

(a): the 2D Fractional Quantum Hall effect (FQHE), where all of the properties (quasiparticle excitations,
thermodynamics, dynamics) are carried by the boundaries, except for
the bulk phonons \cite{FQHE}.

(b): a simple ferromagnet, where the long-range dipolar fields create
a demagnetization field which controls both the thermodynamics and
the spin quasiparticle spectrum no matter how large the system, and
which is controlled by the shape of the system; it also controls the
topological excitations of the system (domain walls, vortices, etc),
in both the classical and quantum regimes \cite{Aharoni1,PCES91,Bark24}.
This example applies in both 2 and 3 dimensions.

We will see below that the same dependence of the vacuum state on
boundaries and topological excitations also obtains in 3-dimensional superfluids ($^{4}$He, $^{3}$He).

The analogy between the 2-superfluid and an EM gauge theory like QED comes from
the mapping \cite{PNAS25,TS12} between the hydrodynamic action
above and a (2+1)-dimensional QED action of the form (with a single
vortex present):
\begin{equation}
S_{\text{V}}\left[\boldsymbol{R}_{\text{V}},\dot{\boldsymbol{R}_{\text{V}}}\right]\;\;=\;\; \int dt\left\{ \left[\frac{1}{2}M_{\text{V}}^{0}\dot{R}_{\text{V}}^{2}-\mathcal{T}\left(\boldsymbol{R}_{\text{V}}\right)\right]+\left[\dot{\boldsymbol{R}}_{\text{V}} \cdot \boldsymbol{A}_{\text{V}}\left(\boldsymbol{R}_{\text{V}},t\right) - V_{\text{V}}\left(\boldsymbol{R}_{\text{V}},t\right)\right]\right\}
 \label{2DV-QED}
\end{equation}

Here $\boldsymbol{\kappa}=\pm\left(h/m_4\right)\hat{\boldsymbol{z}}$
is the circulation quantum, and
$\boldsymbol{R}_{\text{V}}$ is the vortex coordinate. The energy $\mathcal{T}\left(\boldsymbol{R}_{\text{V}}\right)=\frac{1}{2}\int d^{2}r\,\rho_{\text{S}}\left(\boldsymbol{r}-\boldsymbol{R}_{\text{V}}\left(t\right)\right)\boldsymbol{v}_{\text{s}}^{2}\left(\boldsymbol{r},t\right)$
is the superflow energy in the system, which creates a bare
potential acting on the vortex. The gauge field components $\boldsymbol{A}_{\text{V}}$ and
$V_{\text{V}}$ are:
\begin{equation}
\boldsymbol{A}_{\text{V}} \;=\; \boldsymbol{A}_{0}+\boldsymbol{A}_{\text{QP}} \;\;\;
= \;\;\; \frac{1}{2}\rho_{\text{S}}\left(\boldsymbol{R}_{\text{V}}\times\boldsymbol{\kappa}\right)+\frac{\hbar}{2m_{4}}\int d^{2}r\,\boldsymbol{j}_{\text{QP}}\left(\boldsymbol{r},t\right)
 \label{vecA}
\end{equation}
where $\boldsymbol{J}_{\text{QP}}\left(\boldsymbol{r},t\right)$ is
the quasiparticle \underline{{\bf pair}} current, given by
\begin{equation}
\boldsymbol{j}_{\text{QP}}\left(\boldsymbol{r},t\right)=\frac{\hbar}{m_{4}}\left[\eta\left(\boldsymbol{r},t\right)\nabla\Phi\left(\boldsymbol{r},t\right)-\phi\left(\boldsymbol{r},t\right)\nabla\eta\left(\boldsymbol{r},t\right)\right]
 \label{j-QP}
\end{equation}
and where $M_{\text{V}}^{0}$ is the ``bare'' hydrodynamic mass of the
vortex (which also depends on the superflow throughout the system,
so that $M_{\text{V}}^{0}=M_{\text{V}}^{0}\left(\boldsymbol{R}_{\text{V}}\left(t\right)\right)$.
We note that the vortex does \uline{not} couple to single quasiparticles \cite{TS12}, but only to pairs of quasiparticles, at lowest order in
the quasiparticle variables, because the vortex is a quantum soliton
(which in the QED analogy becomes an electric charge). Thus, a photon
in the QED analogy is equivalent to a \uline{pair} of quasiparticles.
Finally, the analogue to the electric potential is
\begin{equation}
V_{\text{V}}\left(\boldsymbol{R}_{\text{V}},t\right)=\frac{\hbar}{2m}\int d^{2}r\,\boldsymbol{j}_{\text{QP}}\left(\boldsymbol{r},t\right)\cdot\nabla\Phi_{\text{V}}\left(\boldsymbol{r}-\boldsymbol{R}_{\text{V}}\left(t\right)\right)
 \label{V-V}
\end{equation}
in which quasiparticle pairs couple to the superflow from the vortex -
this coupling is long-range, ie., even quasiparticle pairs at an arbitrary
distance from the vortex core still couple to it.

The true vacuum of the system is a superposition of states
with different numbers of vortex/anti-vortex pairs,
with vacuum tunneling constantly taking place between these states,
and vacuum fluctuations largely controlled by this tunneling.

One can reveal these vortex/anti-vortex vacuum fluctuations by applying an external superflow
$\boldsymbol{v}_{\text{S}}^{0}\left(\boldsymbol{r}\right)$ to the system. Then for the vortex/anti-vortex pair, {\it{if we ignore completely the effect of the quasiparticles}},
we find an interaction energy
\begin{equation}
\mathcal{T}\left(\boldsymbol{r}_{12}\right)=\frac{\rho_{\text{s}}\kappa^{2}}{4\pi}\ln\left|\frac{r_{12}}{\xi_{0}}\right|-\frac{1}{2}\kappa\rho_{s}\boldsymbol{v}_{\text{s}}^{0}\left(\boldsymbol{r}\right)\cdot\boldsymbol{r}_{12}
 \label{VV-int}
\end{equation}
where $\left|\boldsymbol{r}_{12}\right|$ is the distance between
the virtual vortex and the virtual anti-vortex, and $\xi_{0}$ is
the vortex core radius (the ``healing length'').

	%%%%%%%%%%%%%%%%%%%%%%%%%%%%%
	
		\begin{figure}
		\includegraphics[width=7.4in]{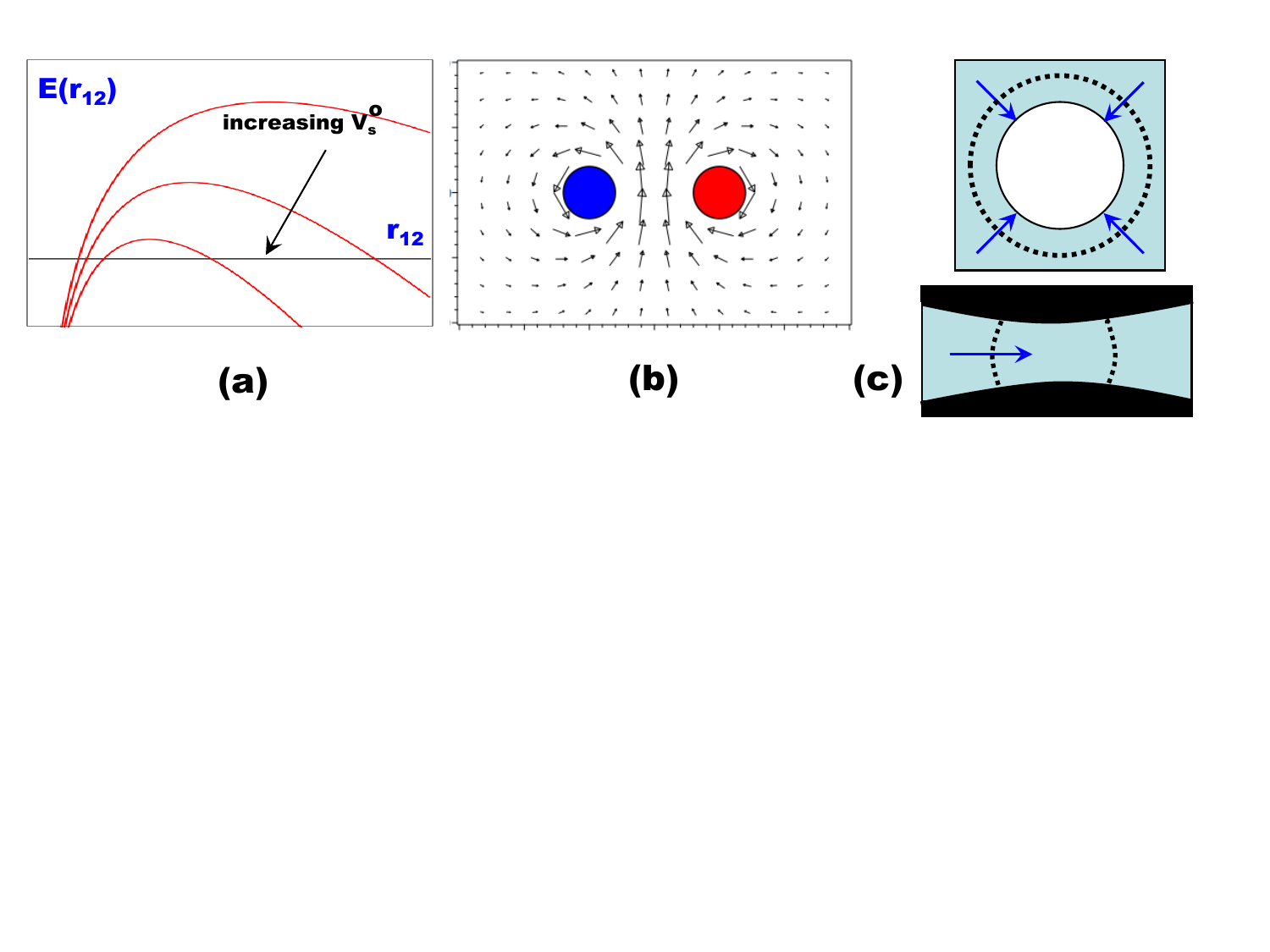}
	\vspace{-8.0cm}
		\caption{\label{fig:TDS4-vortexT} The form of the tunneling potential ${\cal T}({\bf r}_12)$ is shown here in (a) for different external flow velocities $\boldsymbol{v}_{\text{s}}^{0}$. In (b) we show the superflow pattern as the vortex/anti-vortex pair separates. Finally, in (c) we show two other interesting geometries in which one can imagine looking for vacuum tunneling in 2D superfluid films. In the upper figure, superfluid is shown accelerating towards a circular ``drain"; in the figure below, through a constriction. The ``horizons", defined by the critical crossover velocity $\bar{v}_c({\bf r})$ where vacuum tunneling switches on, are shown by dotted lines; and the superflow directions in (b) and (c) are shown by arrows. }
	\vspace{4mm}
	\end{figure}

	%%%%%%%%%%%%%%%%%%%%%%%%%%%%%

We plot (Fig. \ref{fig:TDS4-vortexT})  the way in which the inter-pair vortex potential $\mathcal{T}\left(\boldsymbol{r}_{12}\right)$
varies with $\boldsymbol{r}_{12}$ for different external superflow
velocities $\boldsymbol{v}_{\text{s}}^{0}$ (with $\boldsymbol{v}_{\text{s}}^{0}$
perpendicular to $\boldsymbol{r}_{12}$, so that the Magnus forces
from the superflow, acting equally and oppositely on the two vortices,
pull them apart). At high $T$ vortex
pairs will unbind by thermal activation over the energy barrier - although we note that as
$\left|\boldsymbol{v}_{s}^{0}\right|\rightarrow0,$ the energy barrier blocking this unbinding diverges logarithmically with $|{\bf r}_{12}|$,
meaning that in reality this unbinding could never
occur!

From this argument we see that (i) the Kosterlitz-Thouless
thermal unbinding transition is really only physically meaningful when $\left|\boldsymbol{v}_{s}^{0}\right|$
is large enough - otherwise it will never happen; and (ii) it is not
really a thermodynamic phase transition at all, but rather a rapid \uline{crossover}
between two regimes, one in which there are no unbound pairs, and
the other where they proliferate. This crossover occurs around a critical ``crossover velocity" $\bar{v}_c$. Moreover (iii) traditional Kosterlitz-Thouless
theory, which only takes account of thermal fluctuations out of the vacuum, clearly needs to be revised to incorporate the effect of the very large quantum fluctuations in the vacuum.

In any case we see that the crossover around $\bar{v}_c$ is from what (misleadingly)
seems to be a quiescent
superfluid vacuum state (for low $\left|\boldsymbol{v}_{\text{s}}^{0}\right| < \bar{v}_c$)
to a state where the already rather unstable vacuum has now been completely destabilized (at high
$\left|\boldsymbol{v}_{s}^{0}\right| > \bar{v}_c$). In fact, strictly speaking, for finite $\left|\boldsymbol{v}_{s}^{0}\right|$
, \uline{there is no vacuum state at all}! The system is condemned to remain
forever in a non-equilibrium steady state, in which the superflow field
$\boldsymbol{v}_{s}^{0}$ is constantly
pulling virtual vortex pairs from the vacuum, and making them ``real". At
low $T$, this happens via \uline{vacuum
tunneling} out of the vacuum state (but we note that vacuum tunneling is constantly taking place in the vacuum state already, between different vortex states). The fact that $^{4}$He superfluid can be made in a completely
pure form means that this may be the best available laboratory we have for
looking at vacuum tunneling between topologically different vacua.

The above considerations all assume a uniform superflow. However, as noted in ref. \cite{PNAS25}, one can also introduce a non-uniform background flow field $\boldsymbol{v}_{s}^{0}\left(\boldsymbol{r}\right)$, so that the critical crossover velocity $\bar{v}_c \rightarrow \bar{v}_c({\bf r})$, a position-dependent crossover. We can now, if we fall back on analogical thinking, think of the line surface where $\boldsymbol{v}_{s}^{0}\left(\boldsymbol{r}\right) = \bar{v}_c({\bf r})$ as a ``horizon", where the vacuum instability switches on (note that these ``tunneling horizons" are quite different from the ones discussed in section 3.2 below, which arise when superflow velocities equal phonon velocities or ``slow light" velocities; these are more like ``Cerenkov" horizons". On one side of the tunneling horizon the vacuum is, within a broad range of experimental timescales, stable for all practical purposes. On the other side it is unstable, with vortex/anti-vortex pairs being created in profusion. In Fig. \ref{fig:TDS4-vortexT})(c)  we show two geometries in which one can define tunneling horizons of this kind (we note that one can also have a situation where the superfluid flows fast at a boundary, in which case one has ``{\it extrinsic}" nucleation at the boundary \cite{PNAS25}).

%%%%%%%%%%%%%%%%%%%%%%%%%%%%%%%%%%%%%%%%%%%%%%%%%%%%%%%%%%%
\subsubsection*{2.2.(b): Vortex Avalanches in 2-d Geometry}
%%%%%%%%%%%%%%%%%%%%%%%%%%%%%%%%%%%%%%%%%%%%%%%%%%%%%%%%%%%

As we've seen, one can take an apparently pristine superfluid vacuum
at zero temperature, with no free vortices, and destabilize it by applying a static external superfluid flow field ${\bf v}_s^0({\bf r}$. However, we have also emphasized that the underlying vacuum state is actually {\bf not} pristine, but is actually brimming with incipient vortex/anti-vortex pairs - it is in fact extremely close to instability even without the external flow. It is then no surprise that the picture of vortex pairs, nucleating via vacuum tunneling independently of each other \cite{PNAS25}, is actually not a good approximation. A simple argument shows that the real critical velocity for vacuum tunneling ought to be much lower than that estimated above and in ref. \cite{PNAS25}. This is because the presence of vortex\textbackslash anti-vortex pairs enhances the production
of future pairs under the correct conditions.

In the presence of a single vortex, the effective flow velocity $\boldsymbol{v}_{\text{s}}^{0}$
is altered slightly by the presence of a single nearby vortex. Imagine then that a vortex/anti-vortex pair has just nucleated
and we are now watching a new virtual pair trying to nucleate nearby. It is easy to see that the largest effect of the existing pair on the new pair will be if they all lie along a line perpendicular to ${\bf v}_s^0({\bf r})$. One then sees that either
 (a) The barrier height for the new pair is increased, and the tunneling rate for the new pair is reduced, if their orientation is opposite to that of the existing pair; or (b) if they are parallel, the barrier height is reduced, and the tunneling rate increased,  thereby lowering the critical velocity for the nucleation of the pair.

By the same argument, interactions between virtual vortex/anti-vortex pairs will lower the critical velocity - if the quantum fluctuations are strong enough, they will lower it to zero and the system will never become superfluid at all (as it is, thermal fluctuations prevent ODLRO from developing at any temperature). In $^4$He films, the quantum fluctuations are not quite strong enough to destroy $T=0$ superfluidity, but not too far from it.

This is not the only consequence of the interaction between different vortex/anti-vortex pairs. One should also in principle see ``vortex avalanches". This is a kind of \underline{{\bf quantum avalanche}}, of the kind first described recently  for quantum Ising systems \cite{Bark24}. A quantum avalanche is simply a process occurring at sufficiently low temperature that dissipation is very small, and one remains in the ``quantum regime". Then, when interactions between quantum nucleation events (quantum fluctuations or vacuum tunneling) occur, an avalanche process of similar nucleations then follows, with very little heating - and so the system remains in the quantum regime. This process can then occur almost without limit, with the system still remaining in the quantum regime, so that the entire process is quantum-mechanical. In experiments on a quantum Ising system \cite{Bark24}, quantum avalanche processes involving $\sim 10^{16}$ spins were seen - by far the largest macroscopic quantum tunneling process hitherto seen in the lab.

If the theory is correct, one should certainly be able to see such quantum avalanche processes in 2-d superfluids as well. It will be interesting to look for this experimentally. The critical velocity required to initiate these
vacuum tunneling and subsequent vortex avalanche process will be much lower than the value $\sim 6$ms$^{-1}$ estimated in ref. \cite{PNAS25}, because of the interactions between vortex pairs. Detailed calculations will be given elsewhere.

%%%%%%%%%%%%%%%%%%%%%%%%%%%%%%%%%%%%%%%%%%%%%%%%%%%%%%%%%%%%%%%%%%%%%%%%%%%%%%%%%
\subsection*{2.3. The Vacuum in 3-D Superfluids}
%%%%%%%%%%%%%%%%%%%%%%%%%%%%%%%%%%%%%%%%%%%%%%%%%%%%%%%%%%%%%%%%%%%%%%%%%%%%%%%%%

In three dimensions we have a plethora of superfluids: $^{4}$He,
$^{3}$He, various ultracold BECs, and of course, a huge variety of
charged superconductors (not to mention the interiors of neutron stars
by far the largest contribution to the supply of superfluids and superconductors
in the universe). However most of these are not satisfactory as test
beds for ideas about quantum vacua. Experiments on superfluid $^{3}$He
are difficult - few labs current operating can work below 2.7$^{\circ}$mK.
Fewer still can do experiments on nanoscopic short-lived samples of
cold BEC's at nano Kelvin temperatures. Superconductors are typically
dirty and have many other complications. Here we will focus on $^{4}$He
and BEC's, which have the great advantage of being simple ($^{3}$He
superfluid, with its 18-component order parameter, is endlessly fascinating,
and many experiments have been done to investigate
analogies with cosmological phenomena. However this lies beyond the
scope of this article). So in what follows we first look at bulk $^{4}$He
superfluid, and then at BEC's.

%%%%%%%%%%%%%%%%%%%%%%%%%%%%%%%%%%%%%%%%%%%%%%%%%%%%
\subsubsection*{2.3(a) Bulk $^{4}$He Superfluid:}
%%%%%%%%%%%%%%%%%%%%%%%%%%%%%%%%%%%%%%%%%%%%%%%%%%%%

One can use the same kind of low-energy effective action for 3-d superfluid
$^{4}$He as we described for the 2-d film. Interactions are again strong, and the coherence/healing length is basically unchanged. However, this description is
less useful here, because the 3d excitations are more complex, and
the short-range physics is more important. In the quasiparticle spectrum
one now includes phonons, rotons, and bound roton pairs, as well as
vortex lines \cite{donnellyV,soninV}. The rotons and roton pairs have structure
in the length scale range $4-20\text{\AA}$ ; and the vortex lines
form ``vortex rings'', as well as complicated ``vortex tangles'',
which have very complicated configurations. All this physics has been studied exhaustively over the last 85 years.

In 3D $^4$He, quantum fluctuations are much less pronounced than
in 2d; the vacuum is more ``classical''. We can remove the vorticity
completely in experiments, by applying a heat flush (a ``phonon wind'') to sweep
them out. This leaves thermally excited quasiparticles; at very low
$T$, the rotons disappear, because they are gapped, but just as in
2d, we still have gapless phonons.

	%%%%%%%%%%%%%%%%%%%%%%%%%%%%%
	
		\begin{figure}
	\includegraphics[width=7.2in]{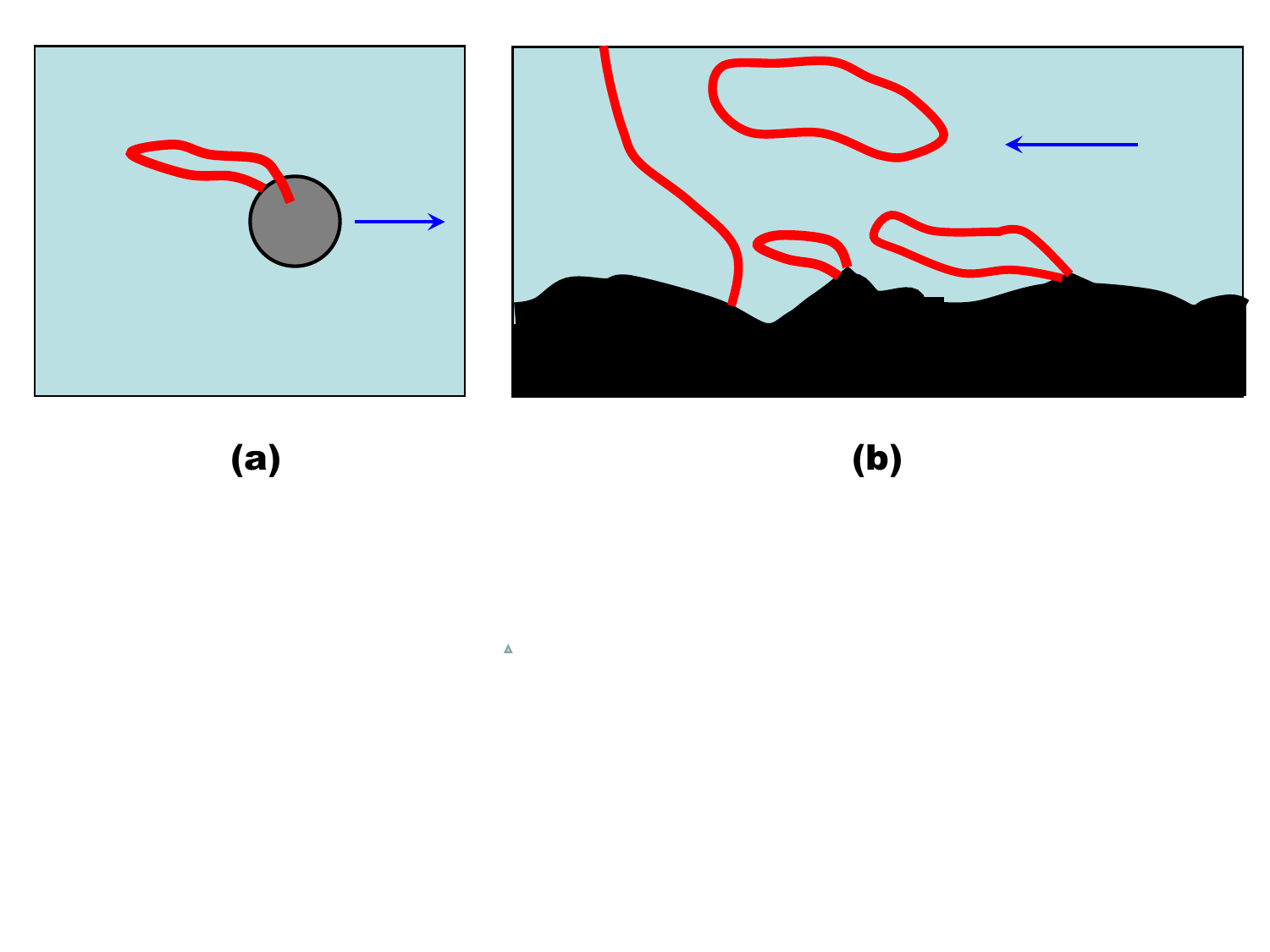}
   \vspace{-7.1cm}
		\caption{\label{fig:TDS5-3Dvortex} Some processes that can produce vortices in 3D $^4$He superfluid. In (a) we see a vortex loop nucleating on the surface of a moving ion (whose motion is denoted by an arrow). In (b) we show vortices nucleating at a boundary, past which the the superfluid in flowing. We also show how the vortex loops can escape from the boundary by forming vortex rings). The vortex loops nucleate around boundary regions where the local boundary curvature and the neighbouring superflow velocity are high. Once formed they can migrate along the surface.  }
	\vspace{4mm}
	\end{figure}

	%%%%%%%%%%%%%%%%%%%%%%%%%%%%%

One can probe the superfluid $^{4}$He
vacuum using neutrons, or by re-injecting either vortices or quasiparticles.
It is quite hard to do the letter in a controllable way in bulk $^{4}$He
superfluid. Vorticity nucleates at boundaries with sharp edges, and
it is hard to do experiments on individual quasiparticles (although
remarkable experiments by Wyatt \cite{wyatt} have succeeded). A new technique using
femtosecond laser pulses has succeeded in creating dense ``fireballs''
of roton pairs \cite{PNAS23}; and one can produce individual vortex rings
by injecting electrons into the superfluid using field emission tips,
and then accelerating them in an electric field \cite{reif}. Depending
on the pressure, the moving ions then reach a critical velocity $\sim$50-60$ms^{-1}$,
where they either emit pairs of rotons, or a vortex loop is nucleated
at the surface of the ion bubble. Such experiments \cite{bowley,
PCES79} give very direct evidence for the tunneling nucleation of vortex rings
in 3d $^{4}$He. But it is \uline{much} harder to understand these
experiments (and other vortex tunneling experiments in the bulk) than
it is for the 2d films previously described, because the variety of
shapes that 3d vortex lines can take  makes any comparison between
theory and experiment very hard (see Fig. \ref{fig:TDS5-3Dvortex}).

%%%%%%%%%%%%%%%%%%%%%%%%%%%%%%%%%%%%%%%%%%%%%%%%%%%%
\subsubsection*{2.3(b) Cold Gas Bose-Einstein Condensates (BECs)}
%%%%%%%%%%%%%%%%%%%%%%%%%%%%%%%%%%%%%%%%%%%%%%%%%%%%

Since the first real experimental realizations of Bose–Einstein condensation (BEC) in dilute alkali gases in 1995 \cite{Anderson1995ScienceBEC,Davis1995PRLBEC}, ultracold atomic condensates have become an important platform for probing macroscopic quantum phenomena. The underlying system consists of alkali atoms in magnetic or optical traps, with interactions, at least in weak coupling, well described by a single $s$-wave scattering length $a_s$. Different confinement geometries (1D, 2D, 3D) can be made by optical lattices or highly anisotropic traps.

In an ideal BEC, all the particles share a common wavefunction, giving rise to a coherent matter wave with phase stiffness (ie., a ``winding energy" associated with phase gradients) across the entire sample. In a true BEC, at sufficiently low temperatures (typically in the nanokelvin regime for atomic BECs), a non-zero fraction $N_o/N$ of atoms occupies the ground state of the external trap, representing a striking manifestation of quantum statistics at the macroscopic scale. As discussed above, for weakly interacting BECs, $N_o/N \sim O(1)$. In atomic BECs, one can vary the coupling strength across several orders of magnitude, from very weak (where $na_s^3 \ll 1$, and the healing/coherence length may be up to several microns in size) up to intermediate values, by tuning the scattering around a Feschbach resonance \cite{zwerger}. However at no point can one approach the very strong coupling regime existing in $^4$He. 

The question of whether or not one really has a BEC in these systems is quite delicate. Often the notion of a ``{\it quasi-condensate}" is used - this simply means that one either has a very small sample, or one can define a small region of the system, in which one can define a local ODLRO with density $n_s({\bf r})$, and some hopefully not too large variation of phase across sample or the region. This can be done for arbitrary dimension - indeed we used it in our description of 2D superfluid films above. However in a trap $n_s({\bf r})$ typically varies strongly across the trapping region, and both thermal and quantum fluctuations become larger as one decreases (i) the dimensionality, and (ii) the strength of the interaction \cite{dalibard,zwerger}. The best approximation to a uniform quasi-condensate comes in 3D with sufficiently strong coupling in ``box" potential wells. 

The numbers are important here. In typical trap experiments, the healing length $\xi_o \sim 0.1 - 0.2~\mu$m, whereas the trap size $L_o \sim 10-40~\mu$m. Thus $\xi_o/L_o < 10^{-2}$ in these experiments. In harmonic traps, where $n_s({\bf r})$ varies strongly across the sample, one is dealing with a fundamentally non-uniform system. However in box potentials, with $\xi_o/L_o \ll 1$, the quasi-condensate can be treated as uniform except near the boundaries. 

In any case, in the same way as described in the previous section on 2D ${ }^4 \mathrm{He}$ films, one can use a low-energy effective action for BEC quasi-condensates. The long-wavelength degrees of freedom are again phonons quasiparticles. As one increases the coupling strength, the quasiparticle dispersion begins to turn over \cite{lopes17,andreev25}, but at no point do we get anything like the spectrum seen in strongly-coupled $^4$He superfluid. 

Under appropriate circumstances (not too strong or not too weak coupling, 3 dimensions), the transition to quasi-condensation can be captured 
by the Gross–Pitaevskii equation (\ref{H-GP}), which encodes both the mean-field 
quasi-condensate and its collective excitations \cite{Pitaevskii2003BECBook,Dalfovo1999RMP}. In these systems, quantum fluctuations are small, and finite traps can strongly suppress thermal fluctuations. 

The situation is less clear for lower dimensions, or for very strong coupling (and in the very weak coupling limit, the requirement that $\xi_o \ll$ system size breaks down). 
In uniform 2D Bose fluids, there is no condensate. In a finite trapped 2D system, a KT transition can be seen. Below $T_K$ one can try to argue that a quasi-condensate exists with well-defined phase coherence across the sample \cite{dalibard,Hadzibabic2006Nature}. This argument is the same as that used to argue for ``local ordering" (but not long-range order) in any 2-D system without long-range forces. Note that in 2D we do \uline{{\bf not}} expect the fluctuations to be small, even for $T < T_K$, unless the system itself is so small that the local order parameter is locked in by boundary conditions. For $T > T_K$ the fluctuations are of course very large, because vortex pairs proliferate. 

In later sections we will discuss how, under specific conditions, these properties of dilute BECs have made them a candidate for analogue quantum field experiments.

%%%%%%%%%%%%%%%%%%%%%%%%%%%%%%%%%%%%%%%%%%%%%%%%%%%%%%%%%%%%%%%%%%%%%%%%%%%%%%%%
\subsection*{2.4. Quantum Mechanics and Gravity}
%%%%%%%%%%%%%%%%%%%%%%%%%%%%%%%%%%%%%%%%%%%%%%%%%%%%%%%%%%%%%%%%%%%%%%%%%%%%%%%%

We now move to ``the real thing'', ie., the deepest problem currently
facing physics. This is the problem of reconciling two apparently
incompatible theories (GR vs. QM/QFT) into a new third theory. We look at this problem
from several points of view in Section V. Here, we just discuss what
one can say about the vacuum in GR and QM/QFT separately.

%%%%%%%%%%%%%%%%%%%%%%%%%%%%%%%%%%%%%%%%%%%%%%%%%%%%
\subsubsection*{2.4(a) The Vacuum in Classical General Relativity}
%%%%%%%%%%%%%%%%%%%%%%%%%%%%%%%%%%%%%%%%%%%%%%%%%%%%

The metric $g^{\mu\nu}\left(x\right)$, with 10 independent components,
is far more complex than anything we have looked at up until now (with
the exception of superfluid $^{3}$He). In spite of attempts \cite{vol-analog}
to make mappings between $^{3}$He superfluid, with its 18-component
order parameter, and a theory in which the standard model is added
to the $10$-component metric field, these two systems are vary different,
and we shall see that this analogy fails. Let's first discuss key
features of the vacuum state in GR.

The metric tensor configuration is sourced by the stress-energy tensor
$T_{\mu\nu}$ in Einstein's field equation,
\begin{equation}
G_{\mu\nu}\left(x\right)=\kappa T_{\mu\nu}\label{eq:Einstein Field Equation}
\end{equation}
where $\kappa=8\pi G/c^{3}$ and $G_{\mu\nu}\left(x\right)=R_{\mu\nu}\left(x\right)-\frac{1}{2}R(x) \, g_{\mu\nu}\left(x\right),$
with $R_{\mu\nu}\left(x\right)$ the Ricci tensor. Many thousands
of solutions are known to this equation \cite{steph09}; it is certain
that these known solutions constitute a miniscule fraction of the
total set of solutions. Those solutions that are discussed in cosmology
and in black hole theory are highly symmetric and extremely simple.
Cosmological models look almost exclusively at homogeneous solutions (whereas a generic
solution of $\text{(\ref{eq:Einstein Field Equation})}$ involves
6 functions of 2 variables and 4 functions of 3 variables). Black
hole theory often only discusses three solutions, viz., the Schwarzchild, the Reissner-Nordstrom, and the Kerr-Newman solutions, these being the unique stationary solutions.

In spite of this, some rather general results are known. In the absence
of any matter, we know that singularities must exist somewhere, and
it is generally believed that these will take the BKL or Mixmaster
form \cite{BKL,mixM}, with metric such that
\begin{equation}
ds^{2}=-c^{2}dt^{2}+R^{2}\left(t\right)\sum_{i,j=1}^{3}e^{\beta_{ij}}\sigma_{i}\sigma_{j}
 \label{mixM}
\end{equation}
where $R\left(t\right)=R_{0}a\left(t\right),$ with $R_{0}$ a constant
and $a\left(t\right)$ a scale factor (so that $a^{3}\left(t\right)\sim$volume), and the differential
forms $\left\{ \sigma_{j}\right\} $ satisfy $d\sigma_{i}=\epsilon_{ijk}\sigma_{j}\wedge\sigma_{k}$
and parameterize rotations on the 3-sphere. Diagonalizing $\beta_{ij}$,
we find $\sum_{j}\beta_{jj}=1$ , and so the dynamics can be written
in terms of a ``$\boldsymbol{\beta}-$particle'' moving in a 2-dimensional
$\boldsymbol{\beta}$-plane, spanned by orthogonal vectors $\beta_{\pm}$
such that $\beta_{+}=-\beta_{33}$ and $\sqrt{3}\beta_{-}=\beta_{11}-\beta_{22}$. As Misner observed \cite{mixM}, the dynamics then assumes the BKL form,
with an infinite sequence of chaotic oscillations accumulating as
the singularity is approached (during which $a\left(t\right)\rightarrow0$).
This behaviour is most easily understood by noting that the $\boldsymbol{\beta}$-particle
is confined to a triangular potential well, with very steep-sided
walls, in the 2-dimensional $\boldsymbol{\beta}$-plane. It bounces off these walls, which themselves are moving outwards. This mapping of the dynamics of the shape of the 3-sphere to the motion of the $\boldsymbol{\beta}$-particle is very helpful; one sees that the 3-sphere suddenly distorts to a new shape each time the $\boldsymbol{\beta}$-particle bounces off one of the walls.

None of this behaviour can be described in any way by phenomena in
superfluid $^{3}$He (where singularities in the flow field, like
the ``boojum'' in $^{3}$He-A, are approached smoothly, with no
oscillations at all). The point of course, is that no field theory,
in condensed matter or anywhere else, has the attractive self-interaction
which exists in the spacetime metric.

There is a further twist to this story, the subject of current research \cite{qingdi25}.
This is that if we add matter fields to the background spacetime,
the dynamics in the $\boldsymbol{\beta}$-plane changes again. Bosonic
fields simply hasten the approach to the singularity, but fermionic
fields have a negative zero-point energy \cite{martin12}, and if one
adds a combination of bosonic and fermionic fields, one finds that
the formation of a singularity is prevented, and the spacetime contraction
reverses to an expansion. 

In any case, even without the quantized matter fields, the vacuum
of GR is very peculiar. The presence of any spacetime curvature leads
to a non-localizable source of stress-energy, which if it is allowed
to grow, or is otherwise focussed (eg., in the collision between two
gravitational shock waves), will ``blow up'' into a singularity.
From a physical standpoint, this is why GR, when quantized, is non-renormalizable:
a single gravitational wave or gravitational pulse with energy $E_{1}$ will create an attractive potential
for another one of energy $E_{2}$, and the attraction diverges
as $E_{1},E_{2}\rightarrow\infty$, and is $\sim\mathcal{O}\left(1\right)$
if $E_{1},E_{2}\sim E_{\text{p}}$, the Planck energy.

Thus the vacuum in GR is already unstable for reasons having nothing
to do with quantum mechanics at all.

%%%%%%%%%%%%%%%%%%%%%%%%%%%%%%%%%%%%%%%%%%%%%%%%%%%%
\subsubsection*{2.4(b) The Vacuum in Quantum Field Theory}
%%%%%%%%%%%%%%%%%%%%%%%%%%%%%%%%%%%%%%%%%%%%%%%%%%%%

Since condensed matter media are already described by QFT's, many
of the features we have seen, notably those connected with vacuum
fluctuations, occur in the same way in relativistic QFT's. However,
there are some key differences which are crucial when we come to discuss Quantum
Gravity (section 5). The most important is the existence of UV divergences. In renormalizable QFT's these are regularized, and any
divergent zero-point energy is eliminated. However this option is
not open to us when incorporating QFT into some quantum gravity theory,
or even when using QFT on a background curved spacetime - see below.

Such UV divergences do not exist in QFT's for condensed matter systems,
because, as already emphasized above, condensed matter QFT's are effective
field theories with built-in UV cut-offs. Typically these cut-offs
arise in a natural way, at energies where the effective $\mathcal{H}$
breaks down as new structures and new physics come into play. Note that
exactly the same point can be made about relativistic QFTs. Any QFT, whether
it is for a non-relativistic superfluid, for ordinary QED or for some attempt at a full quantum-gravitational theory, has some built-in UV cutoff: if this does not appear naturally, it is imposed theoretically in some way. The vacuum state, in the theory, is adjusted accordingly - this is just as true for a relativistic QFT as it is for our discussion of a $H$ atom in a box, at the beginning of this section.

However what is not open to choice is the existence of divergent zero-point vacuum
energy contributions in quantum gravity. This is because they must contribute to the
cosmological constant \cite{martin12}. The contribution is $\sim \Lambda_o^4$, where $
\Lambda_o$ is the UV cutoff. A proper discussion of this must include fermionic as well
as bosonic quantum fields - crucially, the fermions give a negative contribution, and
counteract the attractive contribution of gravitational and bosonic fields. In work to be published elsewhere, we show how we can use this result to give a non-divergent theory \cite{qingdi25}.

\vspace{4mm}

%%%%%%%%%%%%%%%%%%%%%%%%%%%%%%%%%%%%%%%%%%%%%%%%%%%%%%%%%%%%%%%%%%%%%%%%%%%%%%%%%%
%%%%%%%%%%%%%%%%%%%%%%%%%%%%%%%%%%%%%%%%%%%%%%%%%%%%%%%%%%%%%%%%%%%%%%%%%%%%%%%%%%

\section*{3. A Dialogue about Analogues in Physics}

%%%%%%%%%%%%%%%%%%%%%%%%%%%%%%%%%%%%%%%%%%%%%%%%%%%%%%%%%%%%%%%%%%%%%%%%%%%%%%%%%%
%%%%%%%%%%%%%%%%%%%%%%%%%%%%%%%%%%%%%%%%%%%%%%%%%%%%%%%%%%%%%%%%%%%%%%%%%%%%%%%%%%

We have seen that in discussions of the vacuum in condensed matter systems, one can adopt rather different views about the value of analogies between phenomena in physically very different systems. In discussions between two of us (PCES and ST) it has become clear that it would be interesting to air the arguments for and against the use of such analogies. We shall do this in a somewhat Socratic form, in the form of a dialogue between the two of us. The sceptical or ``pessimistic" point of view will be adopted by PCES, and the optimistic antithesis to this view by ST.

	        %%%%%%%%%%%%%%%%%%%%%%%%%%%%%
	
\begin{figure}
		\includegraphics[width=6.2in]{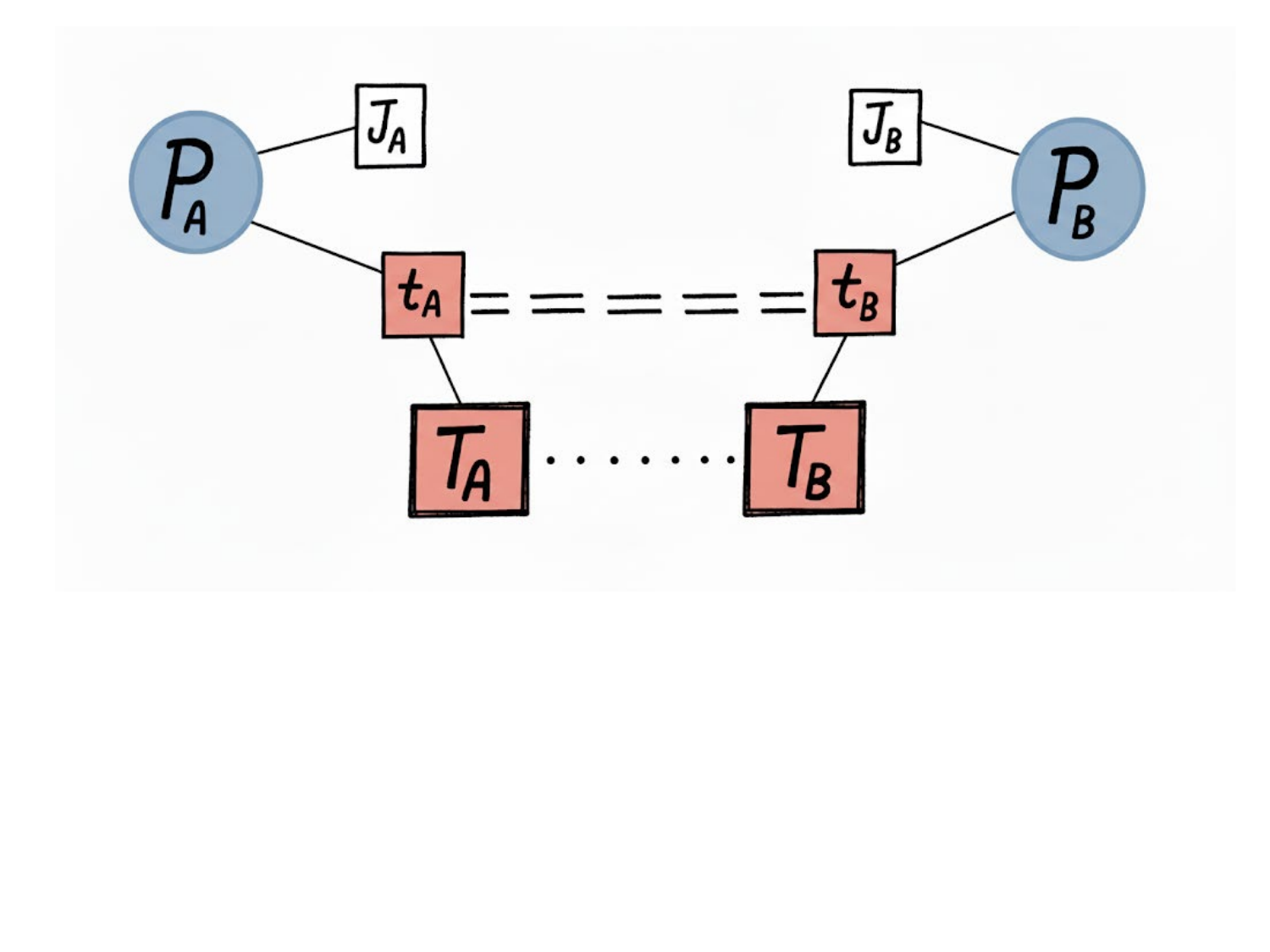}
\vspace{-5.2cm}
		\caption{\label{fig:TDS7-analog} Our rough model of how analogues work in physics. We imagine two quite different physical systems $S_A$ and $S_B$, in which phenomena $P_A$ and $P_B$ are supposed to occur. We describe $S_A$ and $S_B$ using theories $T_A$ and $T_B$; there is assumed to be some sort of (partial) mapping ${\cal M}_{AB}$ between the two theories. We also assume subsidiary theories $t_A$, $t_B$ which are supposed to model $P_A$ amd $P_B$; they are typically simplifications of or approximations to $T_A$ and $T_B$. Often, instead of a mapping ${\cal M}_{AB}$ between $T_A$ and $T_B$, one makes a mapping ${\cal m}_{AB}$ between $t_A$ and $t_B$. Finally, experiments or observations on $P_A$ and/or $P_B$ require interpretive frameworks ${\cal I}_A$ and ${\cal I}_B$ to make the connection between what is seen and how it is to be understood. See text for further explanation. } 
	\vspace{4mm}	
\end{figure}

	            %%%%%%%%%%%%%%%%%%%%%%%%%%%%%

Let's begin by saying, very roughly, what we mean by an analogy in physics. Suppose we imagine two quite distinct physical systems, $S_{\text{A}}$
and $S_{\text{B}}$, and two theories $T_{\text{A}}$ and $T_{\text{B}}$
describing each (See Fig. \ref{fig:TDS7-analog}). We also imagine that we are going to observe, in experiments, some set  $P_{\text{A}}$ of phenomena in  $S_{\text{A}}$, and try to infer from these observations, what may be seen of some other putative set of phenomena
$P_{\text{B}}$ in system  $S_{\text{B}}$. Usually this kind of reasoning is employed when, for some reason or another, one cannot (at least at the time), do experiments or make observations on system  $S_{\text{B}}$. So one is either trying to guess or infer what might be seen on $S_B$. Alternatively, one might be trying to confirm some ideas about $S_B$ by observations of $P_A$ made on $S_A$.  

We observe that this reasoning involves the existence of some mapping ${\cal M}_{AB}$
 - more or less well-defined -  between a theory $T_{\text{A}}$ describing $S_{\text{A}}$
 and another theory $T_{\text{B}}$ describing $S_{\text{B}}$. Since the two theories are
 usually very different, this mapping will almost never be a 1-1 mapping. In fact it is often the case that we instead devise a mapping/analogy  between some model  $t_{\text{A}}$, which is an approximation of some sort to $T_{\text{A}}$, another model $t_{\text{B}}$ which is a similar adaptation to $T_{\text{B}}$.

We also observe that the interpretation ${\cal I}_A$ of any experiment/observation on some phenomenon  $P_{\text{A}}$ is always very complicated. It not only will involve the theory $T_A$ and the model $t_A$ in an essential way, but also a large number of auxiliary theoretical assumptions about the experiment or observation - the way one interprets any experiment or observation always comes ``laden with theory". 

All of this is summarized in Fig. \ref{fig:TDS7-analog}. The idea of the analogue is to then argue that the observation of some phenomenon $P_{\text{A}}$ in system $S_{\text{A}}$
provides evidence that an analogous phenomenon $P_{\text{B}}$ will
occur in system $S_{\text{B}}$.

In this paper we have so far discussed two systems for which analogues can be developed in some detail, viz., (i) the analogy between vortex/anti-vortex tunneling in 2-d superfluids, and vacuum tunneling in either QED or quantum gravity; and (ii) the analogy  between the propagation of phonons in superfluid BECs, or of light in optical media, and the propagation of light or other excitations in curved spacetime near, eg., black holes.

In what follows we first describe how each of these analogues work, and then give a critical discussion of the general question using these examples.

%%%%%%%%%%%%%%%%%%%%%%%%%%%%%%%%%%%%%%%%%%%%%%%%%%%%%%%%%%%%%%%%%%%%%%%%%%%%%%%%%%%
\subsection*{3.1. Example A: 2-d Superfluids and QFT Analogues}
%%%%%%%%%%%%%%%%%%%%%%%%%%%%%%%%%%%%%%%%%%%%%%%%%%%%%%%%%%%%%%%%%%%%%%%%%%%%%%%%%%

We begin with the analogy already mentioned, between the Schwinger effect and vortex/anti-vortex unbinding. The Schwinger effect was originally discussed before modern QFT \cite{heis}, when it was found that strong electric and/or magnetic fields would induce effective photon-photon interactions in the QED Lagrangian. In 1951 Schwinger \cite{schwing51} wrote a remarkable paper - the first non-perturbative calculation in modern QFT - which showed how the QED vacuum would become unstable in strong fields to electron-positron pair production. In related work Dyson \cite{dyson51} pointed out that QED had an essential singularity, as a function of complex coupling constant $\alpha$, with a branch cut extending away from the origin along the negative $\alpha$ axis.

It is immediately obvious what is the parallel between vortex tunneling and the Schwinger
effect. In fact, in (2+1)-D QED with no topological terms in the action,
the form of the electron-positron potential is precisely the same
as $\mathcal{T}\left(\boldsymbol{r}_{12}\right)$, if we substitute
a static electric field $\boldsymbol{E}^0\left(\boldsymbol{r}\right)$ in place of $\boldsymbol{v}_{\text{s}}^{0}\left(\boldsymbol{r}\right)$,
and the 2-D logarithmically attractive $e^{-}/p^{+}$ interaction between charges for the same interaction
between vortex pairs. This seems like a very straightforward and successful
analogy between two theories $T_{1}$ and $T_{2}$, and two different vacuum tunneling
phenomena $P_{1}$ and $P_{2}$ in the two different systems. So, have we found a successful and useful analogy here?

The current theory of the Schwinger effect is confined to 1-loop and 2-loop expansions,  both in $3+1$ dimensions and in $\left(2+1\right)$-D
QED without topological terms. In the latter case the 1-loop result for vacuum
tunneling is \cite{dittrichB}
\begin{equation}
\Gamma_{\text{2+1}}^{\text{QED}}=\frac{\left(eE\right)^{\frac{3}{2}}}{4\pi^{2}}\sum_{n=1}^{\infty}\frac{1}{n^{\frac{3}{2}}}\exp\left\{ -\pi n\frac{m^{2}}{eE}\right\}
 \label{QED-2+1}
\end{equation}

We see immediately
that there are 2 effects contained in the superfluid film calculation that are not included in this result, viz.,

(a): One can assign an effective mass to the vortex pair dynamics
as a function of the separation coordinate $\boldsymbol{r}_{12};$
this is given by $M_{\text{V}}^{\text{pair}}\left(\boldsymbol{r}_{12}\right)=\left(\rho_{\text{s}}\kappa^{2}/4\pi a_{0}^{2}\right)\ln\left|r_{12}/\xi_{0}\right|$, which increases as the vortices separate. Physically, this happens
because as the vortices separate, their mutual dipolar flow field
extends over ever larger regions, and the resulting inertial mass
increases accordingly. The analogous effect in the QED calculation would have to include radiative corrections to all orders - ie., it would need to go to infinite order in a loop expansion. One would have to carefully preserve all the Ward identities in this calculation.

(b): When we couple the tunneling vortex/anti-vortex pair to the quasiparticles,
two things happen. First, there are radiative corrections to the vortex effective mass; and second, the vortex/quasiparticle coupling, even at $T=0$, causes dissipation, thereby reducing the tunneling rate and modifying its dependence on the background $\boldsymbol{v}_{\text{s}}^{0}$. This mechanism, which involves IR divergent processes,  is not in the Schwinger calculation at all.

There are also 2 key features not included in either the Schwinger calculation \cite{dittrichB} or our recent superfluid vortex tunneling calculation \cite{PNAS25}. The first is simply the emission of photons/quasiparticles off to infinity caused by the appearance of the new pair excitations. The second is the effect of interactions between two different vortex/anti-vortex pairs (a 4-way interaction) or between single vortices and a vortex pair (a 3-way interaction). As noted in section 2.2(b), such interactions must be included in any proper calculation, and will lead to important physical processes, including ``quantum avalanche" production of a dense interacting vortex/anti-vortex system, which will show quantum turbulence (recall section 2.2(b) above).

Of course if we are really to see how good the QED/superfuid analogy is, experiments
will have to be done, both in superfluid films (to see the intrincic
vacuum tunneling process) and in QED vacua (to see the Schwinger effect)!
It is probable that the superfluid vacuum tunneling will be seen first. Note that in real vacuum QED, the electron-positron-coupling is weak, so that some of the effects that are crucial in vortex tunneling may be quite small in any experiment on the Schwinger effect, whenever this may be done. If history is any guide, the experiments will certainly be more complicated than the theory, and will bring surprises. One must never forget that the real QED vacuum and the real 2d superfluid vacuum are \underline{{\bf not at all the same}!}

Similar remarks can be made about phenomena in superfluid films and putative analogues in the behaviour of quantum fields in curved spacetime (we noted in ref. \cite{PNAS25}, and reiterated above, how one could simulate processes around a black hole in superfluid films - {\it cf.} Fig. \ref{fig:TDS4-vortexT}(c)). In this analogue, the critical surface $\bar{v}_c$ around which vortex/anti-vortex vacuum tunneling occurs is the analogue of the black hole horizon around which Hawking radiation is supposed to be seen by an external observer. If we pursue this analogy then we can call this a ``vacuum quantum tunnelling horizon". Note that this is not at all the same as the acoustic horizons discussed for BEC analogies with phenomena in curved spacetime (see below), which are entirely classical.

%%%%%%%%%%%%%%%%%%%%%%%%%%%%%%%%%%%%%%%%%%%%%%%%%%%%%%%%%%%%%%%%%%%%%%%%%%%%%%%%%%%
\subsection*{3.2. Example B: BECs and Analogue Gravity}
%%%%%%%%%%%%%%%%%%%%%%%%%%%%%%%%%%%%%%%%%%%%%%%%%%%%%%%%%%%%%%%%%%%%%%%%%%%%%%%%%%

\vspace{2mm}

We now consider possible laboratory tests of quantum field theory in curved spacetime (QFTCS), focussing on ``analogue gravity", which is a framework to study analogues of general relativistic geometries within condensed matter, fluid, and optical systems. The central idea is that as far as the purely kinematic characteristics of QFTCS are concerned, one can to some extent overlook the structure of the underlying system and work with an effective metric, allowing one to apply purely geometric concepts from GR.  One then maps structural properties of QFTCS such as the amplification of fluctuations, the creation of quasiparticles, and mode mixing across horizons, onto excitations in various analog systems.

Analogue models in physics have a long history; as an early example, one can refer to Walter Gordon’s 1923 formulation of an effective metric for light propagating in moving dielectrics  \cite{Gordon1923}, and similar analogical approaches have appeared in various areas of physics since then. In the modern sense, the foundation of the field of analogue gravity dates from Unruh's seminal paper in 1981 \cite{Unruh1981PRL}, in which he shows that sound waves in an irrotational, inviscid fluid propagate according to an effective Lorentzian “acoustic metric", obtaining a ''sonic horizon" if the fluid flow passes from subsonic to supersonic, and phonon emission from that horizon mimics thermal Hawking radiation.

Formally, when one linearizes the equations of motion for small perturbations in a suitable background, the quadratic action for the perturbation field $\phi$ takes the following universal form:
\begin{equation}
S^{(2)}=\frac{1}{2} \int d^4 x \sqrt{-g_{\mathrm{eff}}} g_{\mathrm{eff}}^{\mu \nu} \partial_\mu \phi \partial_\nu \phi,
\end{equation}
so that $\phi$ obeys a Klein–Gordon equation in an effective metric:
\begin{equation}
\square_{g_{\mathrm{eff}}} \phi=\frac{1}{\sqrt{-g_{\mathrm{eff}}}} \partial_\mu\left(\sqrt{-g_{\mathrm{eff}}} g_{\mathrm{eff}}^{\mu \nu} \partial_\nu \phi\right)=0 .
\end{equation}

Hence, phonons in a flowing fluid can be mapped to a quantum field in a curved spacetime, provided that the effective geometry of that curved spacetime can be described by the Painleve-Gullstrand-Lemaitre type metric \cite{Painleve1921}.

Building on this, Jacobson in 1991 emphasized the deep connection between horizon thermodynamics and the underlying
microphysics, introducing the idea that Einstein’s equations themselves may be viewed as an equation of state \cite{Jacobson1991PRL}, and while not experimental, his work motivated the analogue program by showing that the
thermodynamic and the causal-structure aspects of horizons are central, not the full Einstein dynamics. One of the most prolific voices in the field of analogue gravity, Matt Visser, in the 1990s, derived the general form of the acoustic metric for a wide class of fluids, showing how analogue models could mimic black hole, cosmological, and rotating spacetimes \cite{Visser1995CQG}

Multiple research works by Schützhold and Unruh extended the scope of analogue gravity even further, showing that horizons and Hawking-like radiation can also be realized in different fluid \cite{Schutzhold2002PRD}. In these works  the importance of dispersion relations and mode mixing near horizons have been highlighted, and it has been clarified under which conditions analogue systems reproduce Hawking-like particle creation. From a broader perspective, as Hawking and Ellis had already explained \cite{Hawking1973LSS} in 1973, much of the predictive power of GR derives from its causal structure: the light cones, horizons, and the associated kinematics. This insight shed some light on why analogue systems are valuable and how they imitate the causal kinematics of fields in curved spacetimes.

Comprehensive reviews (especially Barceló, Liberati, and Visser \cite{Visser1995CQG}, and more recently by Ralf Schützhold \cite{Schutzhold2025EffectiveSpacetimes}) have framed analogue gravity as a kinematics-not-dynamics correspondence, highlighting the idea that excitations in condensed matter and optical systems propagate according to effective metrics. Analogue systems are valuable precisely because they can imitate these causal and kinematical aspects of curved spacetimes, even though they do not reproduce the full dynamical content of Einstein’s equations. They thereby allow laboratory studies of phenomena such as Hawking radiation, cosmological particle creation, or superradiance under controlled conditions, while leaving the problem of quantum gravity dynamics untouched.  In the following we consider a few examples of these systems, noting that it would not be possible to give all of the references here.

{\it Bose-Einstein condensate as a probe of quantum fields and gravity}:
Bose-Einstein condensates (BEC) are one of the laboratory analogues for relativistic quantum fields, where, their low energy collective excitations can be mapped to an effective Klein–Gordon field in curved spacetime under the right circumstances.

Using the Madelung split of the condensate wave function $\psi$ as $\psi=\sqrt{\rho_{\mathrm{0}}} \exp \{i S / \hbar\}$, it has been shown that in the long wave limit one can find the analogy between sound waves in Bose-Einstein condensates and mass-less and minimally coupled scalar fields in curved space-times \cite{Schutzhold2025EffectiveSpacetimes}

In some of the recent analogue gravity tests in particular in proposals such as the interferometric Unruh detector schemes of Gooding et. al \cite{Gooding2020PRL} it is common to employ a quasi-two-dimensional Bose–Einstein condensate rather than a full
three-dimensional one. The reason is that in a 2D geometry, in the light-matter interaction the effective overlap between probe light and condensate is stronger and easier to control, and the excitations behave more like 'purely scalar' modes, with no transverse modes and no polarization structure. In these models, 2D BECs are realized by applying a strong trapping potential along one direction, creating a 'pancake' condensate \cite{Hadzibabic2006Nature}, and to probe it with laser beams, one can send light through the plane and measure density/phase fluctuations directly (see also ref.  \cite{Gooding2025NondestructiveOptomechanicalBEC}).

In these recent proposals, it is shown how through a mapping of the Gross-Pitaevskii dynamics to an effective Klein–Gordon field, one can examine phenomena such as the Unruh and Hawking effects, vacuum decay, or entanglement harvesting. For the sample of a 2D BEC obeying the two-dimensional Gross-Pitaevskii field equation, one can write small fluctuations of this field as \cite{Unruh2022BlackHolesAcceleration}: 
\begin{equation}
    \psi=\psi_0+\Psi.
\end{equation}

Then, defining the background density of the BEC as $\rho_0=\psi_0^2$, and choosing $V+g \rho_0=0$, one can rewrite the corresponding action in terms of real and imaginary parts as:
\begin{equation}
S=2 \int\left[ \Psi_R^{\dagger} \partial_{t} \Psi_I-\frac{1}{4 M}\left[\left(\nabla \Psi_R\right)^2+\left(\nabla \Psi_I\right)^2\right]-g \rho_0 \Psi_R^2\right] d\sigma dt,
\end{equation}
where considering long wave limit for the dispersion relation, one can obtain the corresponding Lagrangian mimicking Klein Gordon field:
\begin{equation}
    L_{\text{BEC}}=\frac{1}{c_s^2}\left(\partial_t \frac{\Psi_R}{\sqrt{M}}\right)^2-\left(\nabla \frac{\Psi_R}{\sqrt{M}}\right)^2
\end{equation}
where $c_s=\sqrt{g\rho_0/M}$ is the effective light velocity as before; and $\phi(t, \boldsymbol{r})=\Psi_R(t, \boldsymbol{r}) / \sqrt{m}$ is identified as the effective (2+1)dimensional Klein-Gordon field (see \cite{Gooding2025NondestructiveOptomechanicalBEC}. Through this scheme, the condensate functions as a laboratory analogue of a relativistic quantum field.

Employing a BEC as an analogue system for different spacetimes was originally proposed by Fedichev and Fischer in 2003 \cite{FedichevFischer2003PRL}, who showed that an expanding BEC can mimic a de Sitter universe, with the excitations obeying the same equations as fields in curved spacetime, in which the effective phonon metric takes the form:
\begin{equation}
ds^2 = -c_s^2 dt^2 + a^2(t) d\mathbf{r}^2,
\end{equation}
with scale factor
$a(t)$ determined by trap dynamics. They predicted analogue Gibbons–Hawking radiation with temperature:
\begin{equation}
T_{\text{GH}} = \frac{\hbar H}{2\pi k_B}, \qquad H=\frac{\dot{a}}{a},
\end{equation}
emerging as thermal phonon emission from the expanding condensate.

Shortly thereafter, A. Retzker, J. I. Cirac, M. B. Plenio, and B. Reznik, in their 2008 proposal \cite{Retzker2008PRL}, modelled the embedding of accelerated detectors in BECs to probe thermal radiation. They assumed a UDW detector model,  with an interaction Hamiltonian:
\begin{equation}
H_{\text{int}} = \lambda (\sigma^+ + \sigma^-)\phi(t,\mathbf{r}(t)),
\end{equation}
where $\phi$ is the phonon field along the accelerated trajectory $x(t)$, and showed that the transition probability per unit time is thermal. Hence, they recovered the Unruh temperature by verifying Einstein's detailed balance condition between excitation and de-excitation rates of their detector, derived from the BEC correlation functions.

Gooding, Biermann, Erne, Louko, Unruh, Schmiedmayer, and Weinfurtner in their theoretical proposal extended this idea to interferometric Unruh detectors, where a continuous laser probe acts as a quasiparticle detector. They showed that a continuous quasiparticle detector in the form of a laser beam interacting with a BEC can cleanly distinguish the Unruh effect from other sources of noise \cite{Gooding2020PRL}. Their detector’s response function was expressed through the Wightman function of the BEC phonon field:
\begin{equation}
\mathcal{F}(\omega) = \int d\tau \, d\tau' \, e^{-i\omega(\tau-\tau')} G^+(\tau,\tau'),
\end{equation}
with
\begin{equation}
G^+(\tau,\tau')=\langle \phi(\tau)\phi(\tau') \rangle.
\end{equation}

These theoretical ideas connect with a broader analogue-gravity experiments. For instance, the Technion group (Steinhauer, 2016) observed Hawking-like correlations across an analogue horizon in a BEC \cite{Steinhauer2016NatPhys}. Their relevant observable is the density–density correlation function:
\begin{equation}
    G^{(2)}(x,x') = \langle \delta n(x), \delta n(x') \rangle,
\end{equation}
showing that entangled quasiparticle pairs are consistent with analogue Hawking emission. Furthermore, Jaskula et al., in 2012, reported the first experimental realization of an acoustic analog of the dynamical Casimir effect in a Bose–Einstein condensate (BEC), where they observed quasiparticle pair creation in a time-modulated condensate. The mode occupation followed a squeezed-state spectrum:
\begin{equation}
    n_k = \sinh^2 r_k,
\end{equation}
where the squeezing parameter $r_k$ is set by modulation amplitude and frequency \cite{Jaskula2012PRL}. Together, these experiments establish BECs as robust analogues of quantum fields in curved spacetime.

BECs have also been suggested as analogues for false-vacuum decay by Jenkins et al. \cite{Jenkins2024PRD}. In their theoretical proposal, they modeled bubble nucleation and expansion by tunneling through an effective potential barrier in the non-linear Schrödinger dynamics.
Recently, Mann and Gooding have shown how a BEC vacuum can ''harvest” entanglement between two detectors by transferring nonclassical correlations from the condensate to a pair of pulsed laser beams, which highlights the potential use of BEC not only for analogue gravity tests but also for quantum information protocols \cite{Gooding2024NJP}.

Hence, BECs have been used in multiple scenarios to investigate the structure of the vacuum. Taking the approach of mapping Gross-Pitaevskii dynamics to effective Klein–Gordon fields in curved spacetimes, one can examine phenomena such as Unruh and Hawking effects, vacuum decay, and entanglement harvesting. In this way, BECs allow us to test, within the laboratory, features of quantum field theory and quantum gravity that would otherwise be inaccessible.

%%%%%%%%%%%%%%%%%%%%%%%%%%%%%%%%%%%%%%%%%%%%%%%%%%%%%%%%%%%%%%%%%%%%%%%%%%%%%%%%%%%
\subsection*{3.3. Dialogues: For and Against}
%%%%%%%%%%%%%%%%%%%%%%%%%%%%%%%%%%%%%%%%%%%%%%%%%%%%%%%%%%%%%%%%%%%%%%%%%%%%%%%%%%

It will be clear that there is a divergence of opinion between two of us on the use of analogies between phenomena in condensed matter and QFT/gravity. One of us (PCES) is very suspicious of their use, and the other (ST) is an enthusiastic proponent of them. We feel that this is an important question, which only came up in discussions between the two of us while we were in the process of finishing the article. The best way we could think of to air the various issues is in the form of a ``Socratic dialogue". We caution the reader that we have had very little time to prepare this, and we do not think that what is written here is anywhere near the last word on this topic. All we are doing is introducing some of the issues.

\vspace{4mm}

{\bf (A) The Pessimistic View (PCES)}

Unfortunately any reasoning in favour of using analogies is only as good as the links in the chain
between the different components in Fig. \ref{fig:TDS7-analog}, extending from $P_{\text{A}}$ to
$P_{\text{B}}$, as well as the quality of the components themselves.
It is usually the case that the ``mapping'' between $T_{\text{A}}$
and $T_{\text{B}}$ is imprecise and incomplete, and it is never 1-1. The
theories $T_{\text{A}}$ and $T_{\text{B}}$ are themselves often incomplete,
and have their own imprecisions, and the connections that one tries
to establish between $T_{\text{A}}$ and $S_{\text{A}}$, or $T_{\text{B}}$
and $S_{\text{B}}$, usually leave a lot to be desired. A real physical
system is almost always far more complex than the theory being used
to describe it; $T_{\text{A}}$ is usually a crude idealization of
$S_{\text{A}}.$ Moreover - a point often neglected in philosophical
discussions - experiments are never simple, and always come
``laden with theory''. Thus, unless one is very careful, it is easy
to draw false or misleading conclusions about phenomena $\left\{ P_{\text{B}}\right\} $
which one thinks might exist in $S_{\text{B}}$, based on phenomena
$\left\{ P_{\text{A}}\right\}$ which might have been observed in $S_{\text{A}}$.

Consider how things turned out for some of the analogies mentioned earlier:

\vspace{2mm}

  (i) Maxwell's analogy \cite{maxwellB} between the EM field and mechanical systems turned out to be helpful to him in visualizing the field and its consequences. It was actually a pretty good analogy in that the two theories matched each other rather well. However from a physical point of view it turned out to make no sense at all, because of the principle of special relativity; and it was quickly abandoned. A related example, starting even earlier, began from the apparently identical behaviour of waves in fluids and light waves
in vacua (including interference phenomena and diffraction). Use of this analogy began with Huyghens in 1690, in his debate with Newton over the nature of light: he argued that empty space was filled with a fluid-like ``aether'', through which light propagated \cite{huyghens95,newton04}. This argument of course turned out to be completely wrong (as were Newton's 1604 counterarguments!), although Huyghens's ideas led to a large number of predictions (eg., 2-sit interference) that were confirmed. One sees that any analogy between condensed media and spacetime is fraught with pitfalls. There are many more examples like this, some not so well known - in writing the history of physics, we tend to forget the theories that failed.

\vspace{2mm}

   (ii) In the same way, the analogies made since at least the 1940s between curved spacetime in GR and different condensed matter media can look rather awkward. The spacetime metric $g^{\mu\nu}$ has 10 degrees of freedom, and no analogy with, eg., ideal Eulerian hydrodynamics or with simple superfluids (using, eg., Clebsch theory) can possibly hope to capture this. Much has been made of the analogy with $^3$He superfluid, which has an 18-component order parameter - but again, the analogy is very misleading (imagine trying to simulate the approach towards a singularity with a superfluid - all of the BKL/mixmaster complexity would be completely absent!). Current attempts to match ideas in string theory with the ``landscape" picture of spin glass theory seem even more far-fetched.
   
Of course one can employ condensed matter/curved spacetime analogies in a much more limited way. The most commonly used analogy involves optical media and curved spacetime, employed by many since 1920. The best-known use of this was by Penrose, to understand the Raychaudhuri eqtn. and the focusing of null vectors, and to derive his famous results on trapped surfaces and singularities \cite{penrose66,penrose-int}. Here one is analyzing smooth classical spacetime far from any singularities, and using the analogy with a smoothly varying optical medium in the long-wavelength regime. In the same way, Starobinsky and Zeldovich, in their work on acceleration radiation \cite{staro+YaZ} (work which then led Hawking to Hawking radiation), were inspired by an EM analogy. 
   
\vspace{2mm}

    (iii) One analogy that worked rather spectacularly, in a predictive way, was that leading to the Anderson-Higgs boson. By extending the 1957 BCS theory to include long-range Coulomb interactions, both Bogoliubov \cite{bog58} and Anderson \cite{PWA58} understood in 1958 that
these interactions ought to gap certain collective modes in real s-wave superconductors (this prediction was later experimentally verified). Anderson then argued \cite{PWA63} in 1963 that if a similar model applied in particle physics, then one should see the same gapped excitation. In 1964 Higgs took this same model, and applied it directly to make the same prediction \cite{higgs64} for what later became the Anderson-Higgs boson. In the much more complicated  ``standard model'' of particle physics this boson survived. The experimental verification at CERN of the existence of this boson had to wait nearly 50 more years. This story had a happy ending, but as we've just seen, many similar stories do not.

\vspace{2mm}

Let us also consider the 2 examples we discussed in detail above.

\vspace{2mm}

{\it (a) Vortex tunneling/QED analogy}: this is the analogy between vacuum tunneling in 2D superfluids, and the same tunneling in the QED Schwinger effect. At first glance this seems like a fruitful analogy - it has already indicated that the original Schwinger calculation, and subsequent work on it, may have missed out several key effects. It will be interesting to see how the dialogue works between the 2 calculations in the future. But my expectation is that it will not work very well. We are dealing with two very different vacua, and the more one probes them, the more these differences will become obvious. Thus one should be very careful with this analogy, and try to avoid circular argument. My feeling is that while the analogy is interesting, it probably will not lead to new physics. 

\vspace{2mm}

{\it (b) BEC/curved spacetime analogy}: In this case I would argue that if one treats the background superfluid as classical (which is what is being done here), then all that is being done is to compare two classical mathematical models. Thus, eg., experiments on phonons or classical sound waves in flowing fluids are just testing whether well-known classical models of fluid flow match the classical theory of Hawking radiation, and they tell us nothing about how real Hawking radiation would behave around a real black hole.  In fact all they do is confirm that the simple classical causal structure outside black hole horizons is the same as the causal structure in the analogue classical fluid or classical optical system. But these have been deliberately contrived to have the same causal structure! Thus all that has been done is to construct a tautology. 

\vspace{2mm}

Having looked at all these examples, I would now like to ask a key question, viz,

\vspace{2mm}

\uline{{\bf Question}}:{\it \;\; Suppose we have an analogy between two theories $T_A$ and $T_B$, and we then do experiments on the supposedly analogous phenomena $P_A$ and $P_B$, and we find that the analogy \uline{{\bf does not work}} - that predictions made about $P_B$ on the basis of the known phenomenon $P_A$ turn out to be wrong. What then?}

\vspace{2mm}

What {\it ought} to happen, of course, is that the analogy is then rejected. Unfortunately, as history all too frequently shows, this is {\bf not} what happens. Very often we see a quite different behaviour, in which experiments that do not confirm theory are not published. If they are published, they are either ignored, or attempts are made to dismiss them. The criterion for a good experiment then becomes whether or not it agrees with the theory (or in the present case, whether it confirms the analogy).

This is of course a very dangerous tendency; the dialogue between theories then becomes a ``dialogue des sourds", in which circular arguments on both sides prevail.

Examples of this are all too frequent. Without going into detail, I just mention several of them \cite{Scie113}:

- the history of experiments on the Kondo effect in the 1960s and early 1970s, in which every new version of the theory led to experiments which agreed with it. As Anderson remarked a the time, the agreement with each theory was just as good as the last, even though the theoretical predictions and the experiments changed each time! Anderson's recommendation was that the experimental groups simply ignore the theory. But he also remarked that the way the scientific system works made it very hard for these groups to do this - that it was much easier for experimenters to get recognition and funding if they simply confirmed theory. This is probably even more true now than it was then. 

- the lamentable history of experiments on strongly-correlated 2D and and quasi-2D conducting systems. This has led repeatedly to notorious scandals (eg., the ``Batlogg-Schoen" scandal \cite{plasticF}, and numerous more recent scandals) over experimental data purporting to show high-$T_c$ superconductivity, or other exotic states. Common to all these scandals is the suppression of data not supporting the desired theoretical picture, or the outright fabrication of data designed to support some theory. The situation has become so bad in this field that well-known physicists are now unmasking such work, and publishing what they find in journals like Nature, without losing face - so this black cloud may have a silver lining.

- the multiple attempts to show that different quantum computation systems work in the lab, in which the very real effects of decoherence are simply ignored (or treated theoretically using inapplicable models of classical noise).

I have little doubt that most readers of this article will be familiar with similar stories. While understandable (nobody wants to discredit the enormous accomplishments of physics), it is unfortunate that these stories are often suppressed. Sometimes only assiduous science journalists are willing to expose what is really going on in some of these case (although it is now becoming {\it de rigueur} to include material on such cases in introductory university science courses \cite{Scie113}). 

The consequences of neglect can be really bad for the science, even when all parties are being honest. If physicists had spent less time in the century before 1905 trying, in all good conscience, to justify the ``aether analogy", they might have come much earlier to special relativity. In my view the very large effort being made to force fit experiments on BECs and similar systems into analogies with quantum gravity, is not doing either field much good. The study of BECs - which are very interesting systems in their own right - is in no way justified or enhanced by showing that they might look like curved spacetime, and as far as I know, no results obtained on BECs have had any impact on work in classical GR (nor do I expect them to). In the same way, vortex vacuum tunneling in 2D superfluids is not made more interesting by a possible analogy with vacuum tunneling in QED. 2D superfluids are spectacularly interesting systems in their own right (and very likely behave differently from QED). 

To make this clear, let's return to the question above, and ask what what should be the correct reaction of the physics community if experiments give {\it different} results from theory for any of the following 3 analogies (i) the 2D vortex/QED Schwinger analogy; (ii) the analogy between phonon propagation in BECs and classical GR; and (iii) the analogy between fluid experiments and black hole Hawking radiation (assuming this latter is ever observed!). It seems to me that what {\it ought} to happen in these three cases is that

(i) for the vortex/QED analogy, we would reject the analogy, and assume that since QED is well understood, that therefore 2D superfluids are not yet well understood. This \uline{{\it might}} help us understand 2D superfluids better; but not QED. 

(ii) for the BEC/GR analogy, we would {\uline{\it not}} reject classical GR, which for smooth spacetimes far from singularities, is very well understood indeed. We would instead reject our description of BECs, and the analogy with GR. This \uline{{\it might}} help us understand BECs better; but not GR. 

(iii) for the fluid flow/Hawking radiation analogy: since our understanding of classical fluid dynamics is pretty good, especially for smooth flows without turbulence, we would reject the analogy, and assume that our understanding of genuine quantum gravity and of real black holes is just not good enough. It is hard to see how this would help us better understand either phenomenon. 

In other words, if any disagreement occurs, we would reject the analogy. If this is the case, then the only use of the analogy, if it works, is to confirm what we already know, or confirm prejudices we already have. If we accept this, then the analogy has no predictive value at all.

On the other hand, what analogies \underline{{\bf are really good for}} is suggesting \uline{{\bf new physics}}, or new approaches to physical questions. The examples given above, of the Anderson-Higgs boson, and the Penrose-Hawking work on black holes, illustrate this rather well. Nobody would try to push the analogy between charged superconductors and the current version of the standard model very far. But as a means of intuiting important new physics, it worked extremely well.

I therefore conclude that whereas analogies can be of enormous heuristic value in deriving new physics (as in the case of the work done at different times by Anderson, Penrose, Starobinski, Zeldovich, and Hawking) they can also lead to very serious error (as in the case of the aether). In the case of analogies between condensed media and spacetime (as in the QED/superfluid and the BEC/spacetime analogies discussed above) my feeling is that the jury is still out - but that the use of analogies in these two fields runs the clear risk of producing either more errors, or of just restating what we already know.

\vspace{4mm}

{\bf (B) The Optimistic View (ST)} 

Here, I want to first agree that many analogies in physics' history have been misleading, but I would also argue that analogue gravity has already demonstrated itself to be of a different order. I agree with the concern that analogies between condensed matter systems and general relativity must be handled carefully, since the mapping is never one-to-one and can risk overstating what is being tested. On the other hand, however, in my view, this critique can be avoided if one reinterprets the scope of analogue systems and remembers that analogue gravity is not intended to “solve quantum gravity” or to reproduce the Einstein field equations in a laboratory, but rather to probe the robustness of quantum field theory in curved spacetime (QFTCS). 

One can argue that this is similar to the framework that Hawking originally used as well; i.e., he did not quantize gravity or aim to do so, and Hawking radiation was derived from treating matter fields quantum mechanically in a fixed classical spacetime. In other words, by using only minimal assumptions such as the existence of a horizon, causal field propagation, and quantum vacuum fluctuations, and without invoking any quantum gravity effects, Hawking showed that black holes shrink as they emit radiation. The lack of any input from quantum gravity means of course that short lengthscale gravitational phenomena, along with conceptual issues such as the
trans-Planckian problem, are inaccessible to this long wavelength
classical approach. Yet Hawking's result remains one of the most
remarkable successes of theoretical physics.

What makes analogue gravity practical is not the claim that one can reproduce the full dynamics of Einstein’s theory, or solve the Einstein equation, or provide a microscopic theory of quantum gravity. The point is that one can isolate and experimentally probe the kinematical structures of quantum field theory in curved spacetime, and that the strength of analogue gravity lies in its ability to test the kinematical and causal structures that underlie quantum field theory in curved spacetime, rather than in reproducing the full dynamics of general relativity.

As discussed in a recent paper by Ralf Schützhold \cite{Schutzhold2025EffectiveSpacetimes}, one can argue that limitations due to degrees of freedom should not be taken as a barrier; rather, as an outline of what analogue gravity can and cannot do.  For example, the effective sonic or acoustic metric \cite{Unruh1981PRL}, involving degrees of freedom from the flow velocity, the sound speed, and the conformal factor does not span the full $10$ independent components of a $4$D Lorentzian metric, and, we cannot reproduce arbitrary spacetimes but, nevertheless it is reliable enough to simulate many QFTCS phenomena within the subset of metrics that can be realized, and to capture the essential causal structure underpinning Hawking and Unruh effects.

A recent work by Baak, Datta, and Fischer \cite{Baak2023PetrovAnalogue} uses Petrov classification to provide a systematic taxonomy of which algebraic curvature classes can arise in analogue systems. The fact that analogue models correspond to restricted Petrov types highlights that while not in any way a replacement for GR, analog systems remain entirely valid laboratory tests for universality of the physics of horizons, mode mixing, and quantum field kinematics in curved backgrounds.

As discussed in the main body, real experiments in BECs, water waves, and nonlinear optics have reproduced Hawking-like spectra, Unruh-type detector responses, and even cosmological particle creation. One can argue that these are direct observations of the same mathematics. In my view, one of the most exciting aspects of analogue models is their ability to probe the ability of entanglement transfer and partner correlations across analogue horizons, something impossible in astrophysical black holes. The fact that in BECs or optical analogues, as shown by Unruh and Schützhold, and more recently by Mann and Gooding, analogue radiation necessarily involves entangled pairs straddling the horizon, and BEC studies have proposed and demonstrated how these can be extracted via non-local density correlations.

Finally, I believe these results demonstrate why analogue gravity has matured into a genuine research field as it validates in the laboratory the universality of causal-kinematical predictions of QFT in curved spacetime, despite the fact that it leaves the dynamical questions of full quantum gravity for future theory.

\vspace{4mm}

%%%%%%%%%%%%%%%%%%%%%%%%%%%%%%%%%%%%%%%%%%%%%%%%%%%%%%%%%%%%%%%%%%%%%%%%%%%%%%%%%%%
%%%%%%%%%%%%%%%%%%%%%%%%%%%%%%%%%%%%%%%%%%%%%%%%%%%%%%%%%%%%%%%%%%%%%%%%%%%%%%%%%%%

\section*{4. Quantum Mechanics under the Gun}

%%%%%%%%%%%%%%%%%%%%%%%%%%%%%%%%%%%%%%%%%%%%%%%%%%%%%%%%%%%%%%%%%%%%%%%%%%%%%%%%%%%
%%%%%%%%%%%%%%%%%%%%%%%%%%%%%%%%%%%%%%%%%%%%%%%%%%%%%%%%%%%%%%%%%%%%%%%%%%%%%%%%%%%

Let us now turn to a topic on which all three of us are agreed. Up to this point we have assumed that we can use conventional QM and QFT without question. But we now wish to look at genuine quantum gravity, and this means that both conventional GR and conventional QM/QFT must be examined with a critical eye. In section 5 we will argue that it is QM/QFT that must be modified when trying to find a theory of quantum gravity. Here we prepare the ground for this, by recalling key features of QM: our ultimate goal is to see how QM can be modified.

%%%%%%%%%%%%%%%%%%%%%%%%%%%%%%%%%%%%%%%%%%%%%%%%%%%%%%%%%%%%%%%%%%%%%%%%%%%%%%%%%%%
\subsection*{4.1. The Fundamental Objects in QM: Orthodox View}
%%%%%%%%%%%%%%%%%%%%%%%%%%%%%%%%%%%%%%%%%%%%%%%%%%%%%%%%%%%%%%%%%%%%%%%%%%%%%%%%%%%

In the standard texts on QM, one imbibes almost by osmosis the orthodox view of QM, according to which \cite{QM-text,vonN32,bohm51,laloe2}:

{\it (i)}: A non-relativistic system $S$ is described either by a state vector
$\left|\Psi_{s}\right\rangle $ or a density matrix operator $\hat{\rho}_{s}$,
the latter involving some average over an environment $\mathcal{E}$.
These are defined using Hilbert space $\mathcal{H}_{\text{S}}$, spanned
by some set $\{ |\psi_{j}\rangle \} $ of orthonormal
states, so that $|\Psi_S\rangle =\sum_{j}c_{j} |\psi_{j}\rangle$.
Physical quantities are determined by ``measurements''; these are
defined by operators $\{ \hat{O}_{\alpha} \} $ acting in
$\mathcal{H}_{\text{S}}$, so that $\langle O_{\alpha}\rangle  = \langle \Psi_{\text{S}}|\hat{O}_{\alpha}|\Psi_{\text{S}}\rangle $
(or by $\langle \hat{O}_{\alpha}\rangle =\text{Tr}\,\hat{\rho}_{\text{S}}\hat{O}_{\alpha}$
when we describe $\mathcal{S}$ using $\hat{\rho}_{\text{S}}$). We
also define a Hamiltonian operator $\hat{\mathcal{H}}_{\text{S}}$
, so that $\left|\Psi_{\text{S}}\left(t\right)\right\rangle $ evolves
in time according to $\hat{\mathcal{H}}_{\text{S}}\left|\Psi_{\text{S}}\right\rangle =i\hbar\partial_{t}\left|\Psi_{\text{S}}\right\rangle $
, the Schrödinger equation.

{\it (ii)}: Measurements of $\langle \hat{O}_{\alpha}\rangle $ involve
the interaction of $\mathcal{S}$ with some measuring apparatus $\mathcal{M}_{\alpha}$;
if we treat $\mathcal{M}_{\alpha}$ as a quantum system, described
by states $\{ \Phi_{\alpha}^{\left(k\right)} \} ,$then
the measurement process involves entanglement between $\mathcal{S}$
and $\mathcal{M}_{\alpha}$ in such a way that the state of $\mathcal{M}_{\alpha}$
\uline{after} the measurement interaction is correlated with that
of $\mathcal{S}$ \uline{before} the interaction. The simplest example is
an ``ideal measurement'' (called by Pauli \cite{QM-text}, a ``measurement
of the 1st kind''), in which $\mathcal{M}_{\alpha}$ begins in a
state $|\Phi_{\alpha}^{\left(k\right)}\rangle $, and
one has the transition
\begin{equation}
\left|\chi_{\text{in}}\right\rangle \;=\;  |\Psi_{\text{S}}\rangle |\Phi_{\alpha}^{\left(0\right)}\rangle \;\;=\;\; |\Phi_{\alpha}^{\left(0\right)}\rangle \sum_{j}c_{j}|\psi_{j}\rangle  \;\; \underset{\text{"mmt"}}{\longrightarrow} \;\;\sum_{j}c_{j}|\psi_{j}\rangle |\Phi_{\alpha}^{\left(j\right)}\rangle   \qquad\qquad\quad  \text{(1st kind)}
  \label{eq:transition state}
\end{equation}
\\
in which the $|\Phi_{\alpha}^{\left(j\right)}\rangle $
are an orthonormal (sub)set of the states $\mathcal{M}_{\alpha}$
which are in 1-1 correlation with the $\{ |\psi_{j}\rangle \} $
of $\mathcal{S}$. We note that the state $\mathcal{S}$ is left unchanged
(in a non-ideal measurement of the 2nd kind, the state $\mathcal{S}$
would change; indeed $\mathcal{S}$ might even be destroyed), and
that $\mathcal{S}$ and $\mathcal{M}_{\alpha}$ are now entangled.
One also defines a ``reproducible measurement'' or ``state preparation''
if the measurement has a unique outcome, so that we know what the
state of $\mathcal{S}$ is after the measurement (this is of course not the only way that state preparation can occur).

{\it (iii)}: The world is divided into a classical part, in which definite results
and states exist, and the quantum part, described above. At some point
the final state $|\chi_{f}\rangle =\sum_{j}c_{j}|\psi_{j}\rangle |\Phi_{\alpha}^{\left(j\right)}\rangle $
undergoes ``wave-function collapse'' to \uline{one} of the states
$|\chi_{j}\rangle = |\psi_{j}\rangle |\Phi_{\alpha}^{\left(j\right)}\rangle$,
with probability $\left|c_{j}\right|^{2}=p_{j}$; at the same point
the density matrix collapses to the diagonal mixed state $\hat{\rho}_{\text{s}}=\left|\chi_{j}\left\rangle p_{j}\right\langle \chi_{j}\right|$.
From this moment onwards, the measurement has been performed, and
we have a definite result, and the measuring system $\mathcal{M}_{\alpha}$
is in a definite classical state in the classical world. Neither the
ineffable wave-function collapse, nor the behaviour in the classical
world, is describable in QM, or indeed at all in the current theory.

Since the early period of quantum mechanics (up to c.1970) a number of other key ideas have been added to the canon of orthodox QM. (although textbooks have been very slow to catch up!). Amongst these are:

{\it (a)}: The possibility of high-order entanglement and coherence
involving many degrees of freedom. From this vague idea of multi-particle
entanglement were born more precise ideas about macroscopic quantum
coherence \cite{AJL80+MQC} and quantum computation \cite{RPF-QIP}, involving multi-particle
entanglement in an essential way. Although the experimental realization
of genuinely macroscopic coherence,
and of viable quantum computation, still need a lot
of work (see refs. \cite{whaley+AJL,taka09,arndt14}), we do have a much better theoretical
understanding of multi-particle entanglement \cite{multiP,cox18}.

{\it (b)}: The key phenomenon of environmental
decoherence. This is the process by which system $\mathcal{S}$ entangles
with the environment $\mathcal{E}$, so that when one averages over
$\mathcal{E}$ to get a reduced density operator $\rho_{\text{s}}$
for $\mathcal{S}$, coherence and interference phenomenon are lost. Although the idea of decoherence has a very long history (see Appendix), it was not made precise until realistic models for $\mathcal{E}$,
and the coupling between $\mathcal{S}$ and $\mathcal{E}$ became
available, starting with the work of Caldeira and Leggett \cite{cal+AJL83}.
We now have very detailed models describing both
delocalized modes \cite{cal+AJL83} and localized modes \cite{PS00} in the environment $\mathcal{E}$.

	        %%%%%%%%%%%%%%%%%%%%%%%%%%%%%
	
\begin{figure}
		\includegraphics[width=5.2in]{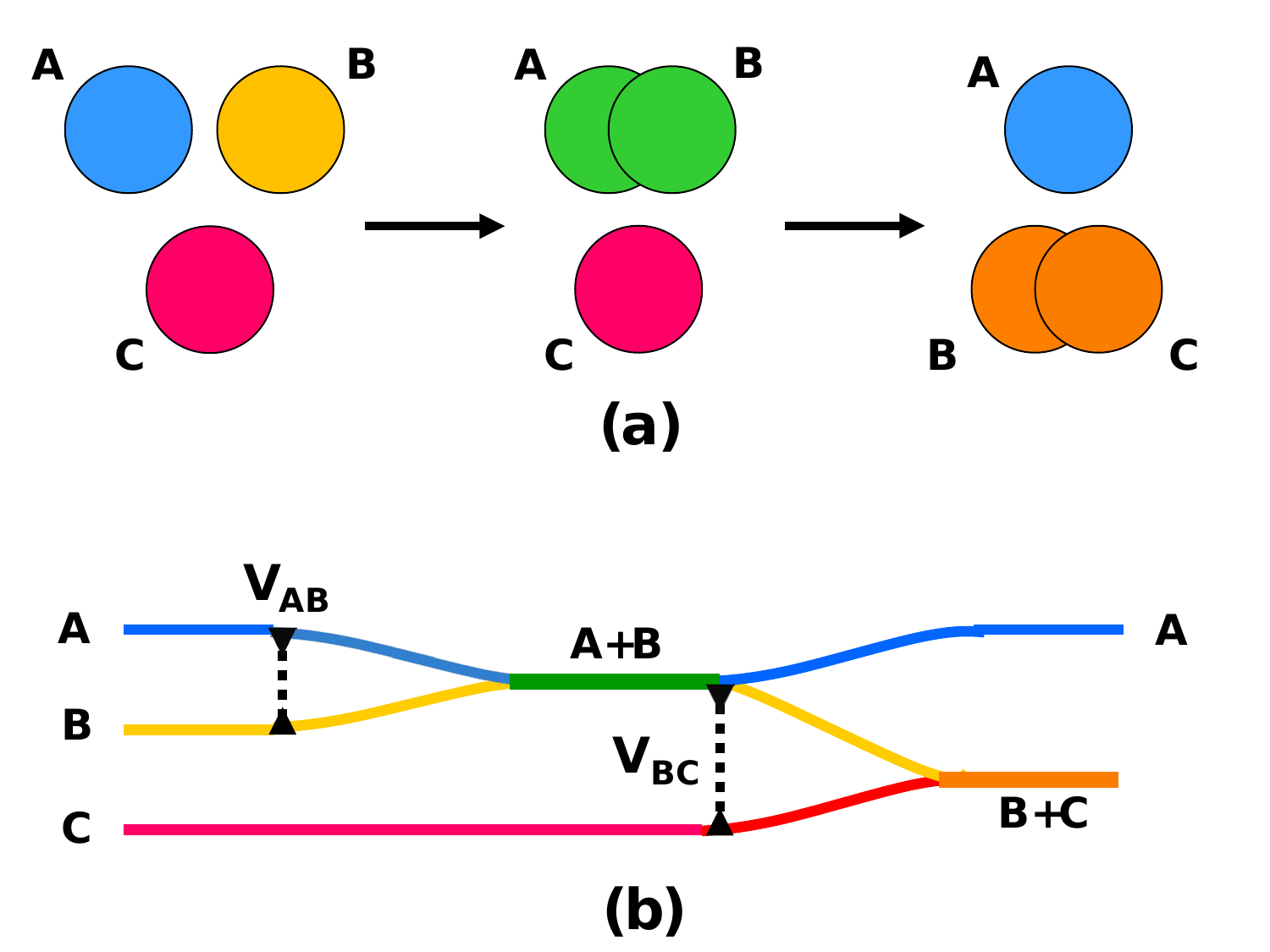}
		\caption{\label{fig:TDS9-monogamy} Illustration of the entanglement process between 3 systems, discussed in the text. In (a) at top we show schematically the 3 different states of entanglement between systems A, B, and C. initially they are unentangled, then A and B are entangled, and finally B and C are entangled. In (b) below, the time evolution of the entanglement is shown schematically, with time along the horizontal axis - the interaction $V_{AB}$ initiates entanglement between A and B, and then the interaction $V_{BC}$ decouples (and disentangles) A and B, and initiates entanglement between B and C.  		} 	
\vspace{5mm}
\end{figure}

	            %%%%%%%%%%%%%%%%%%%%%%%%%%%%%

However, this raises an interesting question: since decoherence is
a one-way process in time, and very difficult to reverse, how are
state preparation and coherent states even possible? We note that
state preparation and independent coherent states evolve quite naturally
both on earth and in intergalactic space, quite independent of any
``intelligent'' intervention. The answer is illustrated in Fig. \ref{fig:TDS9-monogamy};
one can ``disentangle'' two systems A and B from each other, just
by allowing one of them (eg. system B) to interact and entangle with
another system C. Then B and C are entangled, leaving A disentangled
from both of them. This result is an example of ``quantum monogamy''
and happens all the time, throughout the universe. If system $B$ is the environment $\mathcal{E}$, then we see how we can decouple $A$ from the environment by entangling it with system $C$.

From all of this we can see that in orthodox QM, the important theoretical
objects are the state vector $|\Psi_{\text{s}}\rangle $
in Hilbert space $\mathcal{H}_{\text{s}},$ the set of operators $\{ \hat{O}_{\alpha} \} $
which operate on $|\Psi_{\text{s}}\rangle $, and the idea
of a measuring system $\mathcal{M}_{\alpha}$ which, through the process
of wave-function collapse, leads to a projection of $|\Psi_{\text{s}}\rangle $
onto some specific eigenstate of $\hat{O}_{\alpha}$, with a probability
specified by Born's rule.

No fundamental changes occur when we go to QFT; wave-functions are
replaced by wave-functionals, defined again in a Hilbert space, and
we still have measuring systems and wave-functional collapse. In conventional
quantum gravity, things become a little more complicated, since it
is not clear what the correct physical observables are. However,
treated as a gauge theory, conventional quantum gravity is no different
in its structure from, eg., non-Abelian gauge theory.

Up to this point we have not mentioned the path integral formulation of QM. This is because we will take the view that it differs in very important ways from the orthodox
view just given - we discuss it in section 4.3 below.

%%%%%%%%%%%%%%%%%%%%%%%%%%%%%%%%%%%%%%%%%%%%%%%%%%%%%%%%%%%%%%%%%%%%%%%%%%%%%%%%%%%
\subsection*{4.2. Key Problems and Questions in Quantum Mechanics}
%%%%%%%%%%%%%%%%%%%%%%%%%%%%%%%%%%%%%%%%%%%%%%%%%%%%%%%%%%%%%%%%%%%%%%%%%%%%%%%%%%%

Why would anyone doubt the validity of QM at the macroscopic scale? Even now, most physicists believe that there is nothing wrong with quantum mechanics (although this view is certainly not shared by any of the three co-authors of this paper!). 
This question has of course has been controversial; some of the views
on this are as follows:

\vspace{2mm}

{\it (i): \uline{The ``Plain Vanilla'' orthodox view}}: This was formulated long ago \cite{jammer}.
It is an amalgam of Copenhagen and von Neumann arguments, and
appears in well-known textbooks written in the early days of QM \cite{QM-text,londonB38}.
In this view, we know how to use QM and it works, therefore we understand
it and there is nothing to worry about. One simply has to learn the textbook rules involving states in (some) Hilbert space, projections and measurements, and
a fundamental division between the classical and quantum worlds (with
the quantum world defined in terms of objects in the classical world).

There are many problems with this view \cite{bellB}. In the first place, being
able to use something, or follow instructions for its use, does not
mean one understands it! Children and adults with no scientific understanding
at all can use cell phones, laptops, and all manner of other sophisticated
devices with no understanding of how they work. The simple fact is that the orthodox
interpretation has deep and ineluctable problems which could be buried
so long as the domain of QM was restricted to microscopic systems,
or (as in the case of superfluids and superconductors) to systems
described by product states over microscopic degrees of freedom \cite{yang61}.

All of this changed when Leggett pointed out \cite{AJL80+MQC} the possibility that
under very specific circumstances - an absence of decoherence sources
- one could envisage Schrodinger Cat states \uline{{\bf in the lab}}, of form $|\Psi\rangle = \left( a \prod_{j}^{N} |\varphi_{j}^{a}\rangle +b \prod_{j}^{N} |\varphi_{j}^{b}\rangle \right)$
for an $N$-particle system, with $N$ taken to be macroscopic.
Unfortunately, the states of this kind observed so far either have
values of $N$ no larger than $\sim\mathcal{O}\left(10^{4}\right)$
(as with interference experiments with molecules), or have almost
complete overlap between $\left|\varphi_{j}^{a}\right\rangle $ and
$\left|\varphi_{j}^{b}\right\rangle $ (ie., one has both $1-|\langle \varphi_{j}^{a}|\varphi_{j}^{b}\rangle |=\delta_{j}\ll1$, for
all $j$, and also $\sum_{j}\delta_{j} \sim O(1)$). For a proper discussion, see the papers of
Whaley $et.al$ \cite{whaley+AJL}.

If, however, experiments do succeed in preparing systems in which
genuine macroscopic coherence is found, then the orthodox interpretation
(and many other interpretations) of QM will be in big trouble!

\vspace{2mm}

{\it (ii) \uline{The ``Information Theoretic'' interpretation}}:  According to
this view, QM is no more nor less than an account of the information possessed
by experimenters, based on outcomes of measurements (see, eg., Laloe \cite{laloe2}). While
popular in some quarters, this extraordinarily anthropocentric interpretation apparently
abandons any attempt to find a description of a physical world which is
independent of observers and observations - it has been ridiculed
for this reason \cite{bellB}. It is clearly incompatible with any extrapolation
of QM to regions of spacetime that are unobserved, either because
they lie behind horizons or because no measurements have ever been
performed on them.

\vspace{2mm}

{\it (iii) \uline{The ``Multi-Time'' interpretations}}: these include the ``two-time''
or ``weak'' measurement approaches, and began with the well-known work
of Aharonov, Bergmann, and Lebowitz \cite{ABL64}, and in the case of 2-time
measurements, with the key result that for some operator $O_{\alpha}$$\left(x\right)$,
\begin{align}
\chi_{\alpha}\left(2,1|x\right) & =\frac{\left\langle \psi_{2}\left(t_{2}\right)\left|\hat{O}_{\alpha}\left(x\right)\right|\psi_{1}\left(t_{1}\right)\right\rangle }{\left\langle \psi_{2}\left(t_{2}\right)|\psi_{1}\left(t_{1}\right)\right\rangle }\label{eq:2-time measurement}\\
 & =\frac{\int_{1}^{2}\mathcal{D}q\left(t\right)\,e^{\frac{i}{\hbar}S\left[q\right]}O_{\alpha}\left(x\right)}{\int_{1}^{2}\mathcal{D}q\left(t\right)\,e^{\frac{i}{\hbar}S\left[q\right]}}\nonumber
\end{align}
where $\int_{1}^{2}\mathcal{D}q\left(t\right)=\left\langle \psi_{2}\left(t_{2}\right)|q_{2}\left(t_{2}\right)\right\rangle \int_{q\left(t_{1}\right)=q_{1}}^{q\left(t_{2}\right)=q_{2}}\mathcal{D}q\left(t\right)\left\langle q_{1}\left(t_{1}\right)|\psi_{1}\left(t_{1}\right)\right\rangle $.
We see that $\chi_{\alpha}\left(2,1|x\right)$ is just the amplitude
for the system (here a particle) to propagate from the state $\psi_{1}\left(t_{1}\right)$
at time $t_{1}$, to state $\psi_{2}\left(t_{2}\right)$ at time $t_{2}$,
\uline{conditional} on there being a measurement of $\hat{O}_{\alpha}$
performed at $x$ somewhere along the trajectory between $q_{1}$
and $q_{2}$. Alternatively, we can think of it as a measurement of
$\hat{O}_{\alpha}$ at $x$, \uline{conditional} on the system having
started out in state $\left|\psi_{1}\left(t_{1}\right)\right\rangle $
and finished at $\left|\psi_{2}\left(t_{2}\right)\right\rangle $.
This latter interpretation is the one that has been adopted in recent
years \cite{drexel}.

There has been considerable discussion, particularly in the context
of delayed choice \cite{delayZ-RMP} and quantum eraser \cite{QEraser} experiments,
of the extent to which one can think of ideas like prediction and
retrodiction when QM is formulated in this way \cite{retroD}. The same
questions arise when one considers interpretations like the ``transactional
interpretation'' of QM, in which both retarded and advanced signals
propagate between (a) the QM system of interest and (b) both past and future
measuring systems \cite{cramerTr}.

We note that all of these multi-time interpretations still maintain the central role of measurements. As such they are open to the same criticisms as orthodox QM, connected with the division of the world into quantum and classical parts, and the anthropocentric role assigned to measurements.

\vspace{2mm}

{\it (iv) \uline{Deterministic interpretations}}: in all of the approaches to QM just described, one keeps the usual QM paraphernalia of measurements and
the operators associated with them, and one also assumes that the element of \uline{choice} on the part of observers is an essential part of the measurement
process (so that QM probabilities for measurement outcomes are assigned to different choices). The idea that conscious choices are made in experiments is considered
to be more or less axiomatic to the whole business of experimental
science by most physicists. It was built into the QM formalism
from the very beginning (without however, ever making it explicit in the form of an ``axiom"). Choices are made, and as a consequence the
deterministic Schodinger equation is superseded by a probabilistic ``wave function
collapse''. As noted by several authors \cite{choice}, all of the paradoxical
features of QM, along with Bell's inequalities, disappear if one
instead assumes an entirely deterministic theory of QM, where ``choice" does not exist, and where the
Schrodinger equation (or some QFT generalization) is universally valid.

The question of whether one can use a ``wave function for the universe''
is of course intimately bound up with this discussion; the moment
one tries to describe the entire universe by QM/QFT, one can no longer
invoke the existence of external observers, and wave function collapse
becomes moot; we end up with a deterministic universe.

There have been two prominent attempts at a deterministic theory of QM. The first was by Bohm \cite{bohm52}, who reprised older work of de Broglie \cite{deBroglie} to come up with a very interesting {\it non-linear} version of QM, in which a ``quantum potential" $Q({\bf r})$ is added to the Schrodinger eqtn. of motion; this function itself depends on the ``pilot wave" function $\psi({\bf r}, t)$ (hence the non-linearity). However $\psi({\bf r}, t)$
now determines real trajectories for the system - the probabilistic features now come in
entirely traditional fashion, from our lack of knowledge of the initial conditions (hence the
description of Bohm's theory as a ``hidden variable theory"). Equally interesting is
Bohm's view of quantum measurements, in which the hidden variables depend on both the
system ${\cal S}$ and the measuring system ${\cal M}$ in interaction with it.

This brings us to the somewhat controversial ``relative state'' view of QM, which since the influential work of DeWitt \cite{manyW} has been known as the ``many worlds''
interpretation. This appelation is unfortunate, since the original formulation
of Everett \cite{everett} did \underline{{\bf not}} involve a multitude of worlds -
it instead simply demands that the \underline{{\bf single}} world we live in
be described by a single wave-function. Crucially, the special role of measurements
is completely removed from the theory, and wave-function collapse never occurs.
Measurements are then just another physical process occurring in the
universe, and observers like us are physical systems like any other. Clearly a relative state formulation of QM/QFT is appealing to those working in quantum gravity; hence its popularity amongst cosmologists.

It is often argued that the weakness of deterministic approaches is that they do not explain how definite results are obtained from quantum mechanical superpositions, unless one invokes non-linearity (which is what Bohm did \cite{bohm52}). In an Appendix we say a little more about the ideas of Bohm and Everett, and some of the related history, since some of these ideas, and the related history, are misunderstood.

\vspace{2mm}

{\it (v) \uline{The Role of Decoherence and Disentanglement}}: one way to try and bridge the gap between quantum superpositions and definite results has gained currency in recent years. This comes when one considers how different sectors of the universe ``entangle" with, and ``disentangle'' from,
each other. Entanglement of a system ${\cal S}$ with its environment ${\cal E}$, either directly or via ``3rd parties" \cite{SHPMP06} leads to environmental decoherence. Decoherence is a complex process, not
well captured by toy models; but it is known that one can divide the ``environments'' responsible for decoherence into ``oscillator bath'' environments \cite{cal+AJL83}, describing
delocalized modes (phonons, photons, etc.) and ``spin bath'' environments
\cite{PS00}, describing localized modes (defects, dislocations, nuclear
and paramagnetic spins, etc.).

However, systems are also constantly disentangling from their surroundings, and from each other. That this can happen is primarily because of ``quantum monogamy"; If a system ${\cal A}$ is entangled with another system ${\cal B}$ (and ${\cal B}$ might be the environment), we can disentangle them by entangling one of them (eg., ${\cal A}$) with another system ${\cal C}$; this automatically disentangles ${\cal A}$ from ${\cal B}$. This happens quite naturally without intervention from humans or other complex systems - it is happening all over the universe, all the time.

All this is interesting, purely as physics. But there is a school of thought which argues that decoherence also ``solves the measurement problem", by suppressing off-diagonal ``interference" matrix elements in the basis of the system-environment interaction, in the reduced density matrix for the system. That this argument is incorrect has been repeatedly emphasized \cite{deco-false}; one can always in principle (and increasingly in practise) reverse the effects of decoherence, and the interference terms are always ``out there", at least in the framework of orthodox QM. Thus, in the end, decoherence does not help solve the paradoxes of QM.

In the Appendix we also say a little about the history of ideas about decoherence (which has some interesting relations to the history of Bohm and Everett's work).

\vspace{4mm}

We see that none of these attampts to ``interpret" QM is really satisfactory. The only approach that does seem to offer any way out of the QM paradoxes is to try and alter QM; for example, by making it non-linear. There are both pitfalls and promise in this approach, as we will now see.

%%%%%%%%%%%%%%%%%%%%%%%%%%%%%%%%%%%%%%%%%%%%%%%%%%%%%%%%%%%%%%%%%%%%%%%%%%%%%%%%%%%
\subsection*{4.3. Alternative Objects and Structures in QM}
%%%%%%%%%%%%%%%%%%%%%%%%%%%%%%%%%%%%%%%%%%%%%%%%%%%%%%%%%%%%%%%%%%%%%%%%%%%%%%%%%%%

In this sub-section we will argue that the orthodox structure of QM
is not inappropriate for the discussion of certain well-established
physical phenomena, and that there are better ways to do things. In particular we will argue that (i) the path integral formulation of QM, which is \underline{{\bf not the same as}} orthodox QM, captures key features of QM that are not accessible to the orthodox framework; and (ii) that although non-linear theories of QM have problems, one can bypass them (and indeed solve them) in a proper theory of quantum gravity.

%%%%%%%%%%%%%%%%%%%%%%%%%%%%%%%%%%%%%%%%%%%%%%%%%%%%%%
\subsubsection*{4.3 (a) Path Integrals and QM}
%%%%%%%%%%%%%%%%%%%%%%%%%%%%%%%%%%%%%%%%%%%%%%%%%%%%%%

It was first remarked by C Morette \cite{morette} that the path integral formulation of QM is {\it not} equivalent to orthodox QM. Her point was that there are non-local features of QM that exist in path integrals, but not in wave-functions.

	        %%%%%%%%%%%%%%%%%%%%%%%%%%%%%
	
\begin{figure}
		\includegraphics[width=7.2in]{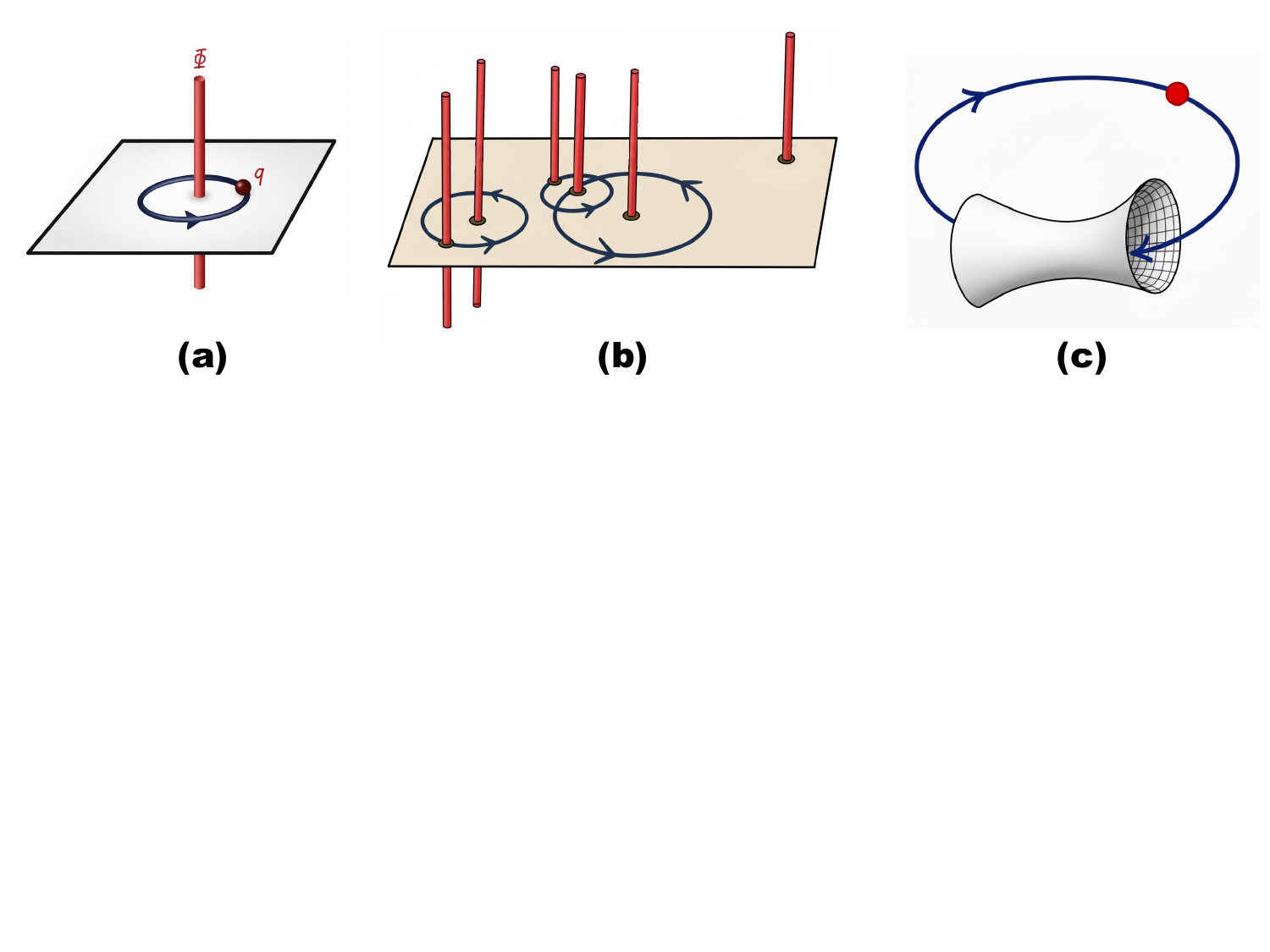}
	\vspace{-8.6cm}
		\caption{\label{fig:TDS10-anyons} Three phenomena which require path integrals for their description. In (a) we show the Aharonov-Bohm effect; motion of a charged particle around the flux tube adds a topological phase to the particle amplitude. In (b) we show how topological phases are incurred when anyons move around each other. In (c) we show a system moving through an achronal region of specetime - here it is a wormhole, and again an extra topological phase is incurred. If the system undergoes a circuit, as shown here, then we have a closed time curve (CTC).    } 	
		\vspace{4mm}
\end{figure}

	        %%%%%%%%%%%%%%%%%%%%%%%%%%%%%

Let's begin with three physical phenomena which make this point very clearly (and which are depicted in Fig. \ref{fig:TDS10-anyons}):

{\it (a) The Aharanov-Bohm effect}: In its original formulation \cite{AhB}, the Aharonov-Bohm effect was formulated as a scattering effect; but it is more interesting (and experimentally relevant) to formulate it in a ring geometry. As first noted by Feynman \cite{RPF-vol3}, it is very simple to formulate it in terms of a sum over paths, with each path accumulating an extra phase $\phi_{AB} = (q/h) \oint d{\bf r} \cdot {\bf A}({\bf r})$ around a ring, where ${\bf A}({\bf r})$ is the vector potential and $q$ the charge. A wave-function has to be defined on an infinite number of sheets; defined as a simple function of ${\bf r},t$ it is multi--valued, and nonsensical. The path integral description is extremely simple \cite{RPF-vol3}, and very natural.

{\it (b) Anyon particles/quasiparticles}: as first shown by Laidlaw and C. Morette
\cite{laidlaw}, in 2 dimensions, particles can show fractional statistics - these are now
called ``anyons" and exist as, eg., quasiparticles in Fractional Quantum Hall (FQHE)
systems \cite{FQHE}. The phase accumulated by moving one anyon around another is fixed but arbitrary; the theory is easily formulated using path integrals but the $N$-anyon wave-function is multi-valued as a function of all of its arguments - again, nonsensical. The path integral description \cite{YSWu84} is simple and illuminating.

{\it (c) Particles in achronal spacetime}: particles or fields passing through achronal regions of spacetime (eg., through a Kerr black hole, or a wormhole) have no Hamiltonian at all - their dynamics can {\it only} be defined using path integrals \cite{CTC}, and this is the only way that we are able to discuss, eg., closed time curves in quantum gravity.

\vspace{2mm}

We see that to try to use orthodox QM for the first 2 examples is misguided, and for the 3rd example, impossible. In all three cases, the non-local features of the QM system require path integrals for their description. All we will need for this paper is contained in the formulas for the ``generating functional" ${\cal Z}_o[{\bf j}]$ and the propagator $K_o(2,1) \equiv K_o({\bf r}_2, {\bf r}_1; t_2,t_1)$ for the system. The simplest example is a particle with coordinate ${\bf r}(t)$, for which we have \cite{CWL3}
\begin{equation}
		{\cal Z}_o[{\bf j}] \;=\; \oint {\cal D} {\bf r}(t) \; e^{i(S_o[{\bf r}, {\bf \dot{r}}] + \int {\bf j}\cdot {\bf r})}
		\label{Zo-QM}
	\end{equation}
where the path integration $\oint {\cal D} {\bf r}(t)$ is taken over a set of closed ``ring paths" and ${\bf j}(t)$ is an arbitrary ``force" applied to the system. The propagator is then:
\begin{equation}
			K_o(2,1) \;=\; \int_1^2 {\cal D} {\bf r}(t) e^{iS_{o}[{\bf r}, {\bf \dot{r}}]}
			\label{Kqm-21}
		\end{equation}
where the path integration $\int_1^2 {\cal D} {\bf r}(t)$ now extends from $1$ to $2$, ie., from ${\bf r}_1(t_1)$ to ${\bf r}_2(t_2)$. One obtains $K_o(2,1)$ from ${\cal Z}_o[{\bf j}]$ by a process of cutting the closed ring path \cite{CWL3}. Note that the action in eqtns. (\ref{Zo-QM}) and (\ref{Kqm-21}) will in general contain topological terms like the Aharonov-Bohm contribution $\phi_{AB}$. 

The key point to emphasize here is that both ${\cal Z}_o[{\bf j}]$ and $K_o(2,1)$ are {\it non-local objects}. They are defined throughout spacetime: the
propagator has information about all of the paths between the initial state $|1\rangle$
and the final state $|2\rangle$, and the object $K_o(2,1)$ is very different from the states at each end of it. In the 3 examples given above, it contains all of the non-local information that is missing from an attempt at a definition in terms of states in Hilbert space.

%%%%%%%%%%%%%%%%%%%%%%%%%%%%%%%%%%%%%%%%%%%%%%%%%%%%%%
\subsubsection*{4.3 (b) Non-Linear QM}
%%%%%%%%%%%%%%%%%%%%%%%%%%%%%%%%%%%%%%%%%%%%%%%%%%%%%%

It is not immediately obvious what one means by a non-linear QM: a large variety have been suggested. One can define it simply as a quantum theory in which the equation of motion for the quantum states evolves non-linearly (so that superpositions are not preserved). A related definition required the Hamiltonian to be a function of the state vector. But these definitions are too restrictive - what if a Hamiltonian can't be defined, or if the theory is not written in terms of state vectors (cf. our discussion immediately above)? The problem becomes particularly acute in dealing with non-linear versions of QFT.

Here we will simply define a non-linear QM as one in which the superposition principle is violated, no matter how the theory is set up.

Examples include (i) the work of Kibble \cite{kibble89}, who introduced non-linearity by coupling the QM state vector $|\psi \rangle$ to the spacetime metric $g^{\mu\nu}(x)$, and letting the  Einstein equations of motion induce the non-linearity in the dynamics of $|\psi \rangle$; and (ii) the work of Weinberg \cite{weinberg89}, who used the standard toolbox of Hilbert space, measurements, etc., but with a built-in dependence of the Hamiltonian on the state vector. It was quickly shown that a broad class of non-linear theories, including Weinberg's,  led to the very undesirable feature of superluminal propagation of information.

It is then extremely interesting to notice that the much earlier 1952 Bohm theory \cite{bohm52} is also a genuinely non-linear theory of QM. Yet no superluminal communication appears in Bohm's theory! Thus remarkably, Bohm's theory not only undermines von Neumann's ``proof" that no consistent hidden variable theory exists \cite{bell66}; it also undermines the more recent claims that non-linear QM theories must necessarily have superluminal signal propagation.

A very careful analysis of this question was given by Polchinski \cite{polch91}, who pointed out that the observables in a non-linear QM which would \underline{\bf not} lead to superluminal communication were those which could be written as a function of the density matrix of the system. Elsewhere we will discuss Bohm's theory in this light.

Notice that none of these analyses applies to a quantum field theory (although Kibble did look at a non-linear dynamics for a scalar field functional \cite{kibble89b}). As Kibble himself remarked, the principal problem with all of these efforts was that they were too closely based on the existing structure of QM, and assumed that the usual apparatus of measurements, Hilbert space, etc., could be taken over to the non-linear theory.

\vspace{3mm}

$\qquad\qquad\qquad\qquad\qquad\qquad\qquad----------------------$

\vspace{3mm}

Let's summarize our point of view on all of this. This is that without the use of alternative structures to describe QM phenomena, it is impossible to solve the paradoxes and contradictions of the subject. In the next section we discuss how path integrals can be used to give a non-linear theory of quantum gravity, which then results in a non-linear QM in the non-relativistic domain. This leads to different predictions for low-energy experiments than the predictions of QM (and, we argue, it deals with some of the problems found in orthodox QM).

\vspace{4mm}

%%%%%%%%%%%%%%%%%%%%%%%%%%%%%%%%%%%%%%%%%%%%%%%%%%%%%%%%%%%%%%%%%%%%%%%%%%%%
%%%%%%%%%%%%%%%%%%%%%%%%%%%%%%%%%%%%%%%%%%%%%%%%%%%%%%%%%%%%%%%%%%%%%%%%%%%%

\section*{5. The Road to Quantum Gravity}

%%%%%%%%%%%%%%%%%%%%%%%%%%%%%%%%%%%%%%%%%%%%%%%%%%%%%%%%%%%%%%%%%%%%%%%%%%%%
%%%%%%%%%%%%%%%%%%%%%%%%%%%%%%%%%%%%%%%%%%%%%%%%%%%%%%%%%%%%%%%%%%%%%%%%%%%%

The most popular approach to quantum gravity is to regard it as a high-energy problem. This makes it hard to test. In this article we adopt a different view, viz., that the key to resolving the conflict between GR and QFT is to be found at low energies. We begin by comparing the high- and low-energy approaches, with their radically different assumptions and methodologies. We then go on to discuss a theory - the CWL theory - which we believe surmounts the main problems confronting a theory of low-$E$ gravity, and which also solves the problems confronting QM that were listed in section 4. Finally, we briefly discuss experiments in quantum gravity.

%%%%%%%%%%%%%%%%%%%%%%%%%%%%%%%%%%%%%%%%%%%%%%%%%%%%%%%%%%%%%%%%%%%%%%%%%%%%%%%%%%
\subsection*{5.1.  High Energy vs. Low-Energy Approaches}
%%%%%%%%%%%%%%%%%%%%%%%%%%%%%%%%%%%%%%%%%%%%%%%%%%%%%%%%%%%%%%%%%%%%%%%%%%%%%%%%%%

In the high-energy  approach, one argues that the key is to be found at the Plank
energy scale. This is a very high energy scale. To get a feeling for the numbers, note that
the Plank mass $M_{\text{P}}=\left(\hbar c/G\right)^{1/2}\simeq 2.18\times10^{-8}$kg,
the Planck energy $E_{\text{p}}=\left(\hbar c^{5}/G\right)^{1/2}\simeq 1.96\times10^{9}$J
$\equiv 1.22\times10^{19}$GeV $\equiv\,1.42\times10^{32}$K, and the
Planck energy density $\rho_{\text{P}}=E_{\text{p}}/\ell_{\text{p}}^{3}\simeq 2.61\times10^{123}$~{GeV}/$m^3$, with the Planck length $\ell_{\text{p}}=\left(\hbar G/c^{3}\right)^{1/2} \simeq 1.616 \times 10^{-35}$m. 

We see that a very small sand grain, with diameter $\sim$ 0.4 mm and having a Planck mass, would if converted to a Planck energy, heat 5 tons of H$_{2}$O
from $0^{\circ}$C to $100^{\circ}$C. To experimentally test theory
at the Planck scale, we would have to deposit this colossal energy
into a volume $\ell_{\text{p}}^{3}$ in size. The resulting energy
density $\rho_{\text{p}}$ is hard to imagine; it is $10^{45}$ times
higher than the highest energy densities realized in particle collisions
in the Large Hadron Collider (LHC), and only imaginable in the very
early universe, at timescales $\sim t_{\text{p}} = 5.39\times10^{-44}$ after the initial singularity (if such a singularity even exists).

	        %%%%%%%%%%%%%%%%%%%%%%%%%%%%%
	
\begin{figure}
		\includegraphics[width=6.2in]{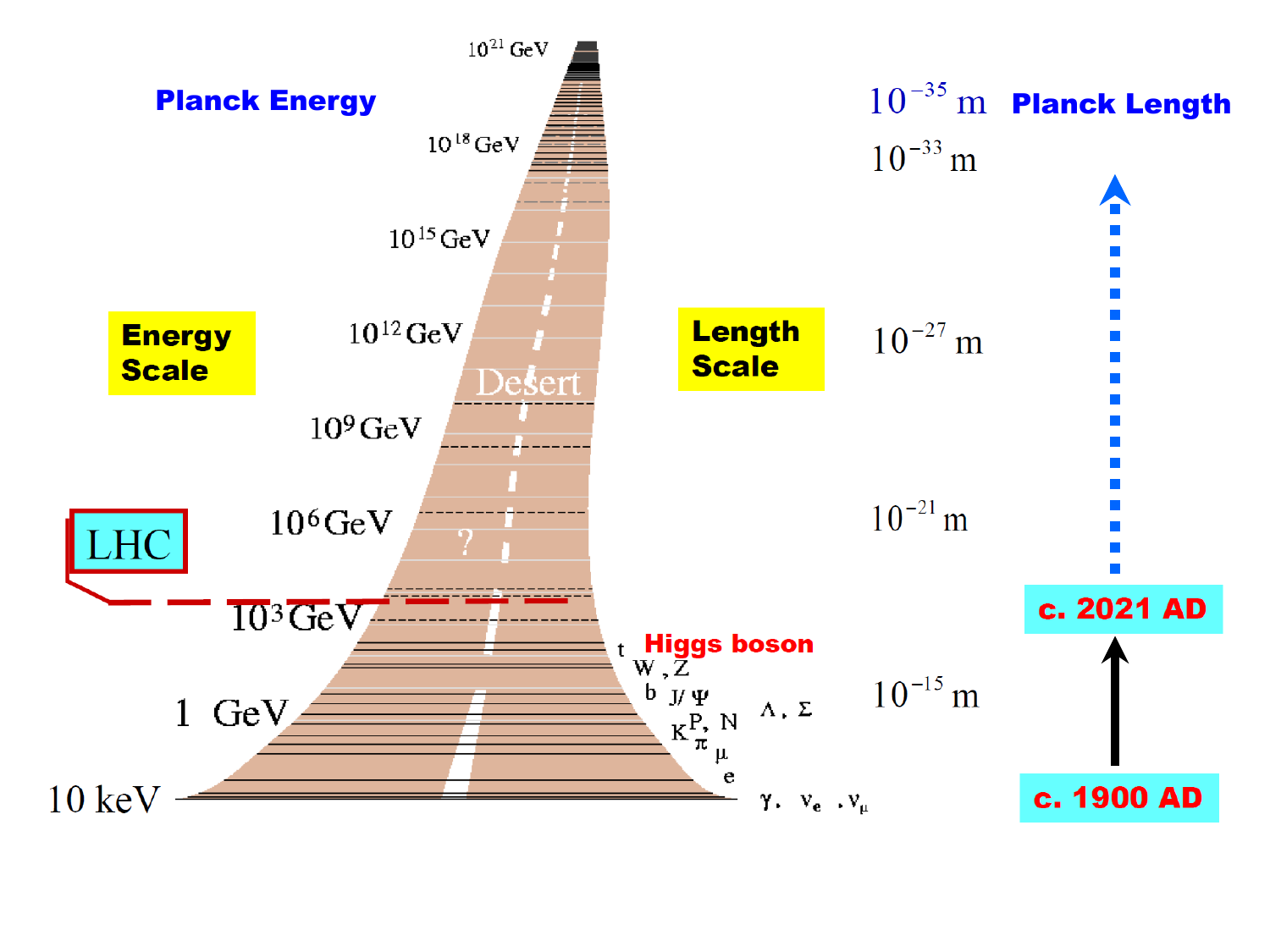}
	\vspace{-1.5cm}
		\caption{\label{fig:TDS11-highRoad}  The high-energy road to quantum gravity. In the last 125 yrs., experiments have gone from roughly 10 keV to centre of mass energies $\sim 1.4 \times 10^4$ GeV, increasing roughly exponentially with time. To get to the Planck scale $E_p \sim 1.22 \times 10^{19}$ GeV will take a long time.   } 	
	\vspace{4mm}
\end{figure}

	            %%%%%%%%%%%%%%%%%%%%%%%%%%%%%

These numbers create a methodological problem for Planck-scale theories. The problem is illustrated in Fig. \ref{fig:TDS11-highRoad}, which depicts the way in which
theory has been tested at progressively higher energies in the last
120 years or so. One sees that if history is a guide, Planck scale theories will be
untestable for another 300 years. The only way out of this
impasse is if the universe itself can provide us with indirect evidence
for one approach or another, in gravitational waves created at the
beginning of the universe - this was in fact Starobinsky's hope.

In the last section we reviewed some of the conceptual problems that arise in QM and QFT. How does the high-energy approach deal with these? Basically it ignores them. To explain the ``quantum/classical division", and the validity of GR at the macroscopic scale, one either invokes some orthodox interpretation like the Copenhagen interpretation, in which this division is simply assumed; or it is argued that QM effects are washed out (eg., by decoherence) once one reaches macroscopic scales. But in any case, the realm of high-energy theories is the ``very small", far from the problems that arise for QM at the macroscopic scale.

Rather than deal with the problems of QM, it is instead argued that some variety of superstring theory is {\it internally consistent}, ie., that it has passed all the consistency tests one might think of (no unphysical sectors in the space of states; satisfaction of all Ward identities and their concomitant conservation laws, along with all the other requisite symmetries; correct behaviour as $\hbar \rightarrow 0$, consistent behaviour as coupling constants $\rightarrow 0$; and so on).  Since one can't do experimental tests of string theory, and one has already dismissed the problems of QM at the macroscopic scale, these arguments of consistency, along with (highly subjective) considerations of mathematical beauty, are regarded as sufficient.

\vspace{3mm}

There is however another methodological hurdle facing any theory of quantum gravity,
which has nothing to do with experimental tests, or with the problems of QM {\it per
se}. This is the question of internal consistency at {\it low energies}. There are
various ways of describing this. In a canonical formulation of the theory, one has to
calculate field commutators; one such is the commutator of the metric field ${\cal C}
^{\mu\nu\alpha\beta}(x,x') = [ g^{\mu\nu}(x), g^{\alpha\beta}(x')]$, which for spacelike
intervals between $x$ and $x'$ must be zero. But, as Wald remarks \cite{wald84}, an operator identity of form ${\cal C}^{\mu\nu\alpha\beta}(x,x') = 0$ cannot depend on what specific gauge one is in (ie., on the metric), and yet the question of whether the interval is spacelike or not depends on the metric. In a quantum  theory, one will in any case have some probability distribution for the different metrics, with both spacelike and timelike intervals.

Another way to think of this argument is to realize that in a path integral for quantum gravity, one must functionally integrate over different metric field configurations, with both spacelike and timelike intervals between 2 specific coordinates $x,x'$, and different lightcones centred on any coordinate for different paths.

\begin{figure}
		\includegraphics[width=7.2in]{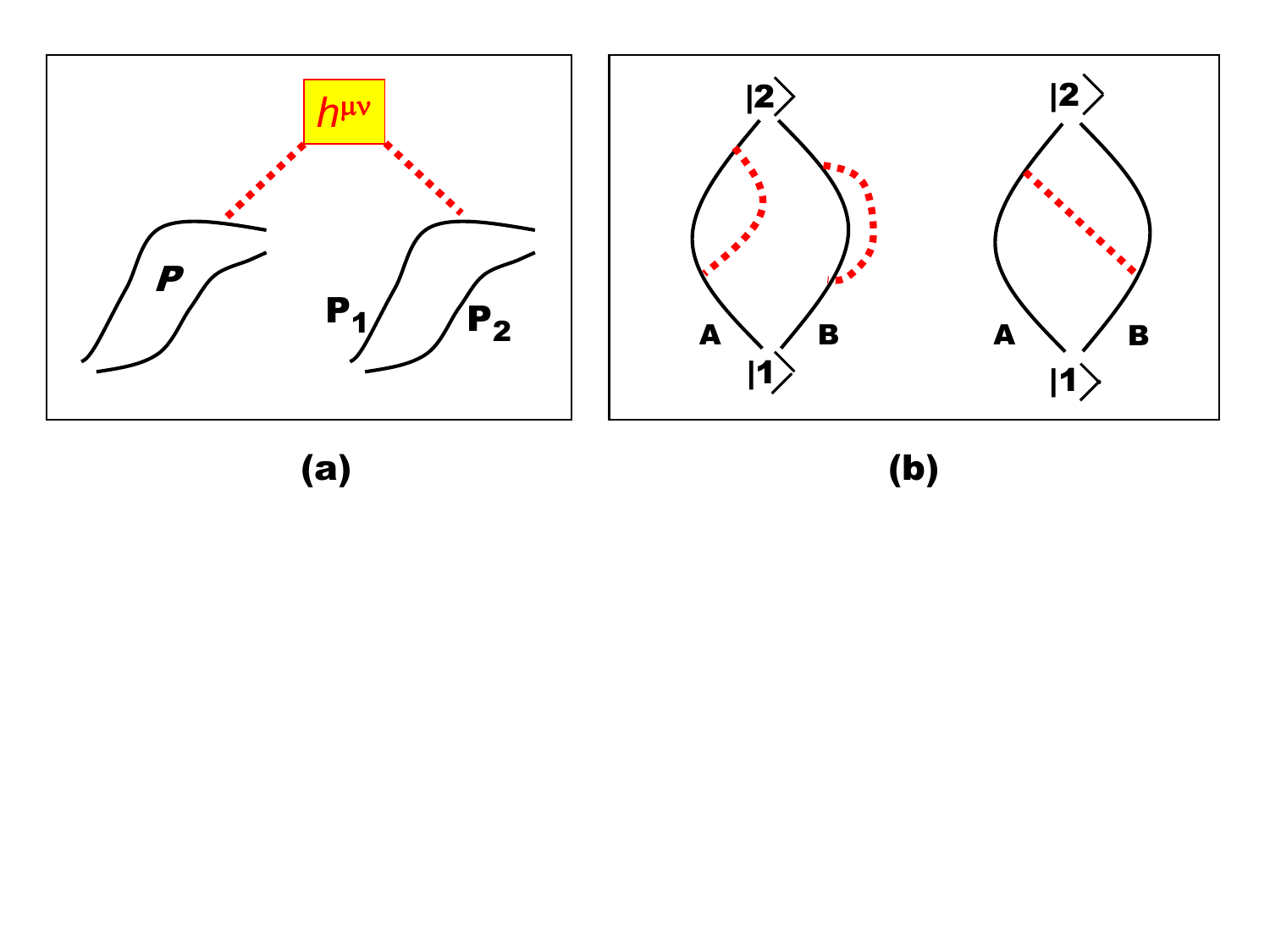}
   \vspace{-7.0cm}
		\caption{\label{fig:TDS12-2path}  Two-path thought experiments in quantum gravity. In (a) we compare a pair of paths for a single particle, with a pair of paths for 2 identical particles. The external gravitational field probe $h^{\mu\nu}(x)$ cannot distinguish them. In (b) we show, at left, a situation in standard QM where a single matter system can propagate along a pair of paths labelled A and B) from an initial state $|1 \rangle$ to a final state $|2\rangle$; each matter path is decorated with graviton lines, which contribute to the gravitational self-energy.  In (b) at right we show the same situation according to CWL theory, in which gravitational interactions between a pair of paths for the \uline{{\it same matter system}} are allowed.  } 	
	\vspace{4mm}
\end{figure}

Yet another way to think about this argument \cite{NJP15} is to imagine a 2-path experiment involving a massive object (Fig. \ref{fig:TDS12-2path}). In conventional QM this would be described by the superposition
\begin{equation}
|\Psi \rangle \;=\; a_1 |\Phi_1; \tilde{\mathfrak{g}}_{(1)}^{\mu \nu}(x) \rangle + a_1 |\Phi_2; \tilde{\mathfrak{g}}_{(2)}^{\mu \nu}(x) \rangle
 \label{GRsup}
\end{equation}
where those gravitational degrees of freedom that are tied to matter, in the gravitational part $|\tilde{\mathfrak{g}}^{\mu \nu}(x) \rangle$ of the state vector, are quantized, and completely entangled with $|\Phi\rangle$. Clearly, once we tie the metric to matter in this way, we must quantize the metric field. But now if we try to evaluate the interference term, we must look at the inner product $\langle\tilde{\mathfrak{g}}_{(1)}^{\mu \nu}(x) | \tilde{\mathfrak{g}}_{(2)}^{\mu \nu}(x') \rangle$. How are we to interpret or evaluate this?

All of these arguments tell us that there is a fundamental problem associated with the application of QM at {\it low energies} to spacetime, if we quantize it - and it becomes much worse for massive bodies.

\vspace{2mm}

	        %%%%%%%%%%%%%%%%%%%%%%%%%%%%%
	
	            %%%%%%%%%%%%%%%%%%%%%%%%%%%%%

So what do we do? An idea that goes back to Feynman \cite{RPF57} is to guess that the gravitational field as formulated in GR is not only incompatible with QM, but that in a proper theory, gravitation will cause a breakdown of QM. We avoid discussing the lengthy history of attempts to find a theory which tries to treat this idea, and instead just go directly to the theory that we favour.

%%%%%%%%%%%%%%%%%%%%%%%%%%%%%%%%%%%%%%%%%%%%%%%%%%%%%%%%%%%%%%%%%%%%%%%%%%%%%%%%%%
\subsection*{5.2. Correlated Worldline Theory}
%%%%%%%%%%%%%%%%%%%%%%%%%%%%%%%%%%%%%%%%%%%%%%%%%%%%%%%%%%%%%%%%%%%%%%%%%%%%%%%%%%

In what follows we focus on one particular candidate for a theory
of quantum gravity which is intended to address the incompatibilities
between QM/QFT and GR. This theory's basic structure is actually rather
rigid, in the sense that there are no adjustable parameters in it,
once one has accepted the initial assumptions.

%%%%%%%%%%%%%%%%%%%%%%%%%%%%%%%%%%%%%%%%%%%%%%%%%%%%
\subsubsection*{5.2(a) Basic Ideas of CWL Theory}
%%%%%%%%%%%%%%%%%%%%%%%%%%%%%%%%%%%%%%%%%%%%%%%%%%%%

In a series of papers (see refs. \cite{CWL1,CWL2}, finishing with ref. \cite{CWL3}) the ``Correlated Worldline'' theory was developed formally (with some of the ideas having been described in earlier papers \cite{NJP15,PCES12}). Here the discussion will be informal:
the ideal is to provide intuition rather than mathematics. We begin
from the view adopted in the previous section, viz., that the fundamental
objects in any quantum theory are paths/histories,
rather than state vectors in Hilbert space. Consider then the thought
experiment in Fig. \ref{fig:TDS12-2path}(a), where we compare (i) a system where we have
two identical particles of mass $M$, moving along 2 paths $q^{\mu}_{1}\left(\tau\right)$
and $q^{\mu}_{2}\left(\tau\right)$, and (ii) a single mass $M$ moving along
a superposition of these paths. In a path integral formulation of
QM, can an external probe gravitational field distinguish these
two cases? Since the field only sees $T_{\mu\nu}\left(x\right)$ ,
the answer is that it cannot: in both cases it sees a sum $\left(T_{\mu\nu}^{\left(1\right)}\left(q_{1}|x\right)+T_{\mu\nu}^{\left(2\right)}\left(q_{2}|x\right)\right)$.
Thus these two situations are \uline{equivalent}, in the same sense
as an accelerated frame and a frame in a gravitational field were
held to be equivalent by Einstein (although Einstein's equivalence
is actually quite misleading \cite{synge}.

As a consequence of this reasoning, because we know that the pair
of masses will interact via gravitation, then there must also be interactions
between a pair of paths A and B for a \uline{single} mass.  In \uline{conventional} quantum gravity,
one is only allowed diagrams like that shown in Fig. \ref{fig:TDS12-2path}(b), in which
a matter line emits and re-absorbs gravitons. However in CWL theory,
diagrams like that in Fig. \ref{fig:TDS12-2path}(c) must occur, in which we show the interaction between paths A and B. A little more thought
\cite{NJP15} then shows that there will be higher-order interactions
between $n$-tuples of matter lines, and a consistent treatment of this
in perturbation theory \cite{CWL3} shows that one can set up a ``diagrammar''
in which all non-zero contributions can be summed.

However the mathematical development of the CWL theory is not limited
to perturbative expansions in powers of the gravitational coupling
$G_{\text{N}}$ (Newton's constant). In fact it turns out that results
for the matter propagator (as well as other quantities like the generating
functional) can be obtained \uline{exactly} in CWL theory, which also
turns out to be a renormalizable theory. Moreover, the theory has
passed every consistency test that has so far been thrown at it. These
include:
\begin{enumerate}
\item If we take $\hbar\rightarrow0$ in CWL theory, we immediately recover
classical GR \cite{CWL2}
\item If we let $G_{\text{N}}\rightarrow0$ in CWL theory, we immediately
recover conventional QM or QFT in flat spacetime \cite{CWL2}.
\item The theory has consistent expansions to all orders in $\hbar$ (about
classical GR) and in $G_{\text{N}}$ (about conventional QM/QFT) \cite{CWL3,CWL2}
\item All relevant Ward identities are satisfied, including those connected
with diffeomorphisms; so all the relevant conservation laws are obeyed \cite{CWL2}.
\item Amongst the invariances following from the Ward identities, one may
also add Lorentz invariance in flat spacetime; and there is no superluminal
propagation of information in CWL theory.
\end{enumerate}
To understand all of this better, it is helpful to first discuss a
specific object in CWL theory, and then look at how the theory can be applied
to experiments.

%%%%%%%%%%%%%%%%%%%%%%%%%%%%%%%%%%%%%%%%%%%%%%%%%%%%
\subsubsection*{5.2(b) Propagator in CWL Theory}
%%%%%%%%%%%%%%%%%%%%%%%%%%%%%%%%%%%%%%%%%%%%%%%%%%%%

In conventional QFT, one has a propagator $K_{0}\left(2,1\right)$
for the propagation of some matter system between states $\left|1\right>$
and $\left|2\right>$. As we have seen above, states in a path integral formulation of QM/QFT can be
formulated as resulting from ``cuts'' in a ring propagator, these
cuts representing the interaction of teh system with some external
``measuring system''. It is helpful, in understanding CWL theory,
to see what we get.

Suppose we have some scalar field $\phi\left(x\right)$, and we wish
to determine its propagator between two configruations $\Phi_{1}\left(x\right)$
and $\Phi_{2}\left(x\right)$, on some background and fixed metric
$g_{0}$. This is just the question asked since the 1960's, about
the dynamics of quantum fields in a \uline{static} curved spacetime.
We write this (letting $\hbar=1$):
\begin{equation}
K_{0}\left(2,1|g_{0}\right)\;=\;\int_{\Phi_{1}}^{\Phi_{2}}\mathcal{D}\phi\,e^{iS_{\phi}\left[\phi,g_{0}\right]}\;=\; e^{i\psi_{0}\left(2,1|g_{0}\right)}  
 \label{eq: scalar field propagator}
\end{equation}
so that $\psi_{0}$ is the total phase accumulated between the two
states, which are assumed to be field configurations on spacelike surfaces
$\Sigma_{1}$ and $\Sigma_{2}$.

If we now go to conventional quantum gravity, this generalizes to 
\begin{align}
K\left(2,1\right) & =K\left(\Phi_{1}\Phi_{2};\mathfrak{h}_{2}^{ab}, \mathfrak{h}_{1}^{ab}\right)\nonumber \\
 & =\int_{h_{1}}^{h_{2}}\mathcal{D}g\,e^{iS_{\text{G}}\left[g\right]}\Delta\left(g\right)\delta\left(\chi^{\mu}\right)K_{0}\left(2,1|g\right)\nonumber \\
 & = \; e^{i\Psi_{0}\left(2,1\right)}
  \label{eq:Scalar field propagator in conventional QG}
\end{align}
where we functionally integrate over $g^{\mu\nu}\left(x\right)$,
using the Einstein action $S_{\text{G}}\left[g\right]$, between induced
metrics $\mathfrak{h}_{1}^{ab}$ and $\mathfrak{h}_{2}^{ab}$ on $\Sigma_{1}$ and $\Sigma_{2}$
respectively. Here $\Delta\left(g\right)$ is the Fadeev-Popov functional
determinant and $\chi^{\mu}\left(g\right)$ is a gauge fixing function.
Finally $K_{0}\left(2,1|g\right)$ is just the scalar field propagator
in some fixed field configuration $g=g^{\mu\nu}\left(x\right).$

The result for CWL theory is rather different. Let's start by defining
the ``conditional'' stress tensor
\begin{equation}
\chi_{\mu\nu}^{\mathfrak{T}}\left(2,1|x,g\right)=\frac{\left\langle \Phi_{2}\left|T_{\mu\nu}\left(x|g\right)\right|\Phi_{1}\right\rangle }{\left\langle \Phi_{2}|\Phi_{1}\right\rangle }
 \label{eq:conditional stress tensor}
\end{equation}
ie., a stress tensor defined at $x$ for a metric $g$, conditional
on the system beginning at state $\left|\Phi_{1}\right>$ and ending
at state $|\Phi_2 \rangle$. Alternatively, we can write it as
\begin{equation}
\chi_{\mu\nu}^{\mathfrak{T}}\left(2,1|x,g\right)=\frac{\int_{\Phi_{1}}^{\Phi_{2}}\mathcal{D}\phi e^{iS_{\phi}\left[\phi,g\right]}T_{\mu\nu}\left(x|g\right)}{\int_{\Phi_{1}}^{\Phi_{2}}\mathcal{D}\phi T_{\mu\nu}\left(x|g\right)}
 \label{eq:conditional stress tensor - path integral}
\end{equation}
ie., a normalized path integral conditional on $T_{\mu\nu}\left(x|g\right)$
being ``measured'' at $x$. Another way to think of this as a ``weak
measurement'' of $T_{\mu\nu}\left(x|g\right)$, or a ``2-time measurement''
(compare ref. \cite{drexel}). Note that $\chi_{\mu\nu}^{\mathfrak{T}}\left(x|g\right)$
is a complex quantity (a conditional amplitude).

The rather remarkable \uline{exact} result in CWL theory is that the
propagator for the scalar field is given by \cite{CWL3},
\begin{equation}
\mathcal{K}_{\text{CWL}}\left(2,1\right) \;=\; \exp i \left[\left(S_{\text{G}}\left[\overline{g}_{21}\right]+\psi_{0}\left(2,1|\overline{g}_{21}\right)\right)\right]
 \label{CWL-exact}
\end{equation}
and the analogue to Einstein's field equation is (again, $\kappa=8\pi G_{\text{N}}/c^{4}$):
\begin{equation}
G_{\mu\nu}\left(\overline{g}_{21}\left(x\right)\right)=\kappa\chi_{\mu\nu}^{\mathfrak{T}}\left(2,1|x,g\right)\label{eq:CWL Einstein Equations}
\end{equation}
where $\overline{g}_{21}\left(x\right)$ is the conditional metric
satisfying,
\begin{equation}
\frac{\delta}{\delta g}\left.\left[S_{\text{G}}\left[g\right]+\psi_{0}\left(2,1|g\right)\right]\right|_{g=\overline{g}_{21}}=0
 \label{g-21-cond}
\end{equation}

Let's understand all of this physically. We have now defined a new
spacetime, a quantum spacetime, which is \uline{complex}; the new
Einstein tensor can also be written as $G^{\mu\nu}\left(\overline{g}_{21}\left(x\right)\right)=R^{\mu\nu}\left(\overline{g}_{12}\right)-\frac{1}{2}R\left(\overline{g}_{12}\right)\overline{g}_{12}^{\mu\nu}$. We see
that $G_{\mu\nu}$ in CWL theory is \uline{not} the same as what one
finds in semiclassical gravity, where
\begin{equation}
G_{\mu\nu}\left[g\left(x\right)\right]=\kappa\left\langle T_{\mu\nu}\left(x|g\right)\right\rangle \label{eq:semiclasical einstein equation}
\end{equation}

Just how different equations (5) and (7) are from each other is seen
by considering again the 2-path experiment \cite{CWL3}. Denoting the two paths again by A and B respectively, we
note that in semiclassical gravity, a slow-moving point object sees the stress
energy expectation value $\left\langle T_{00}\left(x\right)\right\rangle =M\delta\left(\boldsymbol{r}-\frac{1}{2}\left|\boldsymbol{r}_{\text{A}}\left(t\right)+\boldsymbol{r}_{\text{B}}\left(t\right)\right|\right)$
where $\boldsymbol{r}_{\text{A}}\left(t\right)$ and $\boldsymbol{r}_{\text{B}}\left(t\right)$
are the two path coordinates. The semiclassical Einstein equation
then reads,
\begin{equation}
G_{\mu\nu}\left(g\left(\boldsymbol{r},t\right)\right)=\frac{\kappa}{2}\sum_{\alpha}^{\text{A,B}}\left\langle \boldsymbol{r}_{\text{\ensuremath{\alpha}}}\left(t\right)\left|T_{\mu\nu}\left(\boldsymbol{r},t\right)\right|\boldsymbol{r}_{\text{\ensuremath{\alpha}}}\left(t\right)\right\rangle \label{eq:Semiclassical Einstein Tensor}
\end{equation}
which is, of course real on both sides. The conditional correlator
$\chi_{\mu\nu}^{\mathfrak{T}}\left(2,1|x\right)$ is also real in semiclassical
theory:
\begin{equation}
\chi_{\mu\nu}^{\mathfrak{T}}\left(2,1|x\right)=\frac{1}{2}\left(T_{\mu\nu}^{\left(\text{A}\right)}\left(x\right)+T_{\mu\nu}^{\left(\text{B}\right)}\left(x\right)\right)\label{eq:Semiclassical conditional correlator}
\end{equation}
and does not depend on the initial and final states.

On the other hand, if we treat everything quantum-mechanically, we
have a quite different result for the stress energy expectation value:
instead of a result which is peaked at the \uline{mean} position of
the mass, we get $\left\langle T_{00}\left(x\right)\right\rangle =\ M\left[\delta\left(\boldsymbol{r}-\boldsymbol{r}_{\text{A}}\left(t\right)\right)+\delta\left(\boldsymbol{r}-\boldsymbol{r}_{\text{B}}\left(t\right)\right)\right]/2$,
with two peaks on the two different paths. The conditional correlator
is now complex
\begin{equation}
\chi^{\mathfrak{T}}\left(2,1|x\right)=\frac{1}{2}\left[\left(T_{\mu\nu}^{\left(\text{A}\right)}\left(x\right)+T_{\mu\nu}^{\left(\text{B}\right)}\left(x\right)\right)+i\left(T_{\mu\nu}^{\left(\text{A}\right)}\left(x\right)-T_{\mu\nu}^{\left(\text{B}\right)}\left(x\right)\right)\tan\left(\Delta S_{21}^{\text{AB}}\right)\right]\label{eq:CWL conditional correlator}
\end{equation}
where $\Delta S_{12}^{\text{AB}}$ is the phase difference between
paths A and B, along the trajectories from $\left|1\right>$ to $\left|2\right>$.
The resulting field equation is then complex, and is given by $\text{(\ref{eq:CWL Einstein Equations})}$
above.

If we now ask what the propagator for this system is, we find theree
different results for the propagator of our massive object, depending
on which theory we choose; we get $\mathcal{K}\left(2,1\right)=e^{i\Psi_{0}\left(2,1\right)}$
for conventional quantum gravity, $\mathcal{K}_{\text{sc}}\left(2,1\right)=e^{i\chi_{\text{sc}}\left(2,1\right)}$
for semiclassical gravity, and $\mathcal{K}_{\text{CWL}}\left(2,1\right)=e^{i\Phi_{\text{CWL}}\left(2,1\right)}$
for CWL theory. Both $\Psi_{0}\left(2,1\right)$ and $\chi_{\text{sc}}\left(2,1\right)$
are real, but they are not the same; and $\Phi_{\text{CWL}}\left(2,1\right)$
is \uline{complex}.

At first glance this makes one think of a simple decoherence effect,
but this is not actually what is going on - the key is that \uline{{\bf spacetime is now complex}}. This is to be expected - the metric is a quantum object, and so the amplitude for a given spacetime configuration will in general be complex. 
The imaginary part of $\Phi_{\text{CWL}}\left(2,1\right)$ comes entirely
from the interference between paths A and B (see equation $\text{(\ref{eq:CWL conditional correlator})}$
above), and its source can be pinned down to the gravitational interaction
taking place between paths A and B. Thus the imaginary part comes
from a \uline{quantum communication} between the two paths, mediated
by gravitation.

This of course is the way in which CWL theory violates the QM superposition principle, and leads to a non-linear form of QM. The inter-path communication is reminiscent of the description given by Polchinski \cite{polch91} of the physics arising in attempts to generalize QM to non-linear
versions of the theory, which we already discussed above (section 4.3(b)). Polchinski was particularly concerned with the
non-linear version of QM described by Weinberg \cite{weinberg89}, and he
described this communication between different branches or elements of a superposition as an ``Everett telephone''. Amongst other things, Polchinski's work showed that non-linear QM would have problems with causality - typically such theories allow superluminal communication.

% Preview source code from paragraph 0 to 13

It is therefore of considerable interest that CWL theory does \uline{not}
permit superluminal communication. This actually follows from the
proofs given \cite{CWL2} of the Ward identities for the theory, but
can also be shown by simply noting that one may define a complete
set of observables for CWL theory, and we find that the set of operators
$\langle \hat{O}\rangle_{\text{CWL}}$ corresponding to these is in the class
of operators excluded by Polchinski in his demonstration of superluminality \cite{polch91,polch09}.

%%%%%%%%%%%%%%%%%%%%%%%%%%%%%%%%%%%%%%%%%%%%%%%%%%%%
\subsection*{5.3. Experiments and Predictions of CWL Theory}
%%%%%%%%%%%%%%%%%%%%%%%%%%%%%%%%%%%%%%%%%%%%%%%%%%%%

It is interesting to continue with the technical details of CWL theory,
but these have been discussed fairly exhaustively in the original
papers \cite{CWL3,CWL1,CWL2}. What is more interesting is to ask
what kinds of experimental predictions are made by CWL theory. This
is not really a question about ``observables'' (a rather abstract notion
of more interest to mathematicians, at least in the context of conventional
QM).

It is instead a question about real experiments, and about how one
formulates the idea of a measurement in CWL theory.

A number of different kinds of real measurement, and predictions for
experiments, have already been described in CWL theory. These include
(1) optomechanical measurements of mirrors in cavities \cite{CWL4},
two-path experiments \cite{CWL3}, the dynamics of rotating macroscopic
bodies \cite{CWL-rot}, and ``4-path'' experiments \cite{CWL-4path}. Here
we will focus on the first two, which illustrate crucial features
of CWL theory.

%%%%%%%%%%%%%%%%%%%%%%%%%%%%%%%%%%%%%%%%%%%%%%%%%%%%
\subsubsection*{5.3(a): Extended Bodies in CWL Theory}
%%%%%%%%%%%%%%%%%%%%%%%%%%%%%%%%%%%%%%%%%%%%%%%%%%%%

Because any test of CWL theory has to be carried out on a massive
body, we need to work out the way in which different paths for spatially
extended bodies will interact gravitationally. This is a technical
exercise which requires detailed knowledge of the phonons in the body
(here assumed solid). The reason is that all of the gravitational
mass of the body is concentrated in the nuclei, and their motion (ie.,
their paths) are described by phonon collective coordinates. Individual
nuclear paths are then confined to oscillate over a length scale $\xi_{0}\left(T\right)$
, such that
\[
\xi_{0}^{2}\left(T\right)=\frac{\hbar^{2}}{2m}\int\frac{dE}{E}g\left(E\right)\left(1+2n\left(E,T\right)\right)
\]
where $g\left(E\right)$ is the phonon density of states, and $n\left(E,T\right)$
is the Bose distribution function.

	        %%%%%%%%%%%%%%%%%%%%%%%%%%%%%
	
\begin{figure}
		\includegraphics[width=5.2in]{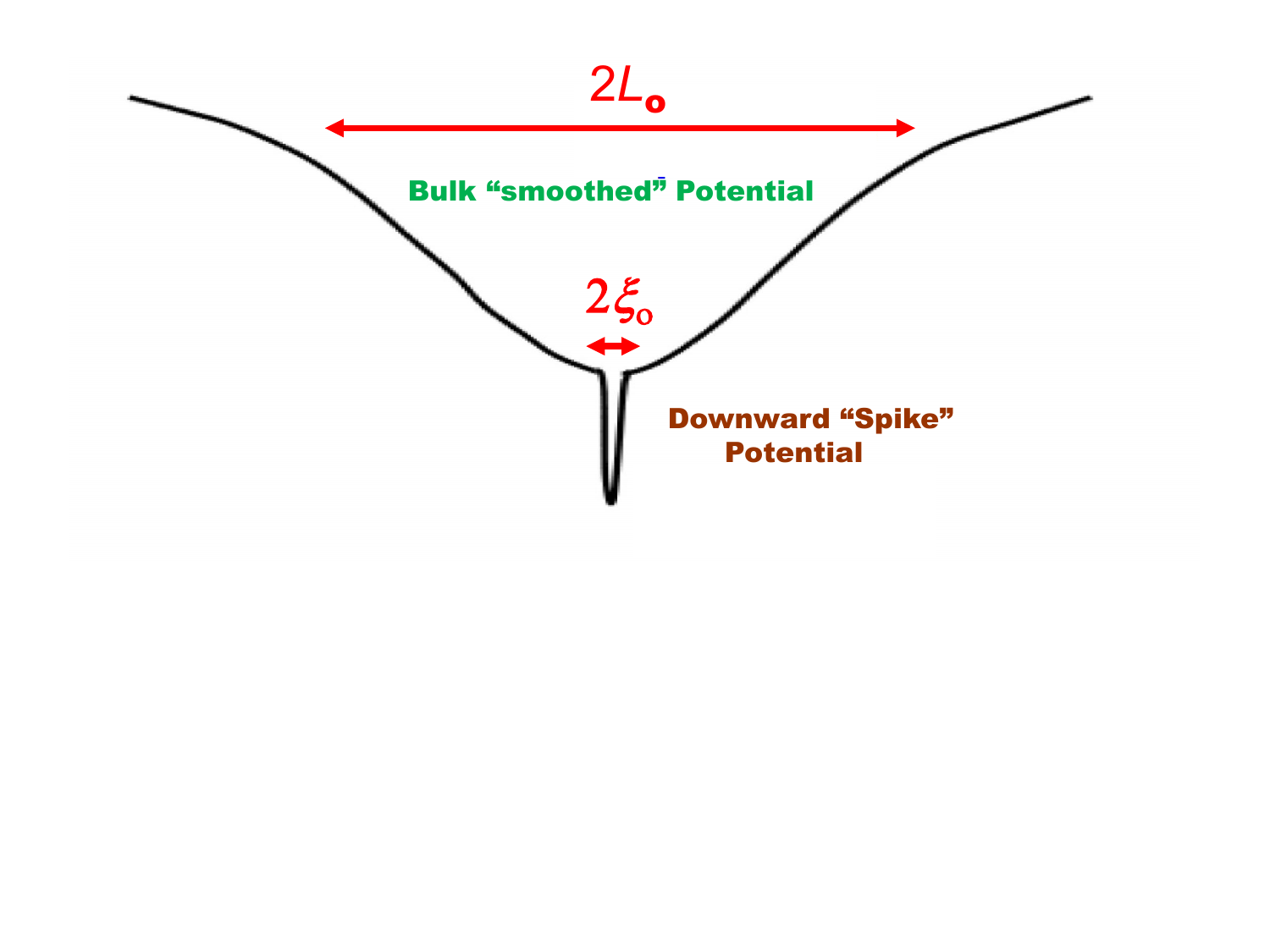}
\vspace{-4.4cm}
		\caption{\label{fig:TDS-F-Uplot}  A plot of the effective gravitational potential $U({\bf Q})$ between paths $\boldsymbol{R}_{0}\left(t\right)$ and $\boldsymbol{R}_{0}^{\prime}\left(t\right)$ of a massive extended body. Here ${\bf Q} \equiv (\boldsymbol{R}_{0}\left(t\right) - \boldsymbol{R}_{0}^{\prime}\left(t\right))$, and $\boldsymbol{R}_{0}\left(t\right)$ and $\boldsymbol{R}_{0}^{\prime}\left(t\right)$ are the centre of mass coordinates of the 2 paths. The sharp downward spike and the broad potential well have widths $2\xi_o$ and $L_o$ respectively (see main text).  } 
		\vspace{4mm}	
\end{figure}

	            %%%%%%%%%%%%%%%%%%%%%%%%%%%%%

The net interaction between a pair of paths for an extended mass is
then given by a function shown in Fig. \ref{fig:TDS-F-Uplot}, which we call $U(\boldsymbol{R}_{0},\boldsymbol{R}_{0}^{\prime})$; here $\boldsymbol{R}_{0}\left(t\right)$ and $\boldsymbol{R}_{0}^{\prime}\left(t\right)$
are the centre of mass coordinates of the body when following these
paths. The downward ``spike'' comes from paths where the nuclei
for one path overlap with the nuclei for the other path; their width
is $\sim2\xi_{0}$, which for typical solids is $\sim10^{-12}$m.
The broad potential well is produced by the interaction of the smoothed
mean distribution of the entire extended mass, on one path, with it's
counterpart on the other - so that this potential well is spread across
a length-scale $L_o$ equal to the size of the extended mass.

Detailed analysis shows that
\begin{enumerate}
\item For typical parameters which obtain
in macroscopic solid objects, one must go to low temperatures for the inter-path
interaction $U(\boldsymbol{R}_{0},\boldsymbol{R}_{0}^{\prime})$
to be relevant. For temperature $T\ll25^{\circ}$K, the two paths
will be closely bound to each other, within a length scale $\sim\xi_{0}$;
their paths will oscillate around each other at frequencies $\sim0.4$Hz,
in a good approximation to a harmonic potential (at the bottom of
the spike).
\item At room temperature, the two paths will move much further apart, and
the spike potential will have very little effect. The relative motion
$\boldsymbol{R}_{0}-\boldsymbol{R}_{0}^{\prime}$ will then be in the broad
smooth potential, with frequency $\sim10^{-3}$Hz.
\end{enumerate}

%%%%%%%%%%%%%%%%%%%%%%%%%%%%%%%%%%%%%%%%%%%%%%%%%%%%
\subsubsection*{5.3(b) Harmonic Experiments in CWL Theory}
%%%%%%%%%%%%%%%%%%%%%%%%%%%%%%%%%%%%%%%%%%%%%%%%%%%%

Consider first experiments on massive oscillating objects like mirrors, in which the centre of mass coordinate ${\bf Q}(t)$ moves in a harmonic potential.
From the results just given, it follows that when $T\ll25^{\circ}$K, or at room temperature, the motion of the relative
coordinate $\boldsymbol{z}\left(t\right)=(\boldsymbol{R}_{0}\left(t\right)-\boldsymbol{R}_{0}^{\prime}\left(t\right))$ between any two CWL paths $\boldsymbol{R}_{0}\left(t\right)$ and $\boldsymbol{R}_{0}^{\prime}\left(t\right)$ is also harmonic. Under these circumstances, one can show \cite{CWL4} that
the motion of the centre of mass coordinate ${\bf Q}(t)$ for
the massive object will be indistinguishable from the corresponding
motion of the oscillator in QM. This is because the CWL corrections - coming from the interactions between all of the different CWL paths - completely decouple from the motion of ${\bf Q}$.

Thus tests of CWL on simple oscillators are pointless; they give the same results as QM. One can try to evade this problem by employing highly non-Gaussian states in the experiment; or by doing pulsed experiments; or by trying to make the potential anharmonic. It remains to be seen if this can be done \cite{CWL4}.

Why is this conclusion important? Simply because any attempt to test
QM at the macroscopic scale with harmonic systems will be unable to
differentiate QM from CWL, and doubtless from other theories as well.
We believe this
point is important in view of claims \cite{LIGO20} that experiments
on the LIGO system have verified that QM works for the motion of 40
kg mirrors. All that these experiments have done is verify that a
certain class of theories can describe the mirror behaviour. That
multiple different theories can describe the same experiment is of
course very common in physics - a well known example is the way in
which the Lorentz-Fitzgerald approach gave identical results to Einstein's
special relativity, for the ensemble of electrodynamic experiments
known at that time.

Thus we can say pretty clearly that the LIGO experiments in no way
prove that QM is valid at the macroscopic scale, and that the mirrors are in a ``macroscopic quantum state". Other experiments will be necessary.

%%%%%%%%%%%%%%%%%%%%%%%%%%%%%%%%%%%%%%%%%%%%%%%%%%%%
\subsubsection*{5.3(c) 2-path Experiments in CWL Theory}
%%%%%%%%%%%%%%%%%%%%%%%%%%%%%%%%%%%%%%%%%%%%%%%%%%%%

One can fairly easily calculate the dynamics for a 2-path experiment in CWL theory, for a simple particle \cite{CWL3}. To adapt these calculations to an extended body is also straightforward. The main result is fairly simple to state. Suppose we have a transition amplitude, in ordinary QM, of form $K_o(2,1) = \langle {\bf r}_2, t_2|e^{-i Ht/\hbar} |\psi_1, t_1 \rangle$, for a system to pass from an `input' state $|\psi_1 \rangle$ to a final state at point ${\bf r}_2$ in real space. Suppose also that the system can follow one or other of 2 paths, labelled A and B.

Then the standard QM amplitude will take the form
\begin{equation}
K_o(2,1) \;=\; \langle {\bf r}_2, t_2|e^{-i Ht/\hbar} |\psi_1, t_1 \rangle \;\;\propto\;\; \sum_{\alpha}^{A,B} K^{(\alpha)}_o(2,1)
 \label{2path}
\end{equation}
where $K^{(\alpha)}_o(2,1)$ is the amplitude to go along path $\alpha$. However the amplitude in CWL theory will take the form
\begin{equation}
{\cal K}_{CWL}(2,1) \; \propto \; e^{i \Theta^{AB}_{21}} K_o(2,1)  \;\;\; \equiv  \;\;\; e^{i S_G [\langle \bar{g}_{21}^{AB} \rangle] }\sum_{\alpha}^{A,B} e^{i S_M[{\bf r}_{\alpha} | \langle \bar{g}_{21}^{AB} \rangle]}
 \label{CWL-2path}
\end{equation}
where the phase $\Theta^{AB}_{21}$ is {\it complex}, as was the conditional correlator for this same 2-path system (see eqtn. (\ref{eq:CWL conditional correlator}) above); the phase $\Theta^{AB}_{21}$ here is incorporated into the complex phase $\Phi_{CWL}(2,1)$ defined previously. We have written the answer in the final form in (\ref{CWL-2path}) so that one can see that the ``conditional" metric field $\langle \bar{g}_{21}^{AB} \rangle$, defined above in eqtn. (\ref{g-21-cond}), comes in. This field is sourced by both paths A and B simultaneously. The matter (ie., the moving massive object) is still propagating quantum-mechanically, in a superposition of paths A and B, in the presence of the same background conditional metric field $\langle \bar{g}_{21}^{AB} \rangle$.

It is very important that the effect of the CWL correlations is to multiply the conventional QM amplitude by the exponential of a global phase. This means the CWL correlations do not select out any one of the paths for special consideration. As we will show elsewhere, this leads to very interesting experimental possibilities,  which there is no space to discuss here. For the 2-path system, the main effect is to alter the form (but not the position) of the light and dark fringes in the 2-path interference [CWL3].

%%%%%%%%%%%%%%%%%%%%%%%%%%%%%%%%%%%%%%%%%%%%%%%%%%%
\subsubsection*{5.3(c) Measurements in CWL Theory}
%%%%%%%%%%%%%%%%%%%%%%%%%%%%%%%%%%%%%%%%%%%%%%%%%%%%

Until now we have discussed specific measurement schemes in CWL theory. But can one devise a general theory of measurements for CWL theory, analogous to that in orthodox QM? This is a topic of some complexity, which has only been discussed briefly so far \cite{NJP15}. The situation in CWL theory is similar to that in any QM theory in which measurements are treated as just another QM process, involving entanglement between a system S and some measurement apparatus ${\cal M}$ (in other words, there is no special ``measurement process", and there is no ``collapse of the wave-function"). 

However CWL theory is very different from QM in that is non-linear: however this non-linearity emerges only once one begins to displace macroscopic objects. Elsewhere we have discussed how big an object is required (see refs. \cite{CWL3,CWL4} and also \cite{NJP15}). But as we have just seen above, a mass displacement is not enough - a body controlled by a harmonic potential will show, in CWL theory, identical behaviour to that in QM, no matter how massive it is - one requires fundamentally non-harmonic systems (of which the 2-path system is a good example).  

Under these circumstances the key feature to pay attention to is the phenomenon of ``path-bunching" \cite{CWL3,CWL4}, whereby different paths for a massive object collapse towards each other onto one dominant trajectory. We note that path-bunching is \uline{{\bf not a decoherence process}}, and can happen quite independently of any environmental couplings, purely because of the graviational interaction between paths. It happens however at a rate determined both by coupling to the external environment and by the internal non-linear dynamics.

For a microscopic object the path-bunching time is astronomically long, and path-bunching is then completely obscured by conventional decoherence and dissipation processes. For a macroscopic object it can be quite short (although not for an oscillator). We now observe that if some microscopic system S then entangles with a macroscopic system ${\cal M}$ which is undergoing or has undergone path punching, then S will also be induced to path bunch with ${\cal M}$ \cite{NJP15}. Again we stress that this is not a process that can in principle happen without decoherence. Since ${\cal M}$ will very easily entangle with other objects, we see that this path-bunching chain will propagate ot a succession of other massive objects, and eventually to everything within the light cone of ${\cal M}$. 

We now see that CWL theory distinguishes between the purely quantum microscopic realm of superpositions, and the classical realm of apparently definite states, not on the basis of some bizarre (and arbitrary) idea like consciousness, or wave-function collapse, but instead on the completely objective feature of path bunching, appropriate to certain kinds of macroscopic object. The objects are distinguished by (i) a large gravitational mass, and (ii) a dynamics which is inherently non-harmonic. 

At this point one can, if one likes, look at different kinds of measurement scheme, and ask which ones satisfy these two criteria. These measurements will then lead to strong departures from orthodox quantum behaviour, and will look, for all practical purposes, like wave-function collapse. However we also see that one can no longer connect measurements in general with path bunching, and that macroscopic states that look quantum-mechanical are possible in CWL theory.

A detailed mathematical treatment of measurements on CWL theory, and the comparison  with QM, will appear elsewhere.

%%%%%%%%%%%%%%%%%%%%%%%%%%%%%%%%%%%%%%%%%%%%%%%%%%%%%%%%%%%%%%%%%%%%%%%%%%%%%%%%%%%%%%
%%%%%%%%%%%%%%%%%%%%%%%%%%%%%%%%%%%%%%%%%%%%%%%%%%%%%%%%%%%%%%%%%%%%%%%%%%%%%%%%%%%%%%

\section*{6) Conclusions}

%%%%%%%%%%%%%%%%%%%%%%%%%%%%%%%%%%%%%%%%%%%%%%%%%%%%%%%%%%%%%%%%%%%%%%%%%%%%%%%%%%%%%%
%%%%%%%%%%%%%%%%%%%%%%%%%%%%%%%%%%%%%%%%%%%%%%%%%%%%%%%%%%%%%%%%%%%%%%%%%%%%%%%%%%%%%%

It is clear from parts of this article that there was an interesting debate going on while it was being written,  between ST and PCES, about the use of analogies in physics. We chose to focus the debate somewhat on 2 key examples, viz., the analogy between 2-d superfluid films and QED, and the analogy between dilute BEC gases and curved spacetime. We note that all of these systems are interesting in their own right. Rather than hide our differences, or attempt to arrive at a synthesis (for which there was no time) we simply chose to air them. The result is very preliminary and we hope to return to this topic. 

Two things we were all agreed on were (i) that QM has serious problems, which one should not try to hide; and (ii) that these cannot be solved unless we modify QM in some way. We pointed out that arguments that are commonly used to dismiss non-linear modifications of QM are not nearly as general as often thought. We also pointed out that there are good reasons to believe an argument first suggested in 1957 by Feynman, that QM may break down for macroscopic systems because of gravity, We discussed a theory (the CWL theory) that implements this idea in a form which passes all theoretical consistency tests - the big test that it now needs to pass is the comparison with experiment. In future it will also be interesting to apply this theory to cosmological phenomena - work in this direction is already underway.  

It is unfortunate that we are unable to discuss the contents of this article with Starobinsky, since his work overlapped at one time or another with all of the themes discussed here, and he was interested in all of them. We hope that this article will help to sustain interest in these topics.

\vspace{4mm}

%%%%%%%%%%%%%%%%%%%%%%%%%%%%%%%%%%%%%%%%%%%%%%%%%%%%%%%%%%%%%%%%%%%%%%%%%%%%%%%%%%%%%%
%%%%%%%%%%%%%%%%%%%%%%%%%%%%%%%%%%%%%%%%%%%%%%%%%%%%%%%%%%%%%%%%%%%%%%%%%%%%%%%%%%%%%%

\section*{7) Acknowledgements}

%%%%%%%%%%%%%%%%%%%%%%%%%%%%%%%%%%%%%%%%%%%%%%%%%%%%%%%%%%%%%%%%%%%%%%%%%%%%%%%%%%%%%%
%%%%%%%%%%%%%%%%%%%%%%%%%%%%%%%%%%%%%%%%%%%%%%%%%%%%%%%%%%%%%%%%%%%%%%%%%%%%%%%%%%%%%%

Several parts of this article were produced by a lengthy and stimulating dialogue between two of us (PCES and ST), which we have tried to reproduce in the paper. One of us (PCES) would like to thank the other two (MJD and ST) for typing all his contributions, as well as doing the figures and references, since he is currently prevented from looking at computer screens. ST thanks W.G. Unruh, S. Weinfurtner, and members of the Weinfurtner group for discussions. MJDR thanks V. Milner and members of the Milner research group for discussions, and he thanks V. Milner for funding. Finally, PCES thanks J. Dalibard, M.O. Scully and C. DeWitt for discussions. This work was funded by NSERC (the National Science and Engineering Research Council) of Canada.

The article began with a collaboration between ST and PCES, and we were later joined by MJDR. Section 1 was written by PCES, section 2 by MJDR and PCES, section 3 by ST and PCES, sections 4 and 5 by PCES, and the appendix by ST and PCES. The article was proof-read by all of us, until we were all agreed on the contents.

%%%%%%%%

\vspace{4mm}

%%%%%%%%%%%%%%%%%%%%%%%%%%%%%%%%%%%%%%%%%%%%%%%%%%%%%%%%%%%%%%%%%%%%%%%%%%%%%%%%%%
%%%%%%%%%%%%%%%%%%%%%%%%%%%%%%%%%%%%%%%%%%%%%%%%%%%%%%%%%%%%%%%%%%%%%%%%%%%%%%%%%%

\section*{Appendix: Some Historical Remarks}

%%%%%%%%%%%%%%%%%%%%%%%%%%%%%%%%%%%%%%%%%%%%%%%%%%%%%%%%%%%%%%%%%%%%%%%%%%%%%%%%%%
%%%%%%%%%%%%%%%%%%%%%%%%%%%%%%%%%%%%%%%%%%%%%%%%%%%%%%%%%%%%%%%%%%%%%%%%%%%%%%%%%%

Two of us (PCES and ST) have spent some time looking at the historical development of  (i) ideas about decoherence, along with ideas about non-linear quantum mechanics and Everett's relative state theory, and (ii) the development of the Aharonov-Bohm effect and the way it was later understood in path integral theory. The is no space here (and currently we have no time, since the deadline for submitting this article is past!), to give more than a quick sketch of some highlights - a proper discussion will have to wait.

%%%%%%%%%%%%%%%%%%%%%%%%%%%%%%%%%%%%%%%%%%%%%%%%%%%%%%%%%%%%%%%%%%%%%%%%%%%%%%%%%%
\subsection*{A.1: Decoherence and Related Topics - Early History}
%%%%%%%%%%%%%%%%%%%%%%%%%%%%%%%%%%%%%%%%%%%%%%%%%%%%%%%%%%%%%%%%%%%%%%%%%%%%%%%%%%

From the very beginning of QM, it was realized that any extrapolation of the theory to the macroscopic scale would imply superpositions and interference at that scale. It is very well known that in the 1930s this led to a divergence between those like Bohr who felt that QM at the microscopic level had to be defined by classical physics at the macroscopic level (and who therefore rejected the use of QM in the classical world) and those like Schrodinger who deliberately extrapolated QM to the macroscopic scale (notably in the form of the Schrodinger Cat) to show the problems that such extrapolations led to. Von Neumann argued that the ``von Neumann chain" of measurement correlations could only be truncated by the introduction of consciousness.

Many of these ideas were summarized in the well known work of London and Bauer in 1938, and later in Bohm's 1951 text, where he analyzed what he called the “destruction of interference in the process of measurement”, i.e., the loss of phase relations when the system becomes correlated with a macroscopic apparatus.

In the 1950s and early 1960s interest grew in what later came to be called ``decoherence", in which the loss of phase coherence came not from interaction with a measuring system but instead with some generic environment. Amongst a large number of papers we can single out the work of Ludwig \cite{ludwig}, Green \cite{green}, and Daneri et al. \cite{daneri}, as well as the extensive discussions at Princeton involving, at different times, Everett and DeWitt \cite{everett,everettL57}, and the group of people surrounding Wigner \cite{wigner60s} (all of this after Bohm had left the USA for Brazil in Oct. 1951).

In 1970 Zeh \cite{zeh70} was able to gather these ideas into a general thesis, viz.,
that decoherence was a generic feature of quantum mechanics, able to explain why one
could not see macroscopic interference. Other papers (see, eg., Simonius
\cite{simonius78} and Peres \cite{peres80}) continued in this vein. But the
situation changed fundamentally with the publications of Leggett and co-workers,
starting in 1978 \cite{AJL78}, and continuing with work in 1980 \cite{AJL80+MQC}
and 1981-83 \cite{cal+AJL83}. This work accomplished two key steps. First, it argued,
contrary to all previous work, that one \uline{{\bf should}} be able to see
macroscopic superpositions in certain systems, if QM worked at the macroscopic scale
- indeed it advocated a search for these in superconducting SQUID devices. Second, it gave for the first time a detailed predictive theoretical framework for how to evaluate the effects of the environment - up to that point there had been no quantitative theory
of any kind.  This work then ushered in what one might call the ``modern period" of
decoherence studies, focussed on detailed predictions for experiments. It is
impossible to imagine any of the subsequent work in fields ranging from quantum
computation to quantum devices without this theoretical framework. Even a superficial review of this modern period would require an entire book. 

It is interesting to see how this early history is bound up with Everett and DeWitt's parallel work. As far as we know, Everett never met Bohm, and Bohm did not visit Princeton again until many years later, well after he had moved to the UK (curiously, Wheeler tried to organize a boycott of Bohm's lecture at Princeton, because of their political differences). However, although he never discussed hidden variables with Bohm, Everett was apparently familiar with Bohm's 1952 theory (regarding it as ``more cumbersome and artificial" than his own theory \cite{everettL57}). 

Apparently neither Everett nor DeWitt properly appreciated the key introduction of non-linearity into QM by Bohm's quantum potential, which fundamentally differentiates Bohm's theory from QM and from Everett's work. The essential point is very simple - if we write the wave function as $\psi ({\bf r}, t) = R({\bf r}, t) \, e^{iS/\hbar}$, and add to the Hamilton-Jacobi eqtn. for $S[{\bf r}, t]$ a new ``quantum potential" given by $Q(\mathbf{r}, t)= -(\hbar^2/2 m)(\nabla^2 R/R)$, along with a ``guiding equation" $m{\bf \dot{r}} = \nabla S$ for the particle, then the system becomes non-linear, and the particle follows a trajectory different from that given by the Hamilton-Jacobi equation without $Q({\bf r}, t)$ (this equation having been found by Schrodinger himself, in the led up to his discover of the Schrodinger wave equation). None of this was acceptable to Everett, for whom QM had to remain linear.  

One thing that would however have united Bohm and Everett very firmly, if they had ever met, was their attitude towards the Copenhagen interpretation and the Bohr-Rosenfeld school. Everett was openly disdainful (regarding it as a ``philosophical monstrosity" \cite{everettL57}). This
sentiment was returned by Rosenfeld (and Bohr's disapproval led Wheeler to withdraw his former enthusiastic support for Everett's ideas). And Bohm's 1952 departure from Copenhagen was complete and irrevocable.

%%%%%%%%%%%%%%%%%%%%%%%%%%%%%%%%%%%%%%%%%%%%%%%%%%%%%%%%%%%%%%%%%%%%%%%%%%%%%%%%%%
\subsection*{A.2: The Aharonov-Bohm effect}
%%%%%%%%%%%%%%%%%%%%%%%%%%%%%%%%%%%%%%%%%%%%%%%%%%%%%%%%%%%%%%%%%%%%%%%%%%%%%%%%%%

Only a few years after leaving Princeton, and by way of Brazil and Israel, Bohm
moved to Bristol in the UK. His peegrinations owed much to M.H.L. Pryce, who along
with Einstein had helped Bohm in his move to Brazil (Pryce having been on sabbatical
in Princeton when Bohm's problems started there). Pryce also helped Bohm get a position in Israel, and then, having become moved from oxford to becoome head of the physics dept. in Bristol, Pryce brought Bohm along with his student Aharonov to Bristol. At Bristol Bohm found a lively intellectual atmosphere, which
included P.K. Feyerabend; the intellectual debates amongst Pryce, Feyerabend, and Bohm, often took place in a weekly seminar led by Feyerabend, and led to a book by Bohm with a related long article by Feyerabend \cite{bohm-causality}, and eventually a book in which both Bohm and Pryce were involved \cite{toulmin62}. Much later, Pryce characterized the discussions at that time \cite{aha-UBC} as ``amongst the most exciting discussions of his life". 

Most famously, it was this atmosphere that produced (amongst other things) the extraordinary Aharonov-Bohm papers on the non-local aspects of QM \cite{AhB}. According to Aharonov \cite{aha-UBC} the formulation of the ``Aharonov-Bohm" effect suggested to him by Bohm (involving an infinitesimal flux tube) was too difficult for him to solve, so he went to Pryce, who then solved it on a blackboard. Pryce however refused to let Aharonov and Bohm put Pryce's name on the first paper, saying that what he had done was ``too trivial". There was considerable scepticism at the time about the effect: according to Pryce \cite{aha-UBC}, Bohr did not believe it existed at all, and a number of other authors were doubtful (including DeWitt \cite{DeWitt62} ).

It is therefore interesting that one of the first people to realize the full power of the Feynman path integral method was C. Morette, who married B. DeWitt in 1951 (and who  thereafter signed herself as either C.M. DeWitt, or as C. DeWitt-Morette, in her publications). Morette also overlapped with Feynman for several months during Feynman's 1951-52 sabbatical in Rio de Janeiro, by which time she had already made an important contribution to the subject of path integrals (in the form of the ``Morette/Pauli/van Vleck" formula). Years later, just before she and Bryce DeWitt left Chapel Hill for Austin Texas, she made a profound contribution to quantum mechanics, by inventing \cite{laidlaw}, along with her student Michael Laidlaw, what are now called ``anyons", ie., particles with fractional statistics. Curiously, according to Laidlaw, at no time during his PhD work did Morette discuss with Laidlaw her views on how one might discuss fractional statistics in terms of path integrals \cite{laidlaw-int}. The full story of how Morette discovered anyons, and the subsequent history, will have to await another article.

%%%%%%%%%%%%%%%%%%%%%%%%%%%%%%%%%%%%%%%%%%%%%%%%%%%%%%%%%%%%%%%%%%%%%%%%%%%%%%%%
%%%%%%%%%%%%%%%%%%%%%%%%%%%%%%%%%%%%%%%%%%%%%%%%%%%%%%%%%%%%%%%%%%%%%%%%%%%%%%%
%%%%%%%%%%%%%%%%%%%%%%%%%%%%%%%%%%%%%%%%%%%%%%%%%%%%%%%%%%%%%%%%%%%%%%%%%%%%%%%
%%%%%%%%%%%%%%%%%%%%%%%%%%%%%%%%%%%%%%%%%%%%%%%%%%%%%%%%%%%%%%%%%%%%%%%%%%%%%%%


\begin{thebibliography}{0}%
\makeatletter
\providecommand \@ifxundefined [1]{%
 \@ifx{#1\undefined}
}%
\providecommand \@ifnum [1]{%
 \ifnum #1\expandafter \@firstoftwo
 \else \expandafter \@secondoftwo
 \fi
}%
\providecommand \@ifx [1]{%
 \ifx #1\expandafter \@firstoftwo
 \else \expandafter \@secondoftwo
 \fi
}%
\providecommand \natexlab [1]{#1}%
\providecommand \enquote  [1]{``#1''}%
\providecommand \bibnamefont  [1]{#1}%
\providecommand \bibfnamefont [1]{#1}%
\providecommand \citenamefont [1]{#1}%
\providecommand \href@noop [0]{\@secondoftwo}%
\providecommand \href [0]{\begingroup \@sanitize@url \@href}%
\providecommand \@href[1]{\@@startlink{#1}\@@href}%
\providecommand \@@href[1]{\endgroup#1\@@endlink}%
\providecommand \@sanitize@url [0]{\catcode `\\12\catcode `\$12\catcode `\&12\catcode `\#12\catcode `\^12\catcode `\_12\catcode `\%12\relax}%
\providecommand \@@startlink[1]{}%
\providecommand \@@endlink[0]{}%
\providecommand \url  [0]{\begingroup\@sanitize@url \@url }%
\providecommand \@url [1]{\endgroup\@href {#1}{\urlprefix }}%
\providecommand \urlprefix  [0]{URL }%
\providecommand \Eprint [0]{\href }%
\providecommand \doibase [0]{https://doi.org/}%
\providecommand \selectlanguage [0]{\@gobble}%
\providecommand \bibinfo  [0]{\@secondoftwo}%
\providecommand \bibfield  [0]{\@secondoftwo}%
\providecommand \translation [1]{[#1]}%
\providecommand \BibitemOpen [0]{}%
\providecommand \bibitemStop [0]{}%
\providecommand \bibitemNoStop [0]{.\EOS\space}%
\providecommand \EOS [0]{\spacefactor3000\relax}%
\providecommand \BibitemShut  [1]{\csname bibitem#1\endcsname}%
\let\auto@bib@innerbib\@empty
%</preamble>
\end{thebibliography}%


\begin{thebibliography}{99}



\bibitem{staro16}     In writing this article we have availed ourselves of 7 hrs of recordings made by PCES in 2016 in Moscow, with A. Starobinsky.




\vspace{2mm}

\bibitem{staro-infl}   A.A. Starobinsky, Pis. Zh. Eksp. Teor. Fiz. {\bf 30}, 719 (1979) [J.E.T.P. Lett. {\bf 30}, 682 (1979)]




\vspace{2mm}

\bibitem{staro+YaZ}   A.A. Starobinsky, S.M. Churilov, Zh. Eksp. Teor. Fiz. {\bf 65},
65 (1973) [J.E.T.P. {\bf 38}, 1 (1974)].  
See also  Ya. B. Zel'dovich, ZhETF Pis. Red. {\bf 14}, 270 (1971)
[JETP Lett. {\bf 14}, 180 (1971)]; and Zh. Eksp. Teor. Fiz. {\bf 62},
2076 (1972) [Sov. Phys. JETP {\bf 35}, 108~, (1972)]




\vspace{2mm}

\bibitem{schwing51}   J. Schwinger, Phys. Rev. {\bf 82}, 664 (1951). See also J. Schwinger, Phys. Rev. {\bf 93}, 615 (1954), and Phys. Rev. {\bf 94}, 1362 (1954)



\vspace{2mm}

\bibitem{PWA63}      P.W. Anderson, Phys. Rev. {\bf 130}, 439 (1963)




\vspace{2mm}

\bibitem{hessF67}   G. B. Hess, W. M. Fairbank, Phys. Rev. Lett. {\bf 19}, 216
(1967)




\vspace{2mm}

\bibitem{onsP56}   O. Penrose, L. Onsager, Phys. Rev. {\bf 104}, 576 (1956). See also O. Penrose, Phil. Mag. {\bf 42}, 1378 (1951)




\vspace{2mm}

\bibitem{yang61}   C.N. Yang, Rev. Mod. Phys. {\bf 34}, 694 (1962)




\vspace{2mm}

\bibitem{PNAS25}   M.J. DesRochers, D. Marchand, P.C.E. Stamp, PNAS {\bf 122}, e2421273122 (2025)




\vspace{2mm}

\bibitem{bog47}   N.N. Bogoliubov, J. Phys USSR {\bf 11}, 23 (1947)




\vspace{2mm}

\bibitem{grossP}    E.P. Gross, Il Nouvo Cimento {\bf 20}, 454 (1961); L.P. Pitaevskii, Sov.Phys. J.E.T.P. {\bf 13}, 451 (1961)




\vspace{2mm}

\bibitem{wilks}       Wilks, ``{\it The Properties of Liqiuid and Solid Helium}", Clarendon Press, Oxford (1967)




\vspace{2mm}

\bibitem{tilley}      D.R. Tilley, J. Tilley, ``{\it Superfluidity and Superconductivity}", 3rd edition (CRC press, 1990)




\vspace{2mm}

\bibitem{rotonP}    A. Griffin, ``{\it Excitations in a Bose condensed liquid}" (Aambridge Univ. Press, 1993).




\vspace{2mm}

\bibitem{PNAS23}   A.A. Milner, P.C.E. Stamp, V. Milner, PNAS {\bf 120}, e2303231120 (2023)




\vspace{2mm}

\bibitem{hallock}   R.B. Hallock, J. Low Temp. Phys. {\bf 205}, 160 (2021)




\vspace{2mm}

\bibitem{LL-FLT}        The form here for the quasiparticle effective Hamiltonian is adapted to a strongly-correlated boson system from the famous Fermi liquid theory formulation of Landau. See, eg., K. Bedell, D. Pines, I. Fomin, J. Low Temp. Phys. {\bf 48}, 417 (1982); or K. Bedell, D. Pines, A. Zawadowski, Phys. Rev. B{\bf 29}, 102 (1984).



\vspace{2mm}

\bibitem{roton3}         D. S. Hirashima, T. Yamashita, F. Iwamoto, Phys. Rev. B{\bf 40}, 10891 (1989)



\vspace{2mm}

\bibitem{landau41}      L. D. Landau, J. Phys. (Moscow) {\bf 5}, 71 {1941); {\it ibid.} {\bf 11}, 91 (1947).



\vspace{2mm}

\bibitem{bowley}         R.M. Bowley, P.V.E. McClintock, F.E. Moss, G.G. Nancolas, P.C.E. Stamp, Phil. Trans. Roy. Soc. A{\bf 307}, 201 (1982).




\vspace{2mm}

\bibitem{reif}          G.W. Rayfield, F. Reif, Phys. Rev. {\bf 136}, A1194 (1964)




\vspace{2mm}

\bibitem{PCES79}        P.C.E. Stamp, P.V.E. McClintock, W.M. Fairbairn, J. Phys. C{\bf 12}, L589 (1979)




\vspace{2mm}

\bibitem{berez72}      V. L. Berezinskii, Sov. Phys. JETP {\bf 34}, 610 (1972).




\vspace{2mm}

\bibitem{KT}           J. M. Kosterlitz, D. J. Thouless, J. Phys. C{\bf 5}, L124 (1973), and J. Phys. C{\bf 6}, 1181 (1973); and J. M. Kosterlitz,
J. Phys. C{\bf 7}, 1046 (1974).





\vspace{2mm}

\bibitem{dalibard}     Z. Hadzibabic, J. Dalibard, Riv. del N. Cimento {\bf 34},  389 (2011)



\vspace{2mm}

\bibitem{TS12}         L. Thompson, P. C. E. Stamp, Phys. Rev. Lett. {\bf 108}, 184501 (2012).




\vspace{2mm}

\bibitem{Ons50}        L. Onsager, Il Nuovo Cimento. {\bf 6} (Suppl 2) (2): 279–287 (1949)




\vspace{2mm}

\bibitem{RPF-vort}     R.P. Feynman,  Prog. Low Temp. Phys. {\bf 1}, 17 (1955)




\vspace{2mm}

\bibitem{SHPMP06}     P.C.E. Stamp, Stud. Hist. Phil. Mod. Phys. {\bf 37}, 467 (2006)




\vspace{2mm}

\bibitem{FQHE}       See, eg., B.I. Halperin, J.K. Jain, ``{\it Fractional Quantum Hall Effects: New Developments}" (World Scientific, 2020)




\vspace{2mm}

\bibitem{Aharoni1}    A. Aharoni, ``{\it Introduction to the Theory of Ferromagnetism}", 2nd edition (Oxford Univ. Press, 1996)




\vspace{2mm}

\bibitem{PCES91}     P.C.E. Stamp, Phys. Rev. Lett. {\bf 66}, 2802 (1991)




\vspace{2mm}

\bibitem{Bark24}    C. Simon, D.M. Silevitch, P.C.E. Stamp, T.F. Rosenbaum, PNAS {\bf 121}, e2315598121 (2024)




\vspace{2mm}

\bibitem{donnellyV}    R. J. Donnelly, ``{\it Quantized Vortices in Helium II}" (Cambridge University Press, 1991).




\vspace{2mm}

\bibitem{soninV}    E. B. Sonin, ``{\it Dynamics of Quantized Vortices in Superfluids}" (Cambridge Unversity Press, 2016)




\vspace{2mm}

\bibitem{wyatt}    M Brown, A F G Wyatt, J. Phys. Condens. Matter {\bf 2}, 5025 (1990); see also
 A F G Wyatt, J. Phys. Condens. Matter {\bf 8}, 9249 (1996); A F G Wyatt, Phys. Scr. {\bf T49}, 59 (1993)





\vspace{2mm}

\bibitem{Anderson1995ScienceBEC}      M.H. Anderson, J.R. Ensher, M.R. Matthews, C.E. Wieman, E.A. Cornell, Science {\bf 269}, 198 (1995)




\vspace{2mm}

\bibitem{Davis1995PRLBEC}       K.B. Davis, M.-O. Mewes, M.R. Andrews, N.J. van Druten, D.S. Durfee, Kurn, W. Ketterle, Phys. Rev. Lett. {\bf 75}, 3969 (1995)




\vspace{2mm}

\bibitem{zwerger}      I. Bloch, Jean Dalibard, W. Zwerger, Rev. Mod. Phys. {\bf 80}, 885 (2008)




\vspace{2mm}

\bibitem{lopes17}      R. Lopes, C. Eigen, A. Barker, K. G. H. Viebahn, M. Robert-de-Saint-Vincent, N. Navon, Z. Hadzibabic, R. P. Smith, Phys. Rev. Lett. {\bf 118}, 210401 (2017)



\vspace{2mm}

\bibitem{andreev25}    S.V. Andreev, Phys. Rev. Lett. {\bf 134}, 166001 (2025)




\vspace{2mm}

\bibitem{Pitaevskii2003BECBook}    L.P. Pitaevskii, S. Stringari, ``{\it Bose-Einstein Condensation}", Oxford Univ. Press (2003)



\vspace{2mm}

\bibitem{Dalfovo1999RMP}    F. Dalfovo, S. Giorgini, L. P. Pitaevskii, S. Stringari, Rev. Mod. Phys. {\bf 71}, 463 (1999)



\vspace{2mm}

\bibitem{Hadzibabic2006Nature}     Z. Hadzibabic, P. Kruger, M. Cheneau, B. Battelier, J. Dalibard, Nature {\bf 441}, 1118 (2006)






\vspace{2mm}

\bibitem{vol-analog}     G.E. Volovik, Phys. Rep. {\bf 351}, 195 (2001)




\vspace{2mm}

\bibitem{steph09}    H. Stephani, D. Kramer, M. MacCallum, C. Hoenselaers, E. Herlt,  ``{\it Exact Solutions to Einstein's Field Equations}", 2nd edition (Cambridge Univ. Press, 2003)




\vspace{2mm}

\bibitem{BKL}       V.A.Belinsky, I.M. Khalatnikov, E.M.Lifshitz, Adv. Phy. {\bf 31}, 639 (1982), and references therein.





\vspace{2mm}

\bibitem{mixM}      C.W.Misner, Phys. Rev. Lett. {\bf 22}, 1071 (1969)





\vspace{2mm}

\bibitem{qingdi25}     Qingdi Wang, Zikun Zhao, P.C.E. Stamp, to be published.





\vspace{2mm}

\bibitem{martin12}    J. Martin, Comptes Rendus Physique {\bf 13}, 566 (2012).





\vspace{2mm}

\bibitem{heis}      W. Heisenberg, H. Euler, Z. Physik {\bf 98}, 714 (1936)





\vspace{2mm}

\bibitem{dyson51}      F.J. Dyson, Phys. Rev. {\bf 85}, 631 (1952)




\vspace{2mm}

\bibitem{dittrichB}    W. Dittrich, H. Gies, ``{\it Probing the Quantum Vacuum}" (Springer, 2000)




\vspace{2mm}

\bibitem{Gordon1923}     W. Gordon, Annalen der Physik {\bf 72}, 421 (1923)





\vspace{2mm}

\bibitem{Unruh1981PRL}      W.G. Unruh, Phys. Rev. Lett. {\bf 46}, 1351 (1981)



\vspace{2mm}

\bibitem{Painleve1921}       P. Painlevé, C. R. Acad. Sci. (Paris)} {\bf 173}, 677 (1921)



\vspace{2mm}

\bibitem{Jacobson1991PRL}     T. Jacobson, Phys. Rev. Lett. {\bf 75}, 1260 (1995)




\vspace{2mm}

\bibitem{Visser1995CQG}       M. Visser, Class. Q. Grav. {\bf 15}, 1767 (1995)





\vspace{2mm}

\bibitem{Schutzhold2002PRD}    R. Sch$\ddot{u}$tzhold, W.G. Unruh, Phys. Rev. D{\bf 66}, 044019 (2002)




\vspace{2mm}

\bibitem{Hawking1973LSS}     S.W. Hawking, G.F.R. Ellis, ``{\it The Large Scale Structure of Space-Time}", Cambridge University Press (1973)






\vspace{2mm}

\bibitem{Schutzhold2025EffectiveSpacetimes}     R. Sch$\ddot{u}$tzhold, preprint, 2025. 





\vspace{2mm}

\bibitem{Gooding2020PRL}       C. Gooding, S. Biermann, S. Erne,  J. Louko, W.G. Unruh, J. Schmiedmayer, S. Weinfurtner, Phys. Rev. Lett. {\bf 125}, 213603 (2020)





\vspace{2mm}

\bibitem{Gooding2025NondestructiveOptomechanicalBEC}      C. Gooding,  C. Bunney, r.D. Cameron,  S. Tajik, S. Erne, S. Biermann, J. Schmiedmayer, J. Louko, W.G. Unruh, S. Weinfurtner, /arXiv 2508.01080 (2025)





\vspace{2mm}

\bibitem{Unruh2022BlackHolesAcceleration}  W.G.  Unruh, J. Low Temp. Phys. {\bf 208},196 (2022)





\vspace{2mm}

\bibitem{FedichevFischer2003PRL}       P.O. Fedichev,  U.R. Fischer, Phys. Rev. Lett.
{\bf 91}, 240407 (2003)




\vspace{2mm}

\bibitem{Retzker2008PRL}       A. Retzker, J.I. Cirac, M.B. Plenio, B. Reznik,
Phys.  Rev.  Lett., {\bf 101}, 110402 (2008)



\vspace{2mm}

\bibitem{Steinhauer2016NatPhys}     J. Steinhauer, Nat. Phys. {\bf 12}, 959 (2016)




\vspace{2mm}

\bibitem{Jaskula2012PRL}     J.C. Jaskula, G.B. Partridge, M. Bonneau, R. Lopes, J. Ruaudel, D. Boiron, C.I. Westbrook,
Phys. Rev. Lett. {\bf 109}, 220401 (2012)





\vspace{2mm}

\bibitem{Jenkins2024PRD}      A. Jenkins, Phys. Rev. D{\bf 109}, 023506 (2024)



\vspace{2mm}

\bibitem{Gooding2024NJP}      C. Gooding,  New J. Phys. {\bf 26}, 105001 (2024)




\vspace{2mm}

\bibitem{Baak2023PetrovAnalogue}       S.-S. Baak, S. Datta, U.R. Fischer, Class. Q. Grav. {\bf 40}, 215001 (2023)







\vspace{2mm}

\bibitem{maxwellB}      J.C. Maxwell, ``{\it A Treatise on Electricity and Magnetism}", vols. I and II (Clarendon, Oxford, 1873)




\vspace{2mm}

\bibitem{huyghens95}    C. Huygens, ``{\it Traité de la Lumière}", (Leiden: Pieter van der Aa, 1690)




\vspace{2mm}

\bibitem{newton04}     I. Newton, ``{it Opticks}" (Royal Society, 1704)



\vspace{2mm}

\bibitem{penrose66}    Penrose discusses the optical analogy in several places - an example is R. Penrose, pp. 259-274 in ``{\it Perspectives in Geometry and Relativity: essays in honour of Vaclav Havaty}", ed. B. Hoffmann (Indiana Univ. Press, 1966). Note that this work was done in 1960-61.




\vspace{2mm}

\bibitem{penrose-int}   Interviews with R. Penrose, conducted between 2013-2019. 




\vspace{2mm}

\bibitem{bog58}       N.N. Bogoloiubov, V.V. Tolmachev, D.V. Shirkov, ``{\it A New Method in the Theory of Superconductivity}" (Consultants Bureau Inc., 1959)




\vspace{2mm}

\bibitem{PWA58}       P.W. Anderson, Phys. Rev. {\bf 112}, 1900 (1958)


\vspace{2mm}

\bibitem{higgs64}    P. Higgs, Phys. Rev. Lett. {\bf 13}, 508 (1964). See also F. Englert, R. Brout, Phys. Rev. Lett. {\bf 13}, 321 (1964); and G. Guralnik. C.R. Hagen, T.W.B. Kibble, Phys. Rev. Lett. {\bf 13}, 585 (1964)




\vspace{2mm}

\bibitem{Scie113}    Many cases of either scientific mistakes or fraud have been discussed in the litereature - we do not attmept a review here. Many universities have now introduced courses which discuss this, either in describing the ``sciemntifc method", or scientific practise. 





\vspace{2mm}

\bibitem{plasticF}   Eugenie S. Reich, ``{\it Plastic Fantastic: How the Biggest Fraud in Physics Shook the Scientific World}" (Pargrave MacMillan, 2009)






\vspace{2mm}

\bibitem{QM-text}   Some of the early texts which defined ``orthodox QM" were: P.A.M Dirac, ``{\it The Principles of Quantum Mechanics}", 1st edition (Oxford Univ. Press, 1930); W. Pauli, ``{\it The General Principles of Quantum Mechanics}", 1st eidtion (Springer 1934); L.D. Landau, E.M. Lifshitz, ``{\it Quantum Mechanics: Non-Relativistic Theory}", 1st English edition (Pergamon, 1958)   




\vspace{2mm}

\bibitem{vonN32}       J. van Neumann, ``{\it Mathematical Foundations of Quantum Mechanics}" (Princeton Univ. Press, 1955), translated from ``{\it Mathematische Grundlagen der Quantenmechanik}" (Springer, 1932).




\vspace{2mm}

\bibitem{bohm51}      D. Bohm, ``{\it Quantum Theory}" (Prentice-Hall, 1951)




\vspace{2mm}

\bibitem{laloe2}      F. Laloe, ``{\it Do we really understand Quantum Mechanics?}", 2nd edition (Cambridge Univ. Press, 2019).



\vspace{2mm}

\bibitem{AJL80+MQC}    A.J. Leggett, Sup. Prog. Th. Phys (Jap) {\bf 69}, 80 (1980); see also A.J. Leggett, Ch. 6, pp. 395-506 in ``{\it Chance and Matter}", Les Houches vol. XLVI (1986), ed. J. Souletie, J. Vannimenus, R. Stora (Elsevier, 1987).




\vspace{2mm}

\bibitem{RPF-QIP}     R.P. Feynman, Optics News, Feb 1985, pp. 11-20; see also R.P. Feynman, ``{\it Lectures on Computation}" (Addison-Wesley, 1996), and P. Benioff, Phys. Rev. Lett. {\bf 48}, 1581 (1982).




\vspace{2mm}

\bibitem{whaley+AJL}    J. I. Korsbakken, K. B. Whaley, J. Dubois, and J. I.
Cirac, Phys. Rev. A{\bf 75}, 042106 (2007); J. Korsbakken, F. Wilhelm, and K. Whaley, Europhys. Lett. {\bf 89}, 30003 (2010); T. J. Volkoff and K. B. Whaley,
Phys. Rev. A{\bf 89}, 012122 (2014).. See also G. C. Knee, K. Kakuyanagi, M.-C. Yeh, Y. Matsuzaki, H. Toida, H. Yamaguchi, S. Saito, A. J. Leggett, and W.
J. Munro, Nat. Commun. {\bf 7}, 13253 (2016); and A. J. Leggett,
arXiv:1603.03992.




\vspace{2mm}

\bibitem{taka09}         S. Takahashi, I. S. Tupitsyn, J. van Tol, C. C. Beedle, D. N. Hendrickson,  P. C. E. Stamp, Nature {\bf 476}, 76 (2011)




\vspace{2mm}

\bibitem{arndt14}        M. Arndt, K. Hornberger, Nat. Phys. 10, 271 (2014).




\vspace{2mm}

\bibitem{multiP}         R. Horodecki, P. Horodecki, M. Horodecki, K. Horodecki, Rev. Mod. Phys. {\bf 81}, 865 (2009)




\vspace{2mm}

\bibitem{cox18}          T. Cox, P.C.E. Stamp, Phys. Rev. A{\bf 98}, 062110 (2018)




\vspace{2mm}

\bibitem{cal+AJL83}    A.O. Caldeira, A.J. Leggett, Phys. Rev. Lett. {\bf 46}, 211 (1981);  A.O. Caldeira, A.J. Leggett, Ann. Phys. {\bf 149}, 374 (1983)




\vspace{2mm}

\bibitem{PS00}         N.V. Prokof'ev, P.C.E. Stamp, Rep. Prog. Phys. {\bf 63}, 669 (2000)




\vspace{2mm}

\bibitem{jammer}      M. Jammer, ``{\it The Philosophy of Quantum Mechanics}" (John Wiley and Sons, 1974)




\vspace{2mm}

\bibitem{londonB38}   F. London, E. Bauer, ``{La Theorie de l'Observation en Mecanique Quantique}" (Hermann et Cie, Paris,, 1938)




\vspace{2mm}

\bibitem{bellB}       J.S. Bell, ``{\it Speakable and Unspeakable in Quantum Mechanics}", 2nd ed. (Cambridge Univ. Press, 2004)




\vspace{2mm}

\bibitem{ABL64}         Y. Aharonov, P.G. Bergmann, J.L. Lebowitz, Phys. {\bf 134}, B1410 (1964)




\vspace{2mm}

\bibitem{drexel}        J. Dressel, M. Malik, F. M. Miatto, A. N. Jordan, and R.W.
Boyd, Rev. Mod. Phys. {\bf 86}, 307 (2014); L. Qin, W. Feng,
and X.-Q Li, Sci. Rep. {\bf 6}, 20286 (2016).                    p.19




\vspace{2mm}

\bibitem{delayZ-RMP}     X.S. Ma, J. Kofler, A. Zeilinger, Rev. Mod. Phys. {\bf 88}, 015005 (2016)




\vspace{2mm}

\bibitem{QEraser}       M.O. Scully, K. Drühl, Phys. Rev. A{\bf 25}, 2208 (1982): and  K. Yoon-Ho, R. Yu, S.P. Kulik, Y.H. Shih, M.O. Scully,  Phys. Rev. Lett. {\bf 84}, 1 (2000); X.S. Ma et al., PNAS {\bf 110}, 1221 (2013)




\vspace{2mm}

\bibitem{retroD}        For a review of ideas about retrocausality in QM, see S.Friederich, P.W. Evans, ``{\it Retrocausality in Quantum Mechanics}", Stanford Encyclopedia of Philosophy (Nov 13, 2023)




\vspace{2mm}

\bibitem{cramerTr}     J.G. Cramer, Rev. Mod. Phys. {\bf 58}, 647 (1986); R.E. Kastner, ``{\it The Transactional Interpretation of Quantum Mechanics: The Reality of Possibility}" (Cambridge Univ. Press, 2012). 




\vspace{2mm}

\bibitem{choice}     Clearly any theory that assumes that QM can be entirely described using the Schrodinger equation, or some other deterministic equation, will make the idea of ``choice" redundant. In recnet years this idea has been renamed ``super-determinism", and discussed by a number of authors - see eg., C.H. Brans,  Int. J. Theor. Phys. {\bf 27} 219 (1988); or S. Hossendfelder, T. Palmer, Frontiers Phys. {\bf 8}, 139 (2020) 




\vspace{2mm}

\bibitem{bohm52}      D. Bohm, Phys. Rev. {\bf 85}, 166 (1952), and {\it ibid} {\bf 85}, 180 (1952). See also D. Bohm, Phys. Rev. {\bf 87}, 389 (1953), and {\it ibid} {\bf 89}, 319 (1953) and {\bf 89}, 459 (1953). 




\vspace{2mm}

\bibitem{deBroglie}   L. De Broglie, Comptes Rendus {\bf 183}, 447 (1926); {\it ibid} {\bf 184}, 273 (1927); and L. de Brogie, J. de Physique et du Radium {\bf 8}, 225 (1927). 




\vspace{2mm}

\bibitem{manyW}      B. DeWitt, N. Graham (eds.), ``{\it The Many World Interpretation of Quantum Mechanics}" (Princeton Univ. Press, 1973), with reprints of several articles by DeWitt; see also B. DeWitt, Physics Today, {\bf 23}, 30 (Sept. 1970)




\vspace{2mm}

\bibitem{everett}     H. Everett, Rev. Mod. Phys. {\bf 29}, 454 (1957); and H. Everett, ``{\it The theory of the universal wave-function}" (PhD thesis, Princeton, 1957), reprinted in B. DeWitt, N. Graham \cite{manyW}.




\vspace{2mm}

\bibitem{deco-false}    That the interaction with the environment did not solve the problem of macroscopic interference terms was of course well understood by, eg., Einstein and Schrodinger in the 1930s. Considerable debate took place in Princeton after the war about this - this is reflected in Bohm's 1951 book \cite{bohm51}, in Everett's work \cite{everett} (see also a letter from Everett to Dewitt in 1957 \cite{everettL57}), and in lecture notes and articles by Wigner in the early 1960s \cite{wigner60s}. For more discussion of this see the Appendix.




\vspace{2mm}

\bibitem{everettL57}     H. Everett, letter to B. DeWitt, May 31 (1957)                     




\vspace{2mm}

\bibitem{wigner60s}     E.P. Wigner, ``{\it Symmetries and Reflections}", Indiana Univ. press (1967), contains several chapters by Wigner on foundational questions in QM. See also E.P. Wigenr, Am. J. PPhys. {\bf 31}, 6 (1963), and E.P. Wigner, The Monist {\bf 48}, 248 (1964).



\vspace{2mm}

\bibitem{morette}     C. Morette-DeWitt, Comm. Math. Phys. {\bf 28}, 47 (1972).




\vspace{2mm}

\bibitem{AhB}        Y. Aharonov, D. Bohm, Phys. Rev. {\bf 115}, 485 (1959); and {\it ibid} {\bf 123}, 1511 (1961)




\vspace{2mm}

\bibitem{RPF-vol3}    R.P. Feynman, R. Leighton, M. Sands, ``{\it The Feynman Lectures on Physics}" (Addison-Wesley, 1965), vol. 2, Ch. 15, and vol. 3, Ch. 21. 




\vspace{2mm}

\bibitem{laidlaw}     M.G.G. Laidlaw, C.Morette-DeWitt, Phys. Rev. D{\bf 3},
1375 (1971); see also the later paper of J.M. Leinaas, J. Myrheim, Nuovo Cim.
{\bf 37}B, 1 (1977)




\vspace{2mm}

\bibitem{YSWu84}     Yong-Shi Wu, Phys. Rev. Lett. {\bf 52}, 2103 (1984)




\vspace{2mm}

\bibitem{CTC}       J.B. Hartle, Phys. Rev. D{\bf 37}, 2818 (1988); Phys. Rev.
D{\bf 38}, 2985 (1988)(1994); and Phys. Rev. D{\bf 49}, 6543; G. Klinkhammer, K.S. Thorne, unpublished; and D.S. Goldwirth, M.J. Perry, T. Piran, K.S. Thorne, Phys.
Rev. D{\bf 49}, 3951 (1994).




\vspace{2mm}

\bibitem{CWL3}      J. Wilson-Gerow, P. C. E. Stamp, Phys. Rev. D{\bf 105},
084015 (2022).




\vspace{2mm}

\bibitem{kibble89}    T.W.B. Kibble, Comm. Math. Phys. {\bf 64}, 73 (1978);
T.W.B. Kibble, S. Randjbar-Daemi, J. Phys A{\bf 13}, 141
(1980)




\vspace{2mm}

\bibitem{weinberg89}   S. Weinberg, Phys. Rev. Lett. {\bf 62}, 485 (1989); S. Weinberg,
Ann. Phys. (NY) {\bf 194}, 336 (1989)




\vspace{2mm}

\bibitem{bell66}      See J.S. Bell, Rev. Mod. Phys. {\bf 38}, 447 (1966)




\vspace{2mm}

\bibitem{polch91}       J. Polchinski, Phys. Rev. Lett. {\bf 66}, 397 (1991);




\vspace{2mm}

\bibitem{kibble89b}     T.W.B. Kibble, Comm. Math. Phys. {\bf 65}, 189 (1979)




\vspace{2mm}

\bibitem{wald84}       R.M. Wald, ``{\it General Relativity}", Univ of Chicago press (1984), section 14.1




\vspace{2mm}

\bibitem{NJP15}      P. C. E. Stamp, New J. Phys. {\bf 17}, 065017 (2015);




\vspace{2mm}

\bibitem{RPF57}      See R. P. Feynman et al., in ``{\it The Role of Gravitation in Physics: Report from the 1957 Chapel Hill Conference}", ed.
C. M. DeWitt, D. Rickles (Max-Planck-Gesellschaft zur
Förderung der Wissenschaften, Berlin, 1957), particularly
session VIII, Secs. 22, 23.




\vspace{2mm}

\bibitem{CWL1}         A. O. Barvinsky, D. Carney, P. C. E. Stamp, Phys. Rev.
D{\bf 98}, 084052 (2018).




\vspace{2mm}

\bibitem{CWL2}         A. O. Barvinsky, J. Wilson-Gerow, P. C. E. Stamp,
Phys. Rev. D{\bf 103}, 064028 (2021).




\vspace{2mm}

\bibitem{PCES12}      P. C. E. Stamp, Phil. Trans. R. Soc. A{\bf 370}, 4429 (2012).




\vspace{2mm}

\bibitem{carney19}     D. Carney, P. C. E. Stamp, J. M. Taylor, Classical
Quantum Gravity {\bf 36}, 034001 (2019).




\vspace{2mm}

\bibitem{synge}        JL Synge, ``{\it Relativity: The General Theory; Pub}", North-Holland (1960). 




\vspace{2mm}

\bibitem{polch09}     On a historical note - it was actually J. Polchinski who first encouraged PCES to publish the CWL ideas in their preliminary form, during a visit by PCES to Santa Barbara in 2009. But at the time Polchinski never told PCES of the paper \cite{polch91} published in 1991!


\vspace{2mm}

\bibitem{CWL4}        J. Wilson-Gerow , Y. Chen, P. C. E. Stamp, Phys. Rev. D{\bf 109}, 064078 (2024)




\vspace{2mm}

\bibitem{CWL-rot}     P. Feldmann, S. Tajik, P.C.E. Stamp, to be published. 




\vspace{2mm}

\bibitem{CWL-4path}     J. Wilson-Gerow, Y. Malicki, M.O. Scully, P.C.E. Stamp, to be published. 





\vspace{2mm}

\bibitem{LIGO20}       Haocun Yu, L. McCuller, et al., Nature {\bf 583}, 43 (2020). to give some context to this work, see K. S. Thorne, in ``{\it The Future of Spacetime}", ed. S.W. Hawking, K. S. Thorne, I. Novikov, T. Ferris, A. Lightman (WW Norton, 2002), pp. 109–152; in particular, note “prediction No. 5”.




\vspace{2mm}

\bibitem{ludwig}       G. Ludwig, Zeit. Physik {\bf 135}, 483 (1953); {\it ibid} {150}, 486 (1958), {\it ibid} {\it 152}, 98 (1958). 




\vspace{2mm}

\bibitem{green}       H.S. Green, Nuovo Cim. {\bf 9}, 880 (1958)




\vspace{2mm}

\bibitem{daneri}     A. Daneri, A. Loinger, G.M. Prosperi, Nucl. Phys. {\bf 33}, 297 (1962), and Nuovo Cimento {\bf 44B}, 119 (1966). 




\vspace{2mm}

\bibitem{zeh70}        H.D. Zeh, Found. Phys {\bf 1}, 69 (1970)                




\vspace{2mm}

\bibitem{simonius78}     M. Simonius, Phys. Rev. Lett. {\bf 40}, 980 (1978)   




\vspace{2mm}

\bibitem{peres80}        A. Peres, Phys. Rev.  D{\bf 22}, 879 (1980)                p.28




\vspace{2mm}

\bibitem{AJL78}       A.J. Leggett, Journal de Physique {\bf 39}, C6-1264 (1978)




\vspace{2mm}

\bibitem{byrne}       For much more detail on Everett's work, and his life, see P. Byrne, ``{\it The Many Worlds of Hugh Everett III: Multiple Universes, Mutual Assured Destruction, and the Meltdown of a Nuclear Family}", Oxford Univ. Press (2010); and also ``{\it Many Worlds? Everett, Quantum Theory, and Reality}", ed. S. Saunders, J. Barrett, A. Kemt, D. Wallace (Oxford Univ. Press, 2010). 




\vspace{2mm}

\bibitem{bohm-causality}      D. Bohm ``{\it Causality and Chance in Modern Physics}" (Routledge, 1957); see also the discussion of P.K. Feyerabend, Brit. J. Phil. Sci. {\bf 10}, 321 (1960).




\vspace{2mm}

\bibitem{toulmin62}   ``{\it Quanta And Reality: A Symposium}", David Bohm, N. R. Hanson, Mary B. Hesse, Nicholas Kemmer, A. B. Pippard, Maurice Pryce (American Research Council, 1962)   




\vspace{2mm}

\bibitem{aha-UBC}    discussions between PCES, Y. Aharonov, and M.H.L. Pryce at UBC (1994)




\vspace{2mm}

\bibitem{DeWitt62}   B. DeWitt, Phys. Rev. {\bf 125}, 2189 (1962)




\vspace{2mm}

\bibitem{laidlaw-int}   Interviews with M. Laidlaw, 9th Nov amd 14 Dec, 2020 (H. Brown also attending). 














\end{thebibliography}
\end{document}